\documentclass[9pt,twocolumn,twoside]{pnas-new}

\templatetype{pnasresearcharticle} 

\usepackage{booktabs}   
\usepackage{multirow}   
\usepackage{array}  

\begin{document}

\title{Dynamics of collective minds in online communities}

\author[a, 1]{Seungwoong Ha}
\author[a,b]{Henrik Olsson}
\author[c]{Kresimir Jaksic}
\author[d]{Max Pellert}
\author[a,b,e]{Mirta Galesic}

\affil[a]{Santa Fe Institute, Santa Fe, NM 87501, USA}
\affil[b]{Complexity Science Hub Vienna, 1080 Vienna, Austria}
\affil[c]{Department of Psychology, University of Zadar, 23000 Zadar, Croatia}
\affil[d]{Barcelona Supercomputing Center, 08034 Barcelona, Spain}
\affil[e]{Vermont Complex Systems Institute, University of Vermont, VT 05405, USA}

\leadauthor{Ha}

\significancestatement{Online communities shape how societies think, argue, and act, yet we still lack tools to understand how their shared “collective minds” evolve and can be influenced. Using data from millions of online discussions, we develop a model that captures how news selection, user behavior, and community feedback jointly shape the shared understandings. The results reveal that small editorial or social shifts can produce lasting changes in public understanding, while others can be rapidly undone. This work introduces a quantitative framework for tracing the dynamics of collective cognition, offering a path to detect manipulation and foster more resilient public discourse in digital societies.}

\authorcontributions{S.H. constructed the model, performed the simulation, and conducted the analysis; K.J. and M.P. contributed to the analysis of the empirical data; H.O. and M.G. collected the data and supervised the project; S.H., H.O., and M.G. wrote the manuscript with input from all authors.}
\authordeclaration{The authors declare no conflict of interest.}
\correspondingauthor{\textsuperscript{1}To whom correspondence should be addressed. E-mail: seungwoong.ha@santafe.edu}


\begin{abstract}
Collective discourse and action are driven by collective minds. These shared semantic representations and related processes shape societal responses to critical societal challenges such as climate change and political upheavals. In online communities, collective minds are susceptible to the influences of editorial practices and community dynamics, making them vulnerable to manipulation. However, understanding these influences is difficult because of the limits of experimenting with and predicting complex social systems. Here, we develop a computational model of collective minds, calibrated and validated with data from 400 million comments across five U.S. online news platforms and a survey. Our model enables us to quantitatively describe and experiment with different editorial agenda-setting practices and aspects of community dynamics to understand how they shape the collective mind. We find that some editorial influences can be reversed relatively rapidly, but others, such as amplification and reframing of certain topics, as well as community influences such as trolling and counterspeech, tend to persist and durably change the collective mind. These findings illuminate ways collective minds can avoid manipulation and pathways for communities to maintain healthy and authentic collective discourse amid ongoing societal challenges.
\end{abstract}

\dates{This manuscript was compiled on \today}
\doi{\url{www.pnas.org/cgi/doi/10.1073/pnas.XXXXXXXXXX}}

\maketitle
\thispagestyle{firststyle}
\ifthenelse{\boolean{shortarticle}}{\ifthenelse{\boolean{singlecolumn}}{\abscontentformatted}{\abscontent}}{}

\dropcap{H}ow communities respond to diverse societal challenges, from economic crises to political upheavals \cite{andre2024narratives,bond2023rise,card2022computational,flottum2017narratives,jing2021characterizing,lee2022storm,MuellerSchwarzJEEA2020}, is shaped by their  shared representations of current issues, continuously modified by a stream of internal and external influences \cite{stanley2s015propaganda,brown2014human,tollefsen2006extended,shteynberg2023theory,jimenez2024effect}. This dynamics is amplified in online communities~\cite{Tausczik2020,vanbavel2024social,zhuravskaya2020political,bak2021stewardship}, where understanding these influences is essential for maintaining healthy discourse and resisting manipulation. However, understanding is hindered by the limits of prediction of complex systems and the inability to conduct counterfactual experiments with human collectives \cite{miller2009complex,hofman2017prediction}.

Here we use the term "collective minds" \cite{tollefsen2006extended, shteynberg2023theory} to describe and model a dynamic system of semantic networks of representations a community is sharing \cite{Collins1975,Griffiths2007,kumar2022critical}, and processes that generate and update these networks over time. This system does not necessarily imply a homogeneous group mind or a single cognitive entity. Rather, collective minds emerge from the dynamic interplay of a semantic network and processes on it. The network representation reflects how issues are connected in a particular community and how they and their relationships change over time as new information is introduced and discussed. The processes on this network include shared attention to different issues, discussions about these issues, updating of relationships between their representations through repeated co-occurrence and forgetting, and feedback from the resulting semantic network to future attention and discourse. 

Different aspects of collective minds have been explored across numerous disciplines, yet our understanding of how various influences shape their dynamics remains limited. Traditional research on collective minds typically involved simple human experiments or observational studies of small groups \cite{shteynberg2023theory,weick1993collective}. While these studies produced valuable qualitative insights, they did not capture the large-scale dynamics of collective processes. Research on semantic networks has mainly investigated structural properties related to individual language and memory \cite{morais2013mapping,hills2009longitudinal}, with less emphasis on collective semantic dynamics arising from community interactions \cite{kumar2022critical,siew2019cognitive}. Beyond individual semantic networks, large-scale textual analyses of semantic and topic networks partially capture collective semantic evolution \cite{steyvers2007probabilistic,vayansky2020review}, although they often do not fully account for the dynamic shaping by internal and external influences. Similarly, generative models of online discussions have examined interactions between user behavior and conversational patterns \cite{aragon2017generative,bollenbacher2021challenges}, but have not investigated the semantic relationships among topics that constitute collective minds. Influences from editorial agenda-setting \cite{Coleman2009,Groeling2013} and community dynamics \cite{Ksiazek2015,buerger2019counterspeech} have typically been studied independently, without modeling and integrating how they jointly shape collective semantic networks \cite{soroka2012gatekeeping}.

\begin{figure*}
    \centering
    \includegraphics[width=0.82\linewidth]{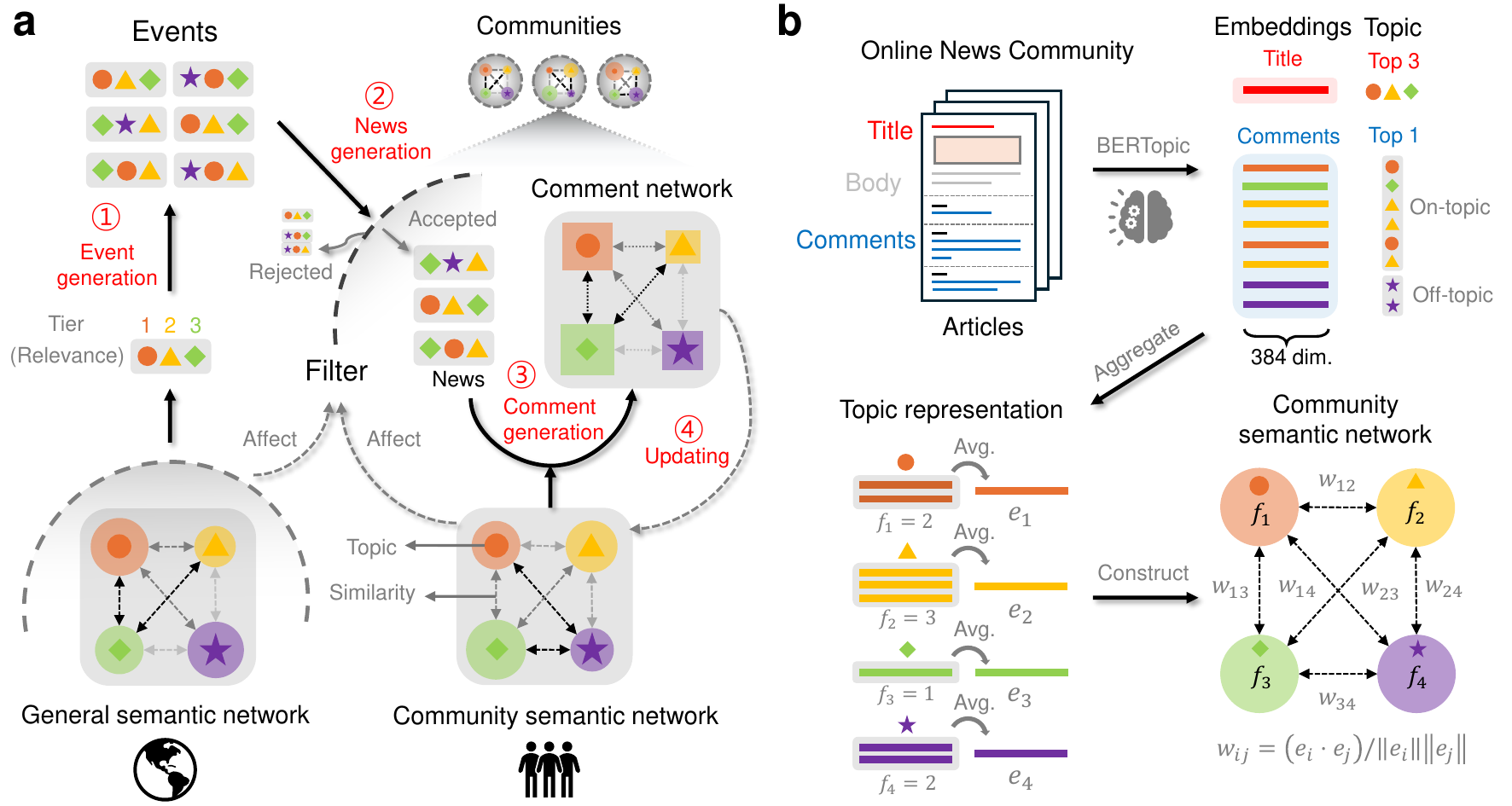}
    \caption{\textbf{Computational model and empirical data.} \textbf{a}, Conceptual illustration of the computational model of collective minds as dynamic semantic networks, developing through a sequence of four processes illustrated by the four numbered steps in red. (1) Event generation: The world that generates events is characterized by the general semantic network. Events are represented by a triplet of topics (here symbolized by geometric shapes), in which the topic that best describes the event is in the first place, followed by two other related but less relevant topics (tiers $1$, $2$, and $3$, respectively). (2) News generation. At each time step, communities are exposed to the same set of events. Each community has an editorial filter that accepts or rejects events, affected by both general and community semantic networks. The accepted events become news published on the news site of the community. (3) Comment generation. The community semantic network responds to this news through its comment section, which is characterized by a network of interrelated topics.(4) Updating. Finally, the community semantic network is updated based on the feedback from the comment network, which will affect the filtering process of the next time step. \textbf{b}, Data collection for calibrating the computational model. First, we gather titles and comments from online news articles, get their BERT embeddings, and use BERTopic to derive topics. We characterize the title by a triplet of topics that best describe it, and each comment by its most relevant topic. For a given time interval, we count the number of comments discussing each topic ($f_i$) and average the embeddings of all such comments to get the topic representation ($e_i$). Finally, we assign weights ($w_{ij}$) for each pair of topics as a cosine similarity of their representations.}\label{fig:1}
\end{figure*}

Here, we investigate how collective minds in online news communities can be influenced by different editorial agenda-setting practices and aspects of community dynamics, and how these influences can be reversed. We develop a computational model that represents collective minds as dynamic semantic networks responding to a constant influx of information (Fig. \ref{fig:1}). The model is grounded in existing knowledge about the dynamics of semantic networks \cite{Siew2019}, dynamics of online communities \cite{baran2010dynamics}, editorial policies \cite{Coleman2009,Perloff2022,Rodrigo-Gines2024,Groeling2013}, and community influences \cite{FIEDLER20142258}. To ensure that it accurately captures real-world dynamics, we calibrate and validate the model using semantic networks derived from longitudinal data comprising millions of comments posted in comment sections of online news sites across the US political spectrum (from Mother Jones, The Atlantic, and The Hill, to Breitbart and Gateway Pundit), verified through a survey ($N = 1{,}022$). Comment sections are a common venue for online citizen engagement.  Over $50\%$ of Americans comment on news articles monthly, and $78\%$ read comments \cite{Stroud2016}. In this way, they express and share their representations of current events, shaping the collective minds of their communities. The comment sections act as microcosms of larger societal debates, although they can sometimes devolve into uncivil battlegrounds \cite{Rossini2022,fraxanetUnpackingPolarizationAntagonism2024a}.

Our model is designed to capture the dynamics of online communities as simply as possible while still reflecting the essential components of these systems: external events, editorial boards, communities, and the feedback loops between them. We do not aim to provide an analytic solution or a predictive model in the narrow sense. Rather, the goal is to provide a flexible framework that can reproduce the primary statistical properties of empirical data and facilitate an exploration of how various influences shape collective minds.

\begin{figure*}
\centering
\includegraphics[width=\linewidth]{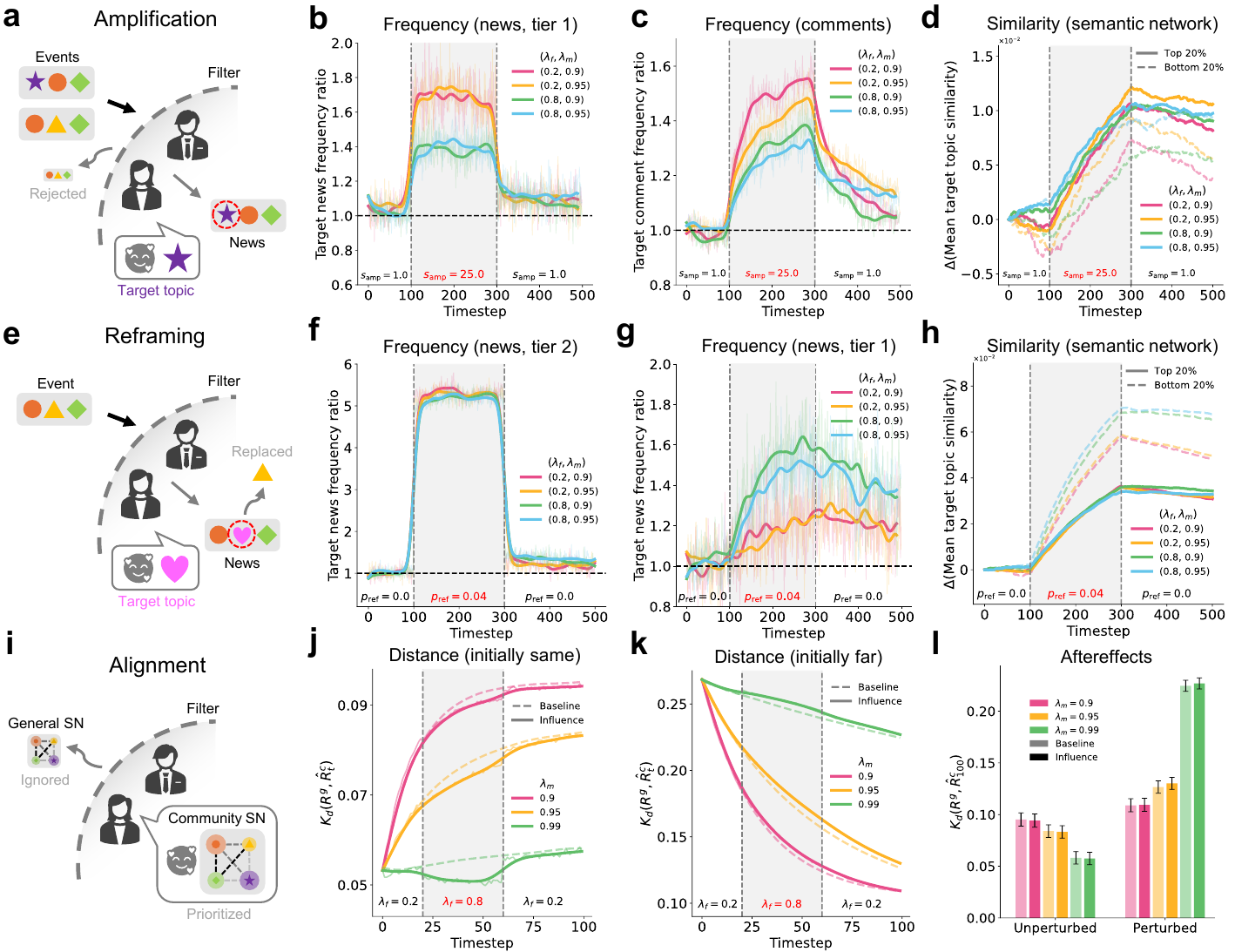}
\caption{\textbf{Impact of editorial influences on the community semantic networks in the model.} \textbf{a-d}, Amplification is modeled as a subjective increase ($s_{\text{amp}}$) in the frequency of the target topic in the general semantic network, as perceived by the editors (\textbf{a}). It increases the target topic's frequency in the news (\textbf{b}) and in the comments (\textbf{c}) across all filter strengths, and this effect persists even after removal. It also increases the similarity between the target topic and other topics, especially for the initially more similar topics (top vs. bottom $20\%$; Fig. \ref{fig:4}d). \textbf{e-h}. Reframing replaces one of the topics in the news that has passed the filter by a target topic (\textbf{e}), with probability $p_{\text{ref}}$. When applied to topics in tier 2 of news, it increases the frequency of that topic in tier 2 (\textbf{f}) and, over time, in tier 1 (\textbf{g}) as well, with both effects persisting after reframing is removed. It also increases the similarity between the target topic and other topics, especially for topics that were initially less similar (\textbf{h}). \textbf{i-l}, Alignment is represented as the strength of the community filter ($\lambda_f$, \textbf{i}). It slows down the movement of the comment network relative to the general one, keeping it in its initial position. When the initial position of the community semantic network is the same as (far from) that of the general one, alignment makes the comment network more (less) similar to the general one (\textbf{j-k}) for all memory strengths ($\lambda_m$), as measured by Kendall-tau rank distance between the networks. The effect quickly disappears once the alignment is removed (\textbf{l}). The error bars indicate $\pm1$ standard deviations across $100,000$ simulations. All ratios and differences are relative to the baselines without external influences. Semi-transparent lines represent raw data, and solid lines indicate denoised data, except in \textbf{d, h}, where all lines represent raw data.}\label{fig:4}
\end{figure*}

We provide quantitative descriptions of how editorial and community-level influences shape collective minds over time (Table \ref{table:1}), which vary with community characteristics. We find that even brief influences can lead to enduring shifts in collective minds, and show that targeted, well-timed actions by platform designers and communities can reduce or even reverse those shifts (Figs. \ref{fig:4} and \ref{fig:5} and Discussion). We finally demonstrate our model's validity by showing that it successfully reproduces dynamics observed in online news communities (Fig \ref{fig:3}).

\section*{Model construction and empirical data collection}

The world we live in generates a continuous stream of a variety of events. The editorial board of each news community curates these events and publishes them as news. Community members post comments about them, creating and evolving their collective mind. We model this process as an interplay of a general semantic network that characterizes the real world where events occur, a community semantic network that captures the characteristics of a specific community and its comments, and a community-specific editorial filter at the interface between the general and community semantic networks (Fig. \ref{fig:1}).  

The collective mind is the system of the community semantic network and internal and external processes that change the network’s structure over time. In the model, the semantic network is a complete graph with topics as nodes, topic frequency as node weights, and inter-topic similarity as edge weights (see Methods). Edge weights are updated using a Hebbian learning-inspired rule: topics that repeatedly co-occur in comments become more strongly associated, while less-reinforced associations decay over time. We cannot directly observe internal semantic networks, so we model the evolution of collective minds through the comments community members make. The semantic networks derived from these expressions serve as proxies for the underlying cognitive representations and processes. Also, it is important to note that comments reflect the views of those who actively post and may differ from the opinions of readers who do not comment. Prior research shows systematic differences between commenters and readers; we therefore interpret comments not as direct measures of community-wide beliefs but as influential expressions that help shape the collective mind on online platforms.

The model involves four processes (Fig. \ref{fig:1}): (1) event generation: real-world events are generated and described as a tuple of $N_w=3$ topics ordered from most to least relevant to the event (tiers $1$, $2$, and $3$, respectively; for instance, adoption of new vaccine policy for COVID-19 might be 'epidemics', 'vaccine', and 'government'), (2) news generation: the editorial filter of each community stochastically determines which events will get posted as news, affected by both general and community semantic networks, (3) comment generation: the community semantic network responds to the news by posting comments, constituting the comment network, and (4) updating of community semantic network: the comment network updates the community semantic network of the next time step through changes in topic frequencies and a Hebbian learning-inspired update of topic associations, in which co-occurring topics become more strongly associated while less reinforced associations decay over time (see Methods and SI Appendix, section 1 for details).

Two main control parameters determine the characteristics of the community. The filter strength, $\lambda_f \in [0, 1]$, effectively functions as a gatekeeper that determines whether the news generation process is affected more by the general semantic network (low $\lambda_f$) or by the community semantic network (high $\lambda_f$), akin to less or more strong echo chambers (for extreme settings where the filter strength is $\lambda_f>1$ beyond the typical range, see SI Appendix, section 2 and Fig.S1). The memory strength, $\lambda_m \in [0, 1]$, controls the decay rate of the community semantic network during the update process. It determines how fast the community semantic network responds to change, effectively acting as its inertia. A high memory strength (e.g., $\lambda_m=1$) results in slower and weaker changes in the community semantic network, while a low memory strength (e.g., $\lambda_m=0$) leads to faster, more pronounced changes.

\begin{table*}[t]
\centering
\caption{Influences on community semantic networks of online news communities, affected model processes, and control parameters.}
\label{table:1}
\begin{tabular}{@{} l l p{0.27\textwidth} l l @{}}
\toprule
\textbf{Category} & \textbf{Influence} & \textbf{Description} & \textbf{Affected model process} & \textbf{Control parameter} \\
\midrule
\multirow{3}{*}{Editorial agenda-setting}
 & Amplification & Emphasizes topics currently deemed less important by the community. & News generation & Amplification strength ($s_\mathrm{amp}$) \\
 & Reframing & Distorts the narrative frame to link a particular topic to unrelated topics. & (Post-)News generation & Reframing probability ($p_\mathrm{ref}$) \\
 & Alignment & Aligns news coverage with topics already emphasized by the community. & News generation & Filter strength ($\lambda_{f}$) \\
\midrule
\multirow{3}{*}{Community dynamics}
 & Trolls & A group of users promotes specific topics, often with malicious intent. & Comment generation & Troll strength ($s_\mathrm{tr}$) \\
 & Counterspeech & Users counteract trolls by increasing the volume of relevant comments. & Comment generation & Counterspeech strength ($s_\mathrm{cs}$) \\
 & Membership turnover & Member replacement alters content and network structure. & Updating & Memory strength ($\lambda_{m}$) \\
\bottomrule
\end{tabular}
\end{table*}

We calibrate our model's initialization and validate its dynamics using comments and article titles posted in five US-based online news communities over 11 years, starting in 2012, including over $400$ million comments and $850$ thousand news articles (SI Appendix, Fig. S2, section 3, and tables S1-2). We use BERTopic \cite{grootendorst2022bertopic}, a topic modeling framework that employs a large language model (BERT) as a latent embedding model and provides topic embeddings and classifications for article titles and comments, which we then use to construct a semantic network for each community.The global semantic network can be represented as the average of the different community semantic networks (see also SI appendix, section 9). The process of data collection and the construction of a community semantic network are shown in Fig. \ref{fig:1}b, and details of the topic modeling approach are described in the Methods and the SI Appendix, section 4. We validate the topic model on a total sample of $N=1022$ participants representative of the U.S. public, using five tasks designed to assess the coherence of the discovered topics, how well comments and article titles are described by the model-assigned topics, and how similar pairs of topics and comments are. The results show a strong alignment between the model and human judgments (see the SI appendix, section 5, for details of each task, and Figs. S3-7 for detailed results).

These rich, quantitative empirical data enable us to construct our computational model based on well-founded choices rather than arbitrary assumptions. A notable example is the way we model comment generation: based on the observation that users typically post more comments related to the topics of the news (i.e., ``on-topic" comments), but that some topics, such as politics and the economy, are always discussed regardless of the news (as modeled by comment multipliers, see SI Appendix, Fig. S8 and Methods). Further empirical insights that motivate our model formulation and functional choices are presented in the SI Appendix, section 6, and in Figs. S9-11.

\section*{Influences on collective minds}

The computational model allows us to explore how editorial agenda-setting and community dynamics affect the collective mind of online news communities with varying levels of filter and memory strength. We implement six different editorial and community influences in Table \ref{table:1} by tuning these parameters or modifying model components (see Method), and observe their effects on community semantic networks relative to baseline dynamics.

\subsection*{Editorial agenda-setting}

News organizations largely determine what topics and challenges are worthy of people’s attention. In journalism research, this is commonly referred to as agenda setting \cite{Coleman2009}. Agenda setting can lead to various biases in the way news items are selected and presented \cite{Groeling2013, Rodrigo-Gines2024}. We investigate three different editorial agenda-setting practices that lead to such biases: amplification, reframing, and alignment.

\subsubsection*{Amplification} 
Consider the case where an editor overestimates the frequency of certain rare events, such as vaccine-related fatalities, that support conspiracy narratives which are not widely shared within the community. This represents amplification through editorial policies that emphasize topics the community currently deems less important. This exemplifies a form of selection bias by commission \cite{Rodrigo-Gines2024}. It can skew the perceived importance and urgency of these events within the community, leading to a distorted community semantic representation relative to the general semantic network.

Amplification is implemented by increasing the target topic multiplier ($s_{\text{amp}}$) that subjectively amplifies the frequency of the target topic in the general semantic network during the filter process (Fig. \ref{fig:4}a), hence exaggerating the target topic's general popularity. We test the amplification effect by setting $s_{\text{amp}}=25.0$ and measuring the ratio of the target topic frequency in news and comments between the influenced and baseline cases across different filter and memory strengths. This first directly affects the topic frequency in the news, and then indirectly affects the comment frequency and inter-topic similarity from the community semantic network. 

When applied, amplification first affects the frequency of the target topic in the news, and, through the news, indirectly affects its frequency in the comments and its similarity to other topics in the community semantic network. These effects are weaker for communities with higher filter strength (blue and green lines in Fig. \ref{fig:4}b and c, time steps $t \in [100, 300]$) because these communities are less sensitive to external influences from the general semantic network. 

Once amplification is removed, communities with high filter and memory strength are more likely to retain their state because they remain strongly influenced by their modified community semantic network (Fig. \ref{fig:4}c, for time steps $t > 300$). In turn, the target topic frequency in news also remains slightly elevated for communities with higher filter and memory strength (blue line, Fig. \ref{fig:4}b). This suggests that an influence such as amplification can have a lasting effect on the community semantic network, especially in communities characterized by strong filtering.

Amplification also increases the similarity between the target topic and all other topics by co-appearing with them more frequently (Fig. \ref{fig:4}d), especially for the initially more similar topics (top vs. bottom $20\%$), which have a higher chance of co-occurring in news.

\subsubsection*{Reframing} 

When an editor consistently links news about vaccines to government control or population surveillance, this might shape and reinforce a biased narrative about epidemics and the ways to counter them. This kind of narrative shift is often referred to as reframing, which involves distorting the narrative frame to link a particular topic to otherwise unrelated topics. While the concept of framing is multifaceted and complex \cite{Druckman2001}, reframing can be seen as a form of presentation bias \cite{Groeling2013}. It occurs when the way information is presented influences the way it is interpreted, often by emphasizing certain aspects over others.

In our model, we implement a simple form of reframing by replacing a topic in one of the three tiers describing an already filtered event by a target topic (Fig. \ref{fig:4}e), with probability $p_{\text{ref}}$. In this example, we replace topics in tier $2$ with the target topic (here, topic 25) with $p_{\text{ref}}=0.04$. Here, changing tier $2$ rather than tier $1$ implies a subtle manipulation of the way an event is portrayed. As for amplification, reframing increases the target topic frequency in the news, and its effect lingers after the removal, more strongly with high filter and memory strength (Fig. \ref{fig:4}f). Notably, as it indirectly increases the frequency of the comment semantic network (as in Fig. \ref{fig:4}g, see SI Appendix, Fig. S12) which in turn affects the filter, the effect also spreads to the main tier $1$, increasing the frequency of the target topic in that tier, especially for communities with high filter strength (Fig. \ref{fig:4}j). These results suggest that even subtle reframing can substantially change the way events are described.

Unlike amplification, reframing ignores the existing semantic structure of both general and community semantic networks, as it affects post-filtering, and is uniformly associated with all other topics. In turn, due to the non-linear nature of the Hebbian learning rule we adopted for weight updates (see Methods), the reframing effect is stronger for topics that are initially less similar to the target topic (Fig. \ref{fig:4}h). 

Our approach to framing focuses on how topics are amplified, substituted, or aligned in the flow of news and comments, and is implemented using topic triplets at the article level. This offers a simplified but interpretable way to study reframing within the dynamics of collective minds. We acknowledge that framing can also involve deeper contextual and cultural dimensions, such as distinguishing between narratives around different conflicts or assigning divergent emotional valences to the same topic. Capturing these layers would require more complex, context-based topic representations, which go beyond the scope of the current study but represent a promising avenue for future extensions.

\subsubsection*{Alignment} 
In cases where the editorial board covers more news about the bad side effects of vaccines because the community shows great interest in it, while intentionally not publishing other news about vaccine safety. This is an example of alignment: editorial policies that line up news coverage with what the community already believes are the most important topics. This can be viewed as selection bias by omission \cite{Rodrigo-Gines2024}, in the sense that certain events in the world are not presented to the community. As a result, individuals are exposed primarily to information that reinforces their existing beliefs, creating a feedback loop that strengthens those beliefs and excludes contradictory information.

In our model, alignment is represented by an increase in the filter strength ($\lambda_f$), which determines the news items selected for publication (Fig. \ref{fig:4}i). Starting from $\lambda_f=0.2$, we investigate how alignment affects the community semantic network as we increase $\lambda_f$ to $0.8$, and how quickly its effect can be reversed by reducing the filter strength again. We investigate the Kendall-tau rank distance between the general semantic network and the comment network (the current instantiation of the community semantic network) in two scenarios. In one scenario, the community semantic network is initially identical to the general one (Fig. \ref{fig:4}j), and in the other, it is initially far from the general one (Fig. \ref{fig:4}k).

In general, without external influences, the initially identical community semantic network diverges from the general semantic network over time due to the stochastic nature of the comment-generation process. In contrast, the initially far community semantic network moves closer to the general semantic network due to the event generation process (Fig. \ref{fig:1}). The resulting baseline behavior is shown by dashed lines in Fig. \ref{fig:4}j-k  (see SI Appendix, Fig. S13 and section 7 for detailed descriptions of this behavior). When an influence is applied, the community semantic network deviates from its baseline behavior (Fig. \ref{fig:4}j-k). The speed of this change increases with $\lambda_m$ and decreases with $\lambda_f$. When the influence is removed, the community semantic network returns to the baseline trajectory, but the speed of this return decreases with $\lambda_m$ and increases with $\lambda_f$. This behavior is consistent with all the influences that we tested.

The alignment works like friction for both, initially the same, and the initially far community semantic networks: it slows down the movement of the community semantic network relative to the general semantic network. Intuitively, this is because the alignment policy forces the community to focus on topics already prevalent in its semantic network, thereby reinforcing existing collective beliefs and resisting change.

This friction-like effect disappears instantly after the alignment is removed, and returns to its original trajectory,  regardless of the memory strength (Fig. \ref{fig:4}l). This fast recovery suggests that although the alignment can maintain the community's semantic network in its current state, its effect is temporary and can be easily nullified by reverting the editorial policy. 

\begin{figure*}
\centering
\includegraphics[width=\linewidth]{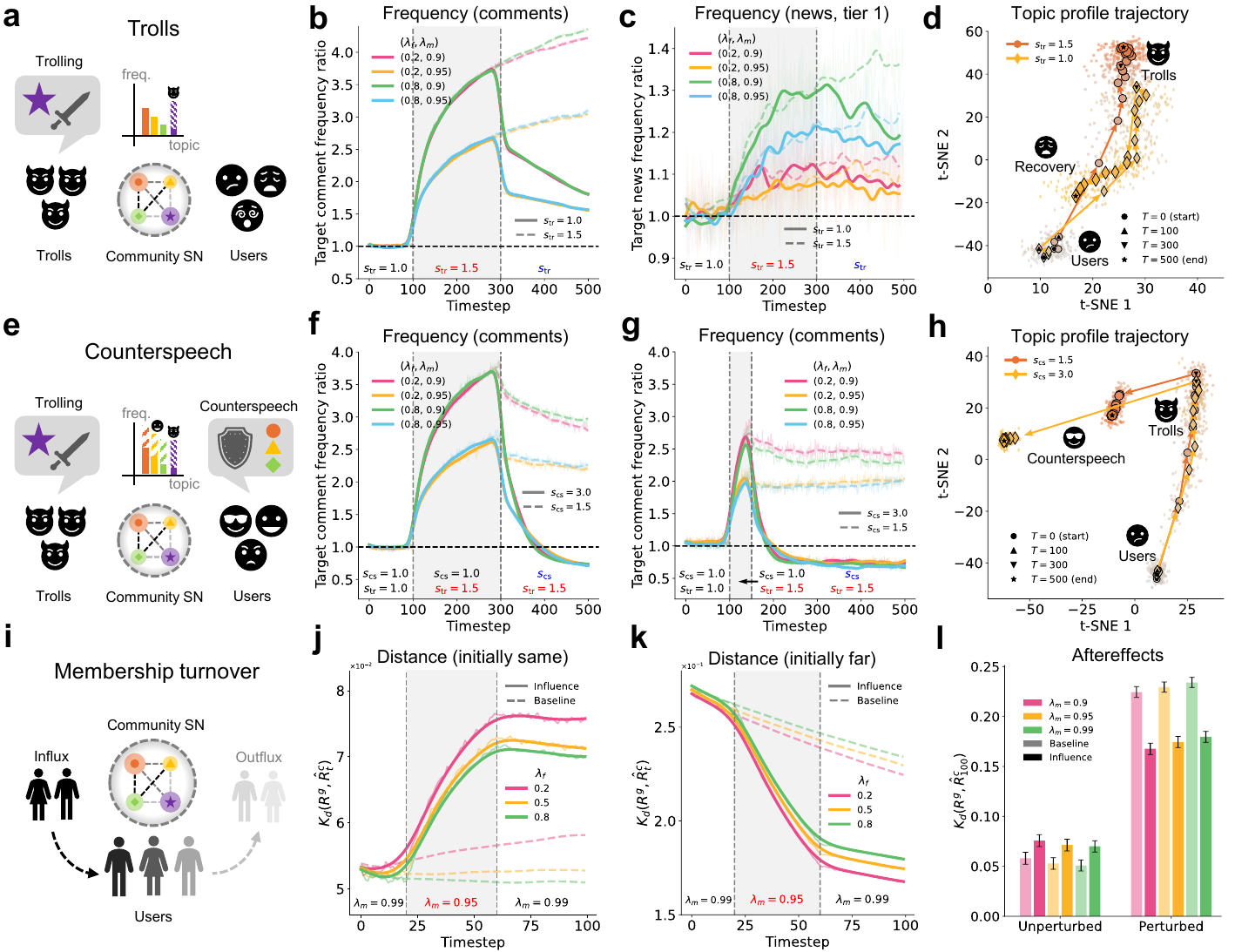}
\caption{\textbf{Impact of community influences on the community semantic networks in the model.} \textbf{a-d}, Trolls are implemented by increasing the frequency of comments discussing a target topic unrelated to the news (\textbf{a}, $s_{\text{tr}}$). They increase the frequency of the target topic in the comment network (\textbf{b}) and in the news (\textbf{c}) for all memory and filter strengths. This effect persists for a long time even after the trolls are removed, but the t-SNE plot of the comment topic profile reveals that eventually, the comment network returns to its original position (\textbf{d}). \textbf{e-h}, Counterspeech is implemented as increasing the frequency of comments related to the news (\textbf{e}, $s_{\text{cs}}$). It reduces the frequency of the target topic promoted by trolls, but it needs to be much stronger than the trolls' influence to completely remove their effect (\textbf{f}). The sooner the counterspeech is introduced, the more effective it is against trolls (\textbf{g}). Unlike the removal of trolls, this does not return the comment network to its original position (\textbf{h}). \textbf{i-l}, Membership turnover is implemented as a decrease in community memory strength ($\lambda_m$, \textbf{i}). It accelerates the movement of the comment network relative to the general semantic network. When the initial position of the community semantic network is the same as (far from) that of the general semantic network, turnover makes the comment network less (more) similar to the general semantic network (\textbf{j-k}) for all memory strengths ($\lambda_m$), as measured by Kendall-tau rank distance between the networks. Once the turnover stops, its effect persists (\textbf{l}). The error bars indicate $\pm1$ standard deviations across $100,000$ simulations. All ratios and differences are relative to the baselines without influences. Semi-transparent lines represent raw data, and solid lines indicate denoised data. For t-SNE plots (\textbf{d}, \textbf{h}), the raw time series of t-SNE coordinates (averaged over $1,000$ simulations) are represented by semi-transparent markers while the smoothed time series (average over $25$ steps) are plotted with larger markers connected by arrows.}\label{fig:5}
\end{figure*}

\subsection*{Community dynamics}

While agenda-setting influences how news items are selected or presented, indirectly affecting the community semantic network, community dynamics affect it more directly. We investigate the effects of membership turnover, trolls, and counterspeech.

\subsubsection*{Trolls} 
Continuing the vaccine example, consider a scenario where a group of users infiltrates a community and begins amplifying claims about vaccine fatality. Even without a factual basis, their coordinated actions can erode public trust in health institutions and steer discourse in divisive directions. Such efforts reflect a broader strategy by trolls—users who deliberately disrupt online. 

Communities suffer from off-topic, inflammatory, or antagonistic messages that often aim to upset or manipulate others \cite{bishop2012psychology}. By promoting specific topics, trolls can manipulate community dynamics, steering discussions toward conflict and division \cite{Cruz2018, Ortiz2020}. In doing so, they can bias the semantic representations of a community in subtle but lasting ways.

We model trolls through an additional troll multiplier ($s_{\text{tr}}$) that amplifies the frequency of comments about the target topic during the comment generation process, regardless of the content of the news, in addition to the comment multiplier (Fig. \ref{fig:5}a). We test the effect of trolls by setting $s_{\text{tr}}=1.5$ and measuring the ratio of the target topic frequency in comments and news between the influenced and baseline cases across different filter and memory strengths.

As expected, the trolls are more effective in intruding on the community semantic networks with lower memory strength, but it takes longer to reverse the damage inflicted on the communities with high memory strength (Fig. \ref{fig:5}b). The frequency of news with the target topic is also affected, with a more pronounced effect in the community with high filter strength, as now the influenced target is the community (Fig. \ref{fig:5}c). Accordingly, the combination of high filter strength and low memory (green line) is particularly vulnerable to trolls, as it is easily affected by the trolls, and editorial influences accelerate the effect. We can visually represent the effect of trolls on the comment network (the current response of the community semantic network) by plotting the trajectories of the comment topic profile in t-SNE space (Fig. \ref{fig:5}d). The trajectory of the comment network experiences a sudden shift when trolls are introduced at $t=100$ (upward triangle), and returns to its original position only if the trolls are removed at $t=300$ (yellow line, downward triangle).

Taken together, these results suggest that, even though trolls do not directly influence the news, the feedback loop from comment sections to editorial decisions eventually indirectly affects the news the community receives. Because of that, trolls' effects can be long-lasting even after they are removed from the community, especially in communities with high filter and memory strength. 

\subsubsection*{Counterspeech} 

In response to exaggerated claims about vaccine fatality introduced by trolls, community members might deliberately promote evidence-based public health information about the benefits of vaccines. This kind of response aligns with what is broadly characterized as counterspeech—a community-initiated effort to address misinformation, incivility, or polarizing content \cite{benesch2018dangerous}. Sometimes evoked by trolls \cite{buerger2019counterspeech}, it often involves providing evidence-based responses, promoting constructive dialogue, encouraging mutual respect among participants, and making more relevant and meaningful contributions while ignoring harmful content. While the forms of counterspeech vary, it generally serves to redirect discussions towards more productive paths, mitigate disruptions, and reinforce shared norms within the community \cite{Friess2021, Garland2022, Rieger2018}.

Among various types of counterspeech that users may put into action, we implement a simple strategy of posting more ``on-topic" comments that are relevant to each news topic, as a counter to trolls' indiscriminate spamming of a single target topic \cite{buerger2021iamhere}. This strategy is implemented by adopting a counterspeech multiplier ($s_{\text{cs}}$) to the comment generation process, amplifying the frequency of on-topic comments (Fig. \ref{fig:5}e) unless it's a troll-targeted topic. 
We test the effect of counterspeech by first applying the influence of trolls at $t=100$ ($s_{\text{tr}}=1.5$), then applying counterspeech at $t=300$ ($s_{\text{cs}}$), while keeping the influence of trolls on. Here, we compare the cases with $s_{\text{cs}}=1.5$ and $3.0$ to see how the strength of counterspeech affects the community semantic network.

While the trolls are still active, counterspeech effectively dilutes the relative target topic frequency in the comments (Fig. \ref{fig:5}f) and in news (SI Appendix, Fig. S12). However, we find that the same strength of multiplier ($s_{\text{tr}}=s_{\text{cs}}=1.5$) is insufficient. A much stronger multiplier ($s_{\text{cs}}=3.0$) is needed to nullify the trolls' effect on the target topic. Furthermore, our findings underscore the importance of timing of influence, especially for weaker counterspeech (Fig. \ref{fig:5}g), as the effect of trolls is suppressed more strongly and quickly when we initiate the counterspeech shortly ($t=150$) after the trolls invade (see SI Appendix, Fig. S12 for early removal of trolls).

Also, trajectory visualization of counterspeech (Fig. \ref{fig:5}h) clearly shows that counterspeech is different from simple removal of trolls: it does not reverse the damage, but instead guides the community semantic network into a different direction proportional to $s_{\text{cs}}$, boosting previously moderately frequent topics at the expense of very frequent and rare topics (SI Appendix, Fig. S14).

\subsubsection*{Membership turnover} 

When a significant number of new members join the community and others leave, the importance and perception of vaccine-related topics in the community can shift considerably. Such membership turnover may substantially reshape the community’s semantic network, as highlighted in \cite{Ransbotham2011}, that turnover in online communities like Wikipedia can have both positive and negative impacts, depending on the balance between new and experienced members.

We can implement the effects of turnover by changing the memory strength ($\lambda_m$), which represents the forgetting of the past collective mind of communities as members change (Fig. \ref{fig:5}i). We investigate how the community semantic network responds to member turnover by decreasing $\lambda_m$ from $0.99$ to $0.95$ for a period, then reverting it. Similar to the amplification, we test the initially same (Fig. \ref{fig:5}j) and far (Fig. \ref{fig:5}k) community semantic networks and measure the Kendall-tau rank distance between the comment topic profile and general semantic network over time.

If a community experiences sudden and frequent membership fluctuations, its community semantic network will be more volatile and prone to random drift or external influences \cite{waller2021quantifying}. This vulnerability is best illustrated by the finding that the community semantic network, when experiencing membership fluctuations, accelerates its movement relative to the general one (Fig. \ref{fig:5}j and k; see SI Appendix, section 7 for more details). Different from amplification, we find that these accelerated trajectories are maintained and not reverted after the termination of turnover, and thus a strong aftereffect remains (Fig. \ref{fig:5}l). This suggests that the community semantic network may be more easily manipulated by external influences when the community is experiencing high turnover \cite{oldemburgo2024twitter}, and that the effect can be long-lasting.

\section*{Model validation}

It is important to distinguish calibration from validation in our approach. Calibration refers to the empirically informed choices of distributional forms and functional assumptions that serve as inputs to the model, such as the frequency, similarity, and comment multiplier distributions (see SI Appendix, Table~S4 and section~6). Validation, in contrast, concerns emergent properties of the model that are not directly imposed by these choices. We validate the model in the following three ways. 

We demonstrate that our computational model accurately captures the statistical distributions observed in real online news communities and successfully reproduces diverse phenomena observed in the empirical data. We check the model's validity in the following three ways.

\begin{figure*}[h!]
    \centering
    \includegraphics[width=0.9\linewidth]{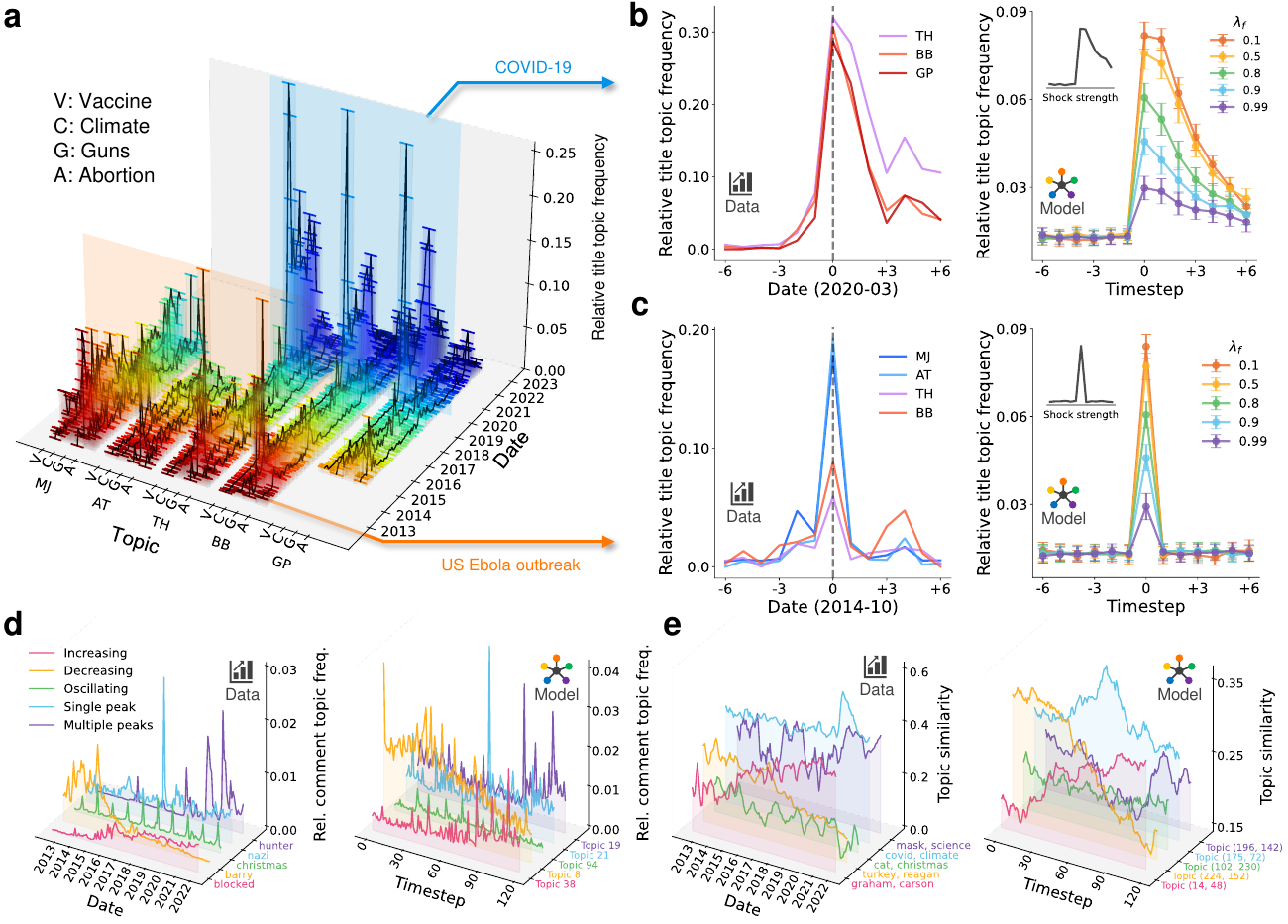}
    \caption{\textbf{Qualitative comparison of data and model output.} \textbf{a}, Examples of empirically observed trends in topic frequencies in article titles (tier 1), for four illustrative topics (Vaccine, Climate, Guns, Abortion) discussed in online news communities Mother Jones (MJ), Atlantic (AT), The Hill (TH), Breitbart (BB), and Gateway Pundit (GP; left panel). We highlight two external shocks associated with high peaks in the title frequency of the Vaccine topic: the COVID-19 pandemic (\textbf{b}, left) and the US Ebola outbreak (\textbf{c}, left). In the model simulation, external shocks were applied to predefined target topics (e.g., vaccine) rather than being assigned retrospectively. This allows a direct comparison of how empirical and simulated communities respond to the same event. \textbf{b-c}, Empirical differences between communities (left panels) can be reproduced in model simulation by tuning the filter strength $\lambda_f$ during the external shock (right panels). The external shock increases the target topic frequency in the general semantic network (insets in right panels). The error bars indicate $\pm1$ standard deviations across $10,000$ simulations. \textbf{d-e}, Selected representative examples of diverse qualitative trends of comment topic frequency (increasing, decreasing, oscillating in time, with single or multiple peaks) (\textbf{d}) and topic similarity (\textbf{e}), observed in the empirical data (left panels) and the corresponding model output (right panels).}\label{fig:3}
\end{figure*}

First, we show that the model can reproduce the topic frequencies in article titles posted in response to external events (Fig. \ref{fig:3}a, b). We model external shocks by increasing the frequency of the relevant topics, and we tune the filter strength $\lambda_f$ to reflect different levels of attention editors pay to the outside world during the shock. The model simulations show varying degrees of reactions that correspond to those observed in real online news communities during the US Ebola outbreak and the COVID-19 pandemic. We also show that model simulations can reproduce qualitative trends in comment topic frequency (Fig. \ref{fig:3}c, d) and topic similarity (Fig. \ref{fig:3}e, f), including increases, decreases, oscillations, and single- or multiple-peak patterns. Unlike prior models of topic popularity that impose life-cycle \cite{wu2007novelty, leskovec2009meme} or built-in periodicity \cite{kobayashi2016tideh} to reproduce such patterns, our model naturally generates these trends from underlying dynamics without requiring explicit constraints.

Second, we quantitatively compare the real data and the model output in terms of the relative topic frequency of news titles and comments (SI Appendix, Fig.~S15a-b), and the topic similarities (SI Appendix, Fig.~S15c). Notably, the model results are time-averaged over 120 time steps, meaning the calibrated distributions are maintained throughout the model's stochastic evolution, implying an emergent outcome rather than a built-in constraint. Each panel shares the same process; it first shows the fitting of the individual empirical data (left, thin dashed lines), choosing a representative exponent for each fitting (left, thick dashed lines), and then compares it with the long-term model output (right, scatter plot) that shows a good match with empirical representative fittings (right, thick dashed lines).

Benefiting from our model design and empirical calibration, the model output is in good agreement with the empirical data, specifically with the relative topic frequencies of news titles (a product of a log and a power-law with tier-specific exponents) and comments (power-law), and the inter-topic similarity distributions (log-normal) across all comparisons. Note that the model results are time-averaged over $120$ time steps, so our model output consistently maintains the empirical distribution throughout its time evolution, without collapses or significant shifts, as we observed in the empirical data.

Third, we investigate the dynamics of the empirical comment network by characterizing it as a comment topic profile, which is a vector that represents the relative comment frequency of each topic at a given time, and track its time-series trajectory (SI Appendix, Fig. S16a and Methods). The dimensionality reduction technique t-SNE \cite{van2008visualizing} reveals that the comment topic profiles of different communities, and with them the community semantic networks, are constantly moving, with each community having different trajectories, initial positions, and speeds. We find that this behavior is well explained by our model with different filters and memory strengths, all of which are attracted to the general semantic network (SI Appendix, Fig. S16b).

\section*{Discussion}

Our computational model illuminates the mechanisms underlying the dynamics of online news communities. Tuning two main parameters — filter strength $\lambda_f$ and memory strength $\lambda_m$ — enables experimentation with editorial and community influences in online news communities with different characteristics, uncovers a number of nontrivial patterns, and helps develop practical recommendations.

Across all influences, we find a common pattern in how key parameters affect the model outcomes. Higher values of the filter strength $\lambda_f$ dampen external shocks but, conversely, intensify their effect when the shock originates within the community. Higher values of memory strength $\lambda_m$ make the community more resilient to the shock but also slow its recovery. The dual nature of each parameter implies that no single optimal parameter set can perfectly address all influences, demonstrating that remedies must be tailored to specific influences based on the community's current characteristics. It also emphasizes the need for collective adaptation, adjusting these parameters as the situation demands. Depending on other properties of a particular real-world community (Table S4), these parameters could lead to different dynamic patterns. Although we believe that our results will generalize across a wide variety of communities, researchers should determine which communities to target based on their own real-world use cases.

Our results on the effects of editorial influences (Fig. \ref{fig:4}) show that the effect of aligning news content with existing community preferences can be removed surprisingly quickly. Comparatively more subtle influences, such as amplification and reframing, can be much more transformative and potentially disruptive than the more obvious alignment. This echoes the findings that amplification can influence members' representations and attitudes \cite{BroockmanKalla2023, Cook2015, Thorson2016}. Also, our results illuminate the difference between two seemingly similar influences that both promote a target topic in news: amplification reinforces pre-existing relationships with other topics, while reframing establishes new connections with previously unrelated topics.   

When it comes to influences due to community dynamics (Fig. \ref{fig:5}), we find that small changes in community membership can have large consequences for collective minds, in line with studies showing that shifts in cultural output are driven by changes in community composition rather than by changes in individual minds \cite{morin2013portraits,underwood2022cohort}. Furthermore, we find that the effect of trolls can be long-lasting even when they are removed, especially in communities with high filter and memory strength. Counterspeech can dilute the effect of trolls, but only when it is much stronger at promoting on-topic discussion than trolls are at promoting their target topics. We also find that it is important to start with counterspeech early on, as the longer the trolls are allowed to influence the discourse, the more difficult it is to nullify their impact. Finally, while both the removal of trolls and responding with counterspeech revert the relative frequency of the topics promoted by the trolls to their baseline, counterspeech response moves the community semantic network in a new direction. 


How and whether communities effectively change their key parameters, such as filter and memory strength, over time, is a promising area for further research. While these parameters are modeling abstractions, they correspond to identifiable real-world processes. The effective filter strength can shift when editorial policies change — for example, when editors are incentivized to maximize engagement metrics such as view counts, they may increasingly favor stories aligned with community preferences, thereby increasing filter strength. Conversely, editorial guidelines mandating balanced coverage of underrepresented topics would decrease it. The effective memory strength can change through membership turnover — for instance, when a community loses members dissatisfied with its editorial direction and attracts a different audience, the incoming members bring different topic preferences, reducing the community's effective memory. Platform design choices also modulate memory strength; for example, comment-ranking algorithms that surface only the most recent or popular comments, rather than preserving the full discussion history, limit exposure to past discourse patterns and thereby reduce memory strength \cite{muchnik2013social,bowing2024news}.

Together, our results suggest practical recommendations to communities on how to protect their genuine collective dynamics. On the level of editorial boards of online communities, regularly reporting detailed metrics on topic frequencies and their interconnections would allow the public and interested parties to detect when amplification, reframing, or disruptive community dynamics are producing persistent shifts in the collective semantic network. For example, the editorial board could transparently track and post statistics on the relative frequencies of different vaccine-related events they observed in the real world, vaccine-related news posted on the site, and topics discussed in news about vaccines. Such disclosures could help mitigate unwanted agenda-setting strategies and incentivize editors to maintain a balanced portrayal of issues. At the community level, maintaining core membership and fostering organized, immediate counterspeech against adversarial influences such as trolling can lead to changes in discourse that better reflect the authentic collective mind. For example, a swift collective response to expose misinformation about vaccines in the community can help counter adversarial attempts to undermine collective well-being.

To develop a useful model of the collective mind emerging from the underlying informational ecosystem of online communities, we deliberately kept the model computationally tractable and interpretable by adopting a simplified structure that focuses on prominent mechanisms capturing the fundamental interplay among external events, editorial agenda-setting, and community dynamics. Despite the stochastic nature of online communities and the role of uncontrollable external shocks, the model reproduces the statistical properties of empirical data (Fig. \ref{fig:3} and SI Appendix, Fig. S15) and provides explanatory insights into the mechanisms at work.

However, the framework is readily extensible to other types of digital platforms and can incorporate additional processes relevant to collective dynamics. On platforms such as Reddit and 4chan, each user can be modeled as a decentralized editor who aligns, amplifies, and reframes news about real-world events in line with their own preferences and the perceived semantic network of their followers. On platforms such as YouTube and TikTok, we can model multiple layers of filters in addition to the users themselves, including platform policies and diverse recommendation algorithms. The model can also be extended to incorporate other aspects of the dynamics of digital platforms, including the effects of several influences at once, the influence of group emotions \cite{goldenberg2020collective}, the interaction between communities, topic-level filter, and memory strengths, as well as feedback from the communities that may alter the general semantic network. In this paper, we examined the effects of different influences in isolation. However, real-world scenarios often involve multiple concurrent and interacting forces. Future explorations of the model could investigate these compound effects by applying combinations of influences simultaneously. We also assumed a unified internal structure within each community and did not model information flow between communities. Incorporating internal heterogeneity and cross-community interactions would allow the model to capture the dynamics of belief diversity, subgroup polarization, and the diffusion across community boundaries.

In summary, this work provides a foundation for a more rigorous understanding of different influences on collective minds. The model can help anticipate changes in community discourse that may result from different editorial policies, shifts in membership, and adversarial influences such as trolling. It also helps anticipate the benefits of editorial and community practices aimed at reducing echo chambers and countering toxic speech, such as more inclusive representation of events in the outside world and the use of counterspeech. Our results reveal the sources of both the fragility and the robustness of collective minds, informing a path toward healthier collective discourse and behavior.

\matmethods{
\subsection*{Collective mind model for online news community}

In this section, we provide a detailed account of the computational model and the procedures undertaken to simulate the dynamics of the collective mind in the online news community. Hyperparameters, functional forms, and additional details on initialization and the rationale for each model choice are provided in the SI Appendix, section 1, and Table S4 to ensure replicability. Note that most of the model settings we employed here are chosen to reflect the empirical findings from our data (see SI Appendix, Fig. S15, where the empirical data and our choice of model parameters are highlighted), and one can freely alter the settings of our framework according to one's data at hand to model different online communities. Besides these configurable settings, our model has exactly two free parameters that characterize the individual community characteristics: the filter strength ($\lambda_f$) and the memory strength ($\lambda_m$).

\subsubsection*{Semantic network construction} In the model, the general semantic network represents the shared structure of meaning in the broader information environment. It is a complete graph where each node is a topic, and each edge reflects how closely two topics are related in meaning. The node weights represent how prominent each topic is in the overall environment, and the edge weights capture the similarity between topics. At each time step $t$, we define the general semantic network $G^g_{t} = (V^g, E^g, F_{t}^g, W_{t}^g)$, consisting of a set of vertices $V^g$, edges $E^g$, normalized frequencies $F_{t}^g$, and edge weights $W_{t}^g$. Vertices represent topics, and edges represent their semantic closeness. The network is complete, with $|E^g| = N(N-1)/2$ edges and no self-loops. Each topic vertex $v_i \in V^g$ has a normalized frequency $f_{i, t}^g \in F_{t}^g$ such that $\sum_i f_{i, t}^g = 1$. A normalized frequency ranking $r_{i, t}^g = \text{rank}(f_{i, t}^g)/N$ is also assigned. Edge weights $w_{ij, t}^g \in W_{t}^g$ represent topic closeness, bounded between 0 and 1. Community-specific semantic networks $G^{k}_{t} = (V^{k}, E^{k}, F{t}^{k}$ where 
$k$ indexes the communities, $W_{t}^{k})$ share vertices and edges with $G^g_{t}$ but can have distinct frequencies and weights. In this study, we fixed our general semantic network for all time steps, $G^g_{t} = G^g$ (except when we applied an external shock in Fig. \ref{fig:3}b). This condition implies that changes in the general semantic network are very slow or nearly constant in the time scale of the simulation. If we wish to extend the model to be more realistic, we may choose not to adopt this condition by assuming a changing general semantic network.

\subsubsection*{Initialization} The general semantic network is initialized with frequencies $f_{i, 0}^g = F_{f}(i)$, where $F_{f}$ is a monotonically decreasing distribution. In our study, $F_{f} \propto i^{-1}$, leading to $f_{i, 0}^g = i^{-1}/C$ where $C = \sum_{i=1}^N i^{-1}$. Edge weights $w_{ij, 0}^g$ are sampled from a log-normal distribution, $F_{w} \propto e^{-\ln^2((x-a)/b)/(2s^2)}$, with parameters $a$, $b$, and $s$ controlling the distribution shape. Community networks are initialized by perturbing the general network’s frequencies and weights, as each community's semantic structure is assumed to closely resemble the shared baseline while differing slightly in topic emphases and associations (see SI Appendix, section~9). For each community $k$, $f_{i, 0}^k = f_{i, 0}^g + \mathcal{N}(0, \sigma_{\text{fp}}) \cdot F_{f}(i)$ and $w_{ij, 0}^k = w_{ij, 0}^g + \mathcal{N}(0, \sigma_{\text{wp}})$, where $\sigma_{\text{fp}}$ and $\sigma_{\text{wp}}$ are standard deviations for frequency and weight perturbations, respectively.

\subsubsection*{Event Generation} At each time step $t$, the general semantic network generates events $X_t = \{x_{1, t}, x_{2, t}, \dots, x_{N_x, t}\}$, where $N_x$ is the number of events per time step. Each event $x_{i, t} = \{v_{z_1}, v_{z_2}, \cdots, v_{z_{N_{w}}}\}$ is composed of $N_w$ topics and each tier of topics sampled with probabilities proportional to $F_{\text{ns}}(r_{z_q, t}^g)$, the event sampling distribution, where $z_q$ indicates $q$-th tier topic index in the event. In this study, we used $F_{\text{ns}} \propto -\ln(r_{i, t}^g)$. Events are generated sequentially, ensuring no topic duplication within an event. Rejection sampling ensures unique topics per event. To introduce temporal correlation, at each time step, the topic frequency distribution is interpolated between the previous one and the newly sampled one with a memory parameter $\lambda_e$. In our simulation, we used $N_x=1000$ and $N_w=3$.

\subsubsection*{Filtering} An editorial board of the community applies a two-stage filter to generate events. The filter criteria depend on a combination of general and community-specific semantic network attributes, weighted by a filter strength parameter $\lambda_f$ as $\bar{f}_{i, t}^{k} = \lambda_f f_{i, t}^{k} + (1-\lambda_f) f_{i, t}^g$ (accordingly, normalized ranking $\bar{r}_{i, t}^{k}$ as well) and $\bar{w}_{ij, t}^{k} = \lambda_f w_{ij, t}^{k} + (1-\lambda_f) w_{ij, t}^g$. The first stage filters events based on normalized frequencies, where events are sampled without replacement proportional to $\prod_{q} (\bar{r}_{{z_q}, t}^{k})^{\alpha_q}$, where $\alpha_q$ denotes the $q$-th tier filter exponent, retaining a fraction $R_1=0.5$ of events. The second stage filters events based on inter-topic similarity $\prod_{q_1, q_2} \bar{w}_{{z_{q_1}}{z_{q_2}}, t}^{k}$, retaining the top $R_2=0.5$ fraction. The final set of filtered events is $X_t^k \subseteq X_t$, representing community news.
Each community applies a two-stage filter to generate events. The filter criteria depend on a combination of general and community-specific semantic network attributes, weighted by a filter strength parameter $\lambda_f$ as $\bar{f}_{i, t}^{k} = \lambda_f f_{i, t}^{k} + (1-\lambda_f) f_{i, t}^g$ (accordingly, normalized ranking $\bar{r}_{i, t}^{k}$ as well) and $\bar{w}_{ij, t}^{k} = \lambda_f w_{ij, t}^{k} + (1-\lambda_f) w_{ij, t}^g$. The first stage filters events based on normalized frequencies, where events are sampled without replacement proportional to $\prod_{q} (\bar{r}_{{z_q}, t}^{k})^{\alpha_q}$, where $\alpha_q$ denotes $q$-th tier filter exponent, retaining a fraction $R_1=0.5$ of events. The second stage filters based on inter-topic similarity $\prod_{q_1, q_2} \bar{w}_{{z_{q_1}}{z_{q_2}}, t}^{k}$, retaining the top $R_2=0.5$ fraction. The final set of filtered events is $X_t^k \subseteq X_t$, representing community news.

\subsubsection*{Comment network generation}To capture community response to filtered news events, we construct the comment semantic network. This network represents a community’s internal structure of expressed meaning. It is a complete graph in which each node is a topic and each edge captures the semantic similarity between topics. Formally, it is defined as $A_t^k = (V, E, \hat{F}_t^k, \hat{W}_t^k)$. This network shares vertices ($V$) and edges ($E$) with other semantic networks, while incorporating frequencies ($\hat{F}_t^k$) and weights ($\hat{W}_t^k$). Note that the frequency of the comment network $\hat{F}_t^k$ corresponds to the empirical comment frequency for each topic.

For each news $x_{i, t}^k$, the comment frequency is derived by first assigning several comments $c_{i, t}^k$ based on a comment number distribution. Based on the current frequency of the community semantic network, a set of comment multipliers $m_{i, t, q}^{k} \sim P_{m, q}(x_{i, t}^{k}, c_{i, t}^{k})$ for each tier (which depends on the tier $q$, community topic ranking for each tier $r_{z_q}^k$, and the comment number $c_{i, t}^k$, see SI Appendix, section 6 and Fig. S8), are sampled to adjust the frequency of "on-topic" comments (applied for $N_w$ topics). Meanwhile, all of the other topic frequencies ($N-N_w$ topics)are scaled by a normalization constant $C_{i, t}^k$. This normalization ensures that the total number of comments remains consistent. The final comment frequency is given by summing all comment topic frequency for all news, $\hat{f}_{j, t}^k = \sum_i \hat{f}_{j, t}^k(x_{i, t}^k)$.

The weight captures co-occurrences of topic pairs within the news for a given period (a month), weighted by the volume of comments under that news. For each pair of topics $(a, b)$, the weight $\hat{w}_{ab, t}^k$ is non-zero only if both topics appear in the same news item. The overall weight at time $t$ is calculated as $\hat{w}_{ij, t}^k = \sum_i c_{i, t}^k \hat{w}_{ij, t}^k(x_{i, t}^k)$.

\subsubsection*{Community semantic network update} To update the community semantic network, we incorporate a feedback mechanism. The topic frequency is updated using a memory strength parameter $\lambda_m$ as  $f_{i, t+1}^k = \lambda_m f_{i, t}^k + (1 - \lambda_m)\frac{\hat{f}_{i, t}^k}{\sum_j \hat{f}_{j, t}^k}$.  
Frequencies are then quantized by rank to preserve the initial distributional structure while reflecting new responses.  The weights are updated using a Hebbian learning-inspired approach \cite{hebb2005organization}:  $w_{ij, t+1}^k = \eta (w_{\max} - |w_{ij, t}^k|) \frac{\hat{w}_{ij, t}^k}{D_t^k} - \gamma w_{ij, t}^k + \epsilon_{ij}$,  
where \( \eta \) is the learning rate, $\gamma$ the (adaptive) decay factor, $w_{\max}$ the weight cap, and $\epsilon_{ij}$ a Gaussian noise term. The normalization factor $D_t^k$ accounts for the number of comments and potential topic pairs.

This iterative process is repeated over $T$ time steps, simulating the dynamic evolution of response and community semantic networks.

\subsection*{Empirical data}
We collected articles and comments from five different online news communities via the comment platform \texttt{Disqus} \cite{Disqus}, which provides all of the news and its comments as raw text. In this study, we considered articles with more than or equal to $10$ comments and filtered out those that did not satisfy the threshold. Aggregated data for a certain period (e.g., 3 days, a week, and a month) represents (1) all of the news posted during the period, and (2) comments made within a short period from the news post date to best represent the collective belief at that period. The period for data collection, the number of articles and comments (both before and after filtering), and their political inclinations are summarized in SI Appendix, Tables S1-2. More details on the preprocessing of empirical data are provided in SI Appendix, section 3. See \cite{bacaksizlar2023group} for more details about the raw comment data. Also, see SI Appendix, section 9, for the discussion on the correspondence between the empirical semantic network and the comment network, as we employed the empirical distribution from the comment network (which is the only direct observables) to construct the model semantic networks.

\subsection*{Topic modeling} 
To extract a community semantic network from the data, we constructed a topic model using data from five online news communities. We used BERTopic \cite{grootendorst2022bertopic}, a topic modeling framework that extracts latent topics from a set of documents, which is a collection of comments in this study. We first fit the model with the sampled subset of the comments and further classified all comments with the fitted model. For the model fitting, a total of $2$ million comments were sampled ($400$ thousand comments per community). We employed the SBERT\cite{reimers2019sentence} model (all-MiniLM-L6-v2), a pre-trained transformer-based language model, to extract the embeddings of these sample comments, and then the embeddings are clustered by first applying UMAP \cite{mcinnes2018umap} for the dimensionality reduction and then HDBSCAN \cite{mcinnes2017hdbscan} for the clustering to get the classification of each comment. We performed a two-stage grid search on the hyperparameter space of BERTopic to find the topic model that best represents the topic space while keeping other settings as default. As a result, the final topic model used in this study has $228$ distinct topics, such as vaccine, climate, and taxes. In addition, we also constructed separate, community-specific topic models to verify the empirical distribution by cross-validation (see SI Appendix, section 8 and Fig. S17 for the empirical data distribution from community-specific topic models). Details of the topic model construction and hyperparameter grid search are described in the SI Appendix, section 4 and Tables S5-6, and its top $10$ topics are summarized in SI Appendix, Table S7.

The topics are characterized by the number of comments discussing that topic at the given time point ($f_i$) and the average of the embeddings of such comments (topic representation, $e_i$). The similarity between topics is calculated as a cosine similarity of their topic representations. In the empirical data, we find that most of the topic similarities (more than $99.8\%$) are greater than $0$.

\subsection*{Validation survey}

To validate the quality of the topic models used to construct the semantic networks, we conducted a survey ($N = 1,022$) with participants representative of the U.S. public, recruited from the social experiment platform \texttt{Prolific} \cite{palan2018prolific}. The survey included five tasks, each targeting a distinct aspect of topic model quality: (T1) \textit{word intrusion}, testing whether individual model-generated topics are perceived as coherent; (T2) \textit{topic assignment for comments}, testing whether comments from online news communities can be correctly matched to the model-assigned topic; (T3) \textit{topic assignment for article titles}, testing whether titles can be accurately linked to their corresponding topics; (T4) \textit{topic similarity based on topic descriptions}, assessing whether cosine similarities between topic embeddings align with human judgments of semantic similarity; and (T5) \textit{topic similarity based on comment content}, evaluating whether cosine similarities between BERT embeddings of comments reflect perceived semantic closeness. The results show a strong alignment between model-generated structures and human judgments. Participants reliably identified the intruding word in assessed topics (T1), and they were able to match comments to the correct topic (T2) at rates substantially above chance for both tasks. Participants also consistently selected the model-assigned topic that best matched the content of each article title (T3). In both pairwise similarity tasks (T4 and T5), participants' ratings correlated above $0.5$ with cosine similarities between the corresponding embeddings. Across all tasks, results were comparable for local and global topic models across all communities (see SI Appendix, section 5 and Figs. S3-7 for more details and the results).

\subsection*{Testing influences on the collective minds}

For testing the effect of influences on the collective minds, the community semantic network has $250$ topics and is updated for $T=100$ (alignment and membership turnover) or $T=500$ (other influences) time steps, where in each time step, $N_x=1,000$ events are generated (after the filtering, this equates to roughly a day's worth of news). For the influences with target tier (amplification, reframing, trolls, and counterspeech) and target tier (reframing), influence toward the target topic ($25$) and target tier ($2$) is applied from time step $t=100$ to $t=300$ and then removed to examine the model response before and after the influence. We can easily expand this test scenario with more extensive influences by setting multiple target topics and tiers.

\subsection*{Comment topic profile visualization}
In this study (particularly in Fig. \ref{fig:5} and Fig. S3), we characterize the state of the semantic network at a given time step $t$ by comment topic profile $\mathbf{f}_t$ as follows, 

\begin{equation}
    \mathbf{f}_t = \frac{1}{{\sum_j \hat{f}_{j, t}}} \left[ \hat{f}_{1, t}, \hat{f}_{2, t}, \cdots, \hat{f}_{N, t} \right],
\end{equation}

where $\hat{f}_{i, t}$ denotes the comment frequency of topic $i$ at time $t$. Note that these values are not sorted according to their respective frequency ranking in time, but rather according to the order of the predefined index. In practice, we chose this topic order based on the comment frequency ranking of the overall data. We then visualize the trajectory of the comment topic profile by applying t-SNE\cite{van2008visualizing}, a dimensionality reduction technique that projects high-dimensional data into a two-dimensional space while preserving the local structure of the data, to the time series of $\mathbf{f}_t$. From its initial $228$ ($250$ for the model output) dimension, we first reduced the dimension by taking the projection to the first $50$ principal component. Then we applied t-SNE to further reduce the dimension to $2$ for visualization.

\subsection*{Denoising of the model output}
In Fig. \ref{fig:4} and \ref{fig:5}, some of the time series from the model output are denoised for better visibility and plotted with thick lines. The denoising is performed by applying $1$-D total variation denoising \cite{rudin1992nonlinear} to the raw time series data. Given the raw time series $(x_1, x_2, \dots x_N)$, the denoising is performed by minimizing the functional  $y$, $J(y) = E(x, y) + \lambda V(y)$ for denoised time series $(y_1, y_2, \dots y_N)$, where $E(x, y) = \frac{1}{n} \sum_n (x_n - y_n)^2$ denotes the sum of squared differences between the raw and denoised time series, and $V_y = \sum_n{x_{n+1}-x_{n}}$ denotes the total variation of the raw time series. In this study, we employed $\lambda=0.4$ for all visualizations.

\subsection*{Ethics statement}
The survey was approved by the UNM IRB, protocol no. 2250030447. All participants provided informed consent before participation. The acquisition of news data from Disqus \cite{Disqus} is exempt from requiring IRB approval, because the data were obtained within the terms of use of the Disqus API, they displayed their privacy policy prominently at the top of the commenting section of each site, and the comments are publicly available and used in a completely anonymized form.

}

\showmatmethods{} 

\section*{Data Availability}
The empirical data from online news communities is freely available to download through the Disqus API \cite{Disqus}. The data from the human survey results are available from the corresponding author upon reasonable request. All other data and code used in this study for analysis are available at \url{https://github.com/nokpil/collmind} \cite{data2025zenodo}.

\acknow{This work has been supported by the ERC Advanced Grant COLLADAPT 101140741, the  Austrian  Research  Promotion  Agency FFG grant 873927, and the Austrian Science Fund FWF OPUS grant 10.55776/PIN2028524. Also, generous grants from the Siegel Family Endowment and Omidyar Network supported this research. We thank Timo Damm, Peter Dodds, Laurent Hebert-Dufresne, Byungwhee Lee, Juniper Lovato, Victor Poulsen, Juniper Rodriguez, Valentin Ruppert, Peter Steiglechner, Will Tracy, and Hyejin Youn for helpful discussions and valuable comments on earlier versions of the manuscript.}

\showacknow{} 


\bibliography{Nature}

@dataset{data2025zenodo,
  author       = {Ha, Seungwoong and Olsson, Henrik and Jaksic, Kresimir and Pellert, Max and Galesic, Mirta},
  title        = {[dataset] Data and Code for "Dynamics of Collective Mind in Online Communities", doi:10.5281/zenodo.17930413},
  publisher    = {Zenodo},
  year         = {2025},
  doi          = {10.5281/zenodo.17930413},
  url          = {https://doi.org/10.5281/zenodo.17930413}
}

@article{fraxanetUnpackingPolarizationAntagonism2024a,
  title = {Unpacking Polarization: {{Antagonism}} and Alignment in Signed Networks of Online Interaction},
  shorttitle = {Unpacking Polarization},
  author = {Fraxanet, Emma and Pellert, Max and Schweighofer, Simon and G{\'o}mez, Vicen{\c c} and Garcia, David},
  editor = {Moro, Esteban},
  year = {2024},
  month = nov,
  journal = {PNAS Nexus},
  volume = {3},
  number = {12},
  issn = {2752-6542},
  doi = {10.1093/pnasnexus/pgae276},
  urldate = {2025-01-09},
  copyright = {https://creativecommons.org/licenses/by/4.0/},
  langid = {english}
}

@article{morin2013portraits,
  title={How portraits turned their eyes upon us: Visual preferences and demographic change in cultural evolution},
  author={Morin, Olivier},
  journal={Evolution and Human Behavior},
  volume={34},
  number={3},
  pages={222--229},
  year={2013},
  publisher={Elsevier}
}

@article{underwood2022cohort,
  title={Cohort succession explains most change in literary culture},
  author={Underwood, Ted and Kiley, Kevin and Shang, Wenyi and Vaisey, Stephen},
  journal={Sociological Science},
  volume={9},
  pages={184--205},
  year={2022}
}

@article{FIEDLER20142258,
title = {Influence of community design on user behaviors in online communities},
journal = {Journal of Business Research},
volume = {67},
number = {11},
pages = {2258-2268},
year = {2014},
issn = {0148-2963},
doi = {https://doi.org/10.1016/j.jbusres.2014.06.014},
url = {https://www.sciencedirect.com/science/article/pii/S0148296314002112},
author = {Marina Fiedler and Marko Sarstedt}
}

@article{baran2010dynamics,
  title={The dynamics of online communities in the activity theory framework},
  author={Baran, Bahar and Cagiltay, Kursat},
  journal={Journal of Educational Technology \& Society},
  volume={13},
  number={4},
  pages={155--166},
  year={2010},
  publisher={JSTOR}
}

@article{Siew2019,
  author  = {Cynthia S. Q. Siew and Dirk U. Wulff and Nicole M. Beckage and Yoed N. Kenett},
  title   = {Cognitive Network Science: A Review of Research on Cognition through the Lens of Network Representations, Processes, and Dynamics},
  journal = {Complexity},
  volume  = {2019},
  number  = {1},
  pages   = {2108423},
  year    = {2019},
  doi     = {10.1155/2019/2108423},
  url     = {https://doi.org/10.1155/2019/2108423}
}

@article{zhuravskaya2020political,
  author    = {Zhuravskaya, Ekaterina and Petrova, Maria and Enikolopov, Ruben},
  title     = {Political Effects of the Internet and Social Media},
  journal   = {Annual Review of Economics},
  year      = {2020},
  volume    = {12},
  pages     = {415--438},
  doi       = {10.1146/annurev-economics-081919-050239},
  url       = {https://www.annualreviews.org/content/journals/10.1146/annurev-economics-081919-050239},
  publisher = {Annual Reviews},
  issn      = {1941-1391}
}

@article{vanbavel2024social,
  author    = {Van Bavel, Jay J. and Robertson, Claire E. and del Rosario, Kareena and Rasmussen, Jesper and Rathje, Steve},
  title     = {Social Media and Morality},
  journal   = {Annual Review of Psychology},
  year      = {2024},
  volume    = {75},
  pages     = {311--340},
  doi       = {10.1146/annurev-psych-022123-110258},
  url       = {https://www.annualreviews.org/content/journals/10.1146/annurev-psych-022123-110258},
  publisher = {Annual Reviews},
  issn      = {1545-2085}
}

@article{Thorson2016,
  author    = {Emily Thorson},
  title     = {Belief Echoes: The Persistent Effects of Corrected Misinformation},
  journal   = {Political Communication},
  volume    = {33},
  number    = {3},
  pages     = {460--480},
  year      = {2016},
  publisher = {Routledge},
  doi       = {10.1080/10584609.2015.1102187},
  url       = {https://doi.org/10.1080/10584609.2015.1102187},
  eprint    = {https://doi.org/10.1080/10584609.2015.1102187}
}

@inbook{Cook2015,
  author    = {John Cook and Ullrich Ecker and Stephan Lewandowsky},
  title     = {Misinformation and How to Correct It},
  booktitle = {Emerging Trends in the Social and Behavioral Sciences},
  pages     = {1--17},
  year      = {2015},
  publisher = {John Wiley \& Sons, Ltd},
  isbn      = {9781118900772},
  doi       = {10.1002/9781118900772.etrds0222},
  url       = {https://onlinelibrary.wiley.com/doi/abs/10.1002/9781118900772.etrds0222}
}

@book{brown2014human,
  title={The human capacity for transformational change: harnessing the collective mind},
  author={Brown, Valerie A and Harris, John A},
  year={2014},
  publisher={Routledge}
}

@article{hofman2017prediction,
  author    = {Hofman, Jake M. and Sharma, Amit and Watts, Duncan J.},
  title     = {Prediction and explanation in social systems},
  journal   = {Science},
  volume    = {355},
  number    = {6324},
  pages     = {486--488},
  year      = {2017},
  doi       = {10.1126/science.aal3856},
  url       = {https://www.science.org/doi/abs/10.1126/science.aal3856},
  eprint    = {https://www.science.org/doi/pdf/10.1126/science.aal3856}
}

@article{BroockmanKalla2023,
  author    = {David E. Broockman and Joshua L. Kalla},
  title     = {Consuming Cross-Cutting Media Causes Learning and Moderates Attitudes: A Field Experiment with Fox News Viewers},
  journal   = {The Journal of Politics},
  volume    = {0},
  number    = {0},
  pages     = {000-000},
  year      = {0},
  doi       = {10.1086/730725},
  url       = {https://doi.org/10.1086/730725}
}

@article{Druckman2001,
  author  = {James N. Druckman},
  title   = {The Implications of Framing Effects for Citizen Competence},
  journal = {Political Behavior},
  year    = {2001},
  date    = {2001-09-01},
  volume  = {23},
  number  = {3},
  pages   = {225-256},
  issn    = {1573-6687},
  url     = {https://doi.org/10.1023/A:1015006907312},
  doi     = {10.1023/A:1015006907312}
}

@book{miller2009complex,
  title={Complex adaptive systems: an introduction to computational models of social life: an introduction to computational models of social life},
  author={Miller, John H and Page, Scott E},
  year={2009},
  publisher={Princeton university press}
}

@article{benesch2018dangerous,
  title={Dangerous speech: a practical guide},
  author={Benesch, Susan and Buerger, Cathy and Glavinic, Tonei and Manion, Sean and Bateyko, Dan},
  journal={Dangerous Speech Project},
  year={2018}
}

@article{tollefsen2006extended,
  title={From extended mind to collective mind},
  author={Tollefsen, Deborah Perron},
  journal={Cognitive systems research},
  volume={7},
  number={2-3},
  pages={140--150},
  year={2006},
  publisher={Elsevier}
}

@article{shteynberg2023theory,
  title={Theory of collective mind},
  author={Shteynberg, Garriy and Hirsh, Jacob B and Wolf, Wouter and Bargh, John A and Boothby, Erica J and Colman, Andrew M and Echterhoff, Gerald and Rossignac-Milon, Maya},
  journal={Trends in Cognitive Sciences},
  volume={27},
  number={11},
  pages={1019--1031},
  year={2023},
  publisher={Elsevier}
}

@article{weick1993collective,
  title={Collective mind in organizations: Heedful interrelating on flight decks},
  author={Weick, Karl E and Roberts, Karlene H},
  journal={Administrative Science Quarterly},
  pages={357--381},
  year={1993},
  publisher={JSTOR}
}

@article{grootendorst2022bertopic,
  title={BERTopic: Neural topic modeling with a class-based TF-IDF procedure},
  author={Grootendorst, Maarten},
  journal={arXiv preprint arXiv:2203.05794},
  year={2022}
}

@article{mcinnes2017hdbscan,
  title={hdbscan: Hierarchical density based clustering.},
  author={McInnes, Leland and Healy, John and Astels, Steve and others},
  journal={J. Open Source Softw.},
  volume={2},
  number={11},
  pages={205},
  year={2017}
}

@article{mcinnes2018umap,
  title={Umap: Uniform manifold approximation and projection for dimension reduction},
  author={McInnes, Leland and Healy, John and Melville, James},
  journal={arXiv preprint arXiv:1802.03426},
  year={2018}
}

@inproceedings{moulavi2014density,
  title={Density-based clustering validation},
  author={Moulavi, Davoud and Jaskowiak, Pablo A and Campello, Ricardo JGB and Zimek, Arthur and Sander, J{\"o}rg},
  booktitle={Proceedings of the 2014 SIAM international conference on data mining},
  pages={839--847},
  year={2014},
  organization={SIAM}
}

@article{reimers2019sentence,
  title={Sentence-bert: Sentence embeddings using siamese bert-networks},
  author={Reimers, Nils and Gurevych, Iryna},
  journal={arXiv preprint arXiv:1908.10084},
  year={2019}
}

@book{hebb2005organization,
  title={The organization of behavior: A neuropsychological theory},
  author={Hebb, Donald Olding},
  year={2005},
  publisher={Psychology press}
}

@article{palan2018prolific,
  title={Prolific. ac—A subject pool for online experiments},
  author={Palan, Stefan and Schitter, Christian},
  journal={Journal of behavioral and experimental finance},
  volume={17},
  pages={22--27},
  year={2018},
  publisher={Elsevier}
}

@inbook{Coleman2009,
  author = {Coleman, R. and McCombs, M. and Shaw, D. and Weaver, D.},
  title = {Agenda setting},
  booktitle = {The handbook of journalism studies},
  pages = {167--180},
  publisher = {Routledge},
  year = {2009}
}

@article{Collins1975,
  author = {Collins, A. M. and Loftus, E. F.},
  title = {A spreading-activation theory of semantic processing},
  journal = {Psychological Review},
  volume = {82},
  number = {6},
  pages = {407--428},
  year = {1975}
}

@article{Friess2021,
  author = {Friess, D. and Ziegele, M. and Heinbach, D.},
  title = {Collective civic moderation for deliberation? Exploring the links between citizens’ organized engagement in comment sections and the deliberative quality of online discussions},
  journal = {Political Communication},
  volume = {38},
  number = {5},
  pages = {624--646},
  year = {2021}
}

@article{Garland2022,
  author = {Garland, J. and Ghazi-Zahedi, K. and Young, J. G. and Hébert-Dufresne, L. and Galesic, M.},
  title = {Impact and dynamics of hate and counter speech online},
  journal = {EPJ Data Science},
  volume = {11},
  number = {1},
  pages = {3},
  year = {2022},
  doi = {10.1140/epjds/s13688-021-00314-6}
}

@article{Griffiths2007,
  author = {Griffiths, T. L. and Steyvers, M. and Tenenbaum, J. B.},
  title = {Topics in Semantic Representation},
  journal = {Psychological Review},
  volume = {114},
  number = {2},
  pages = {211--244},
  year = {2007}
}

@incollection{steyvers2007probabilistic,
  title={Probabilistic topic models},
  author={Steyvers, Mark and Griffiths, Tom},
  booktitle={Handbook of latent semantic analysis},
  pages={439--460},
  year={2007},
  publisher={Psychology Press}
}

@article{Groeling2013,
  author = {Groeling, T.},
  title = {Media bias by the numbers: Challenges and opportunities in the empirical study of partisan news},
  journal = {Annual Review of Political Science},
  volume = {16},
  number = {1},
  pages = {129--151},
  year = {2013}
}

@article{Ksiazek2015,
  author = {Ksiazek, T. B.},
  title = {Civil interactivity: How news organizations' commenting policies explain civility and hostility in user comments},
  journal = {Journal of Broadcasting \& Electronic Media},
  volume = {59},
  number = {4},
  pages = {556--573},
  year = {2015}
}

@article{Ortiz2020,
  author = {Ortiz, S. M.},
  title = {Trolling as a collective form of harassment: An inductive study of how online users understand trolling},
  journal = {Social Media + Society},
  volume = {6},
  number = {2},
  pages = {2056305120928512},
  year = {2020}
}

@article{Perloff2022,
  author = {Perloff, R. M.},
  title = {The fifty-year legacy of agenda-setting: Storied past, complex conundrums, future possibilities},
  journal = {Mass Communication and Society},
  volume = {25},
  number = {4},
  pages = {469--499},
  year = {2022}
}

@article{Rieger2018,
  author = {Rieger, D. and Schmitt, J. B. and Frischlich, L.},
  title = {Hate and counter-voices in the internet: Introduction to the special issue},
  journal = {SCM Studies in Communication and Media},
  volume = {7},
  number = {4},
  pages = {459--472},
  year = {2018}
}

@article{vayansky2020review,
  title={A review of topic modeling methods},
  author={Vayansky, Ike and Kumar, Sathish AP},
  journal={Information Systems},
  volume={94},
  pages={101582},
  year={2020},
  publisher={Elsevier}
}

@article{siew2019cognitive,
  title={Cognitive network science: A review of research on cognition through the lens of network representations, processes, and dynamics},
  author={Siew, Cynthia SQ and Wulff, Dirk U and Beckage, Nicole M and Kenett, Yoed N},
  journal={Complexity},
  volume={2019},
  number={1},
  pages={2108423},
  year={2019},
  publisher={Wiley Online Library}
}

@article{hills2009longitudinal,
  title={Longitudinal analysis of early semantic networks: Preferential attachment or preferential acquisition?},
  author={Hills, Thomas T and Maouene, Mounir and Maouene, Josita and Sheya, Adam and Smith, Linda},
  journal={Psychological Science},
  volume={20},
  number={6},
  pages={729--739},
  year={2009},
  publisher={SAGE Publications Sage CA: Los Angeles, CA}
}

@article{morais2013mapping,
  title={Mapping the structure of semantic memory},
  author={Morais, Ana Sofia and Olsson, Henrik and Schooler, Lael J},
  journal={Cognitive Science},
  volume={37},
  number={1},
  pages={125--145},
  year={2013},
  publisher={Wiley Online Library}
}

@article{kumar2022critical,
  title={A critical review of network-based and distributional approaches to semantic memory structure and processes},
  author={Kumar, Abhilasha A and Steyvers, Mark and Balota, David A},
  journal={Topics in Cognitive Science},
  volume={14},
  number={1},
  pages={54--77},
  year={2022},
  publisher={Wiley Online Library}
}

@article{Rossini2022,
  author = {Rossini, P.},
  title = {Beyond incivility: Understanding patterns of uncivil and intolerant discourse in online political talk},
  journal = {Communication Research},
  volume = {49},
  number = {3},
  pages = {399--425},
  year = {2022}
}

@article{Stroud2016,
  author = {Stroud, N. J. and Scacco, J. M. and Curry, A. L.},
  title = {The presence and use of interactive features on news websites},
  journal = {Digital Journalism},
  volume = {4},
  number = {3},
  pages = {339--358},
  year = {2016}
}

@article{Tausczik2020,
  author = {Tausczik, Y. and Huang, X.},
  title = {Knowledge generation and sharing in online communities: Current trends and future directions},
  journal = {Current Opinion in Psychology},
  volume = {36},
  pages = {60--64},
  year = {2020}
}

@article{Rodrigo-Gines2024,
  author    = {Rodrigo-Gin{\'e}s, F. J. and Carrillo-de-Albornoz, J. and Plaza, L.},
  title     = {A systematic review on media bias detection: What is media bias, how it is expressed, and how to detect it},
  journal   = {Expert Systems with Applications},
  volume    = {237},
  pages     = {121641},
  year      = {2024},
  doi       = {https://doi.org/10.1016/j.eswa.2023.121641}
}

@article{Ransbotham2011,
  author    = {Ransbotham, S. and Kane, G. C.},
  title     = {Membership turnover and collaboration success in online communities: Explaining rises and falls from grace in Wikipedia},
  journal   = {MIS Quarterly},
  pages     = {613--627},
  year      = {2011}
}

@article{Cruz2018,
  author    = {Cruz, A. G. B. and Seo, Y. and Rex, M.},
  title     = {Trolling in online communities: A practice-based theoretical perspective},
  journal   = {The Information Society},
  volume    = {34},
  number    = {1},
  pages     = {15--26},
  year      = {2018}
}

@article{van2008visualizing,
  title={Visualizing data using t-SNE.},
  author={Van der Maaten, Laurens and Hinton, Geoffrey},
  journal={Journal of machine learning research},
  volume={9},
  number={11},
  year={2008}
}

@article{rudin1992nonlinear,
  title={Nonlinear total variation based noise removal algorithms},
  author={Rudin, Leonid I and Osher, Stanley and Fatemi, Emad},
  journal={Physica D: nonlinear phenomena},
  volume={60},
  number={1-4},
  pages={259--268},
  year={1992},
  publisher={Elsevier}
}

@article{flottum2017narratives,
  title={Narratives in climate change discourse},
  author={Fl{\o}ttum, Kjersti and Gjerstad, {\O}yvind},
  journal={Wiley Interdisciplinary Reviews: Climate Change},
  volume={8},
  number={1},
  pages={e429},
  year={2017},
  publisher={Wiley Online Library}
}

@article{card2022computational,
  title={Computational analysis of 140 years of US political speeches reveals more positive but increasingly polarized framing of immigration},
  author={Card, Dallas and Chang, Serina and Becker, Chris and Mendelsohn, Julia and Voigt, Rob and Boustan, Leah and Abramitzky, Ran and Jurafsky, Dan},
  journal={Proceedings of the National Academy of Sciences},
  volume={119},
  number={31},
  pages={e2120510119},
  year={2022},
  publisher={National Academy of Sciences}
}

@article{jing2021characterizing,
  title={Characterizing partisan political narrative frameworks about COVID-19 on Twitter},
  author={Jing, Elise and Ahn, Yong-Yeol},
  journal={EPJ data science},
  volume={10},
  number={1},
  pages={53},
  year={2021},
  publisher={Springer Berlin Heidelberg}
}

@article{aragon2017generative,
  title={Generative models of online discussion threads: state of the art and research challenges},
  author={Arag{\'o}n, Pablo and G{\'o}mez, Vicen{\c{c}} and Garc{\'\i}a, David and Kaltenbrunner, Andreas},
  journal={Journal of Internet Services and Applications},
  volume={8},
  pages={1--17},
  year={2017},
  publisher={Springer}
}

@article{bollenbacher2021challenges,
  title={On the challenges of predicting microscopic dynamics of online conversations},
  author={Bollenbacher, John and Pacheco, Diogo and Hui, Pik-Mai and Ahn, Yong-Yeol and Flammini, Alessandro and Menczer, Filippo},
  journal={Applied Network Science},
  volume={6},
  pages={1--21},
  year={2021},
  publisher={Springer}
}

@article{bond2023rise,
  title={The rise of presidential eschatology: Conspiracy theories, religion, and the January 6th insurrection},
  author={Bond, Bayleigh Elaine and Neville-Shepard, Ryan},
  journal={American Behavioral Scientist},
  volume={67},
  number={5},
  pages={681--696},
  year={2023},
  publisher={SAGE Publications Sage CA: Los Angeles, CA}
}

@techreport{andre2024narratives,
  title={Narratives about the Macroeconomy},
  author={Andre, Peter and Haaland, Ingar and Roth, Christopher and Wiederholt, Mirko and Wohlfart, Johannes},
  year={2024},
  institution={SAFE Working Paper}
}

@article{lee2022storm,
  title={Storm the capitol: linking offline political speech and online Twitter extra-representational participation on QAnon and the January 6 insurrection},
  author={Lee, Claire Seungeun and Merizalde, Juan and Colautti, John D and An, Jisun and Kwak, Haewoon},
  journal={Frontiers in Sociology},
  volume={7},
  pages={876070},
  year={2022},
  publisher={Frontiers Media SA}
}

@article{jimenez2024effect,
  title={The effect of content moderation on online and offline hate: Evidence from Germany's NetzDG},
  author={Jim{\'e}nez Dur{\'a}n, Rafael and M{\"u}ller, Karsten and Schwarz, Carlo},
  journal={Available at SSRN 4230296},
  year={2024}
}

@article{MuellerSchwarzJEEA2020,
author = {Müller, Karsten and Schwarz, Carlo},
title = "{Fanning the Flames of Hate: Social Media and Hate Crime}",
journal = {Journal of the European Economic Association},
volume = {19},
number = {4},
pages = {2131-2167},
year = {2020},
month = {10},
issn = {1542-4766},
doi = {10.1093/jeea/jvaa045},
url = {https://doi.org/10.1093/jeea/jvaa045},
eprint = {https://academic.oup.com/jeea/article-pdf/19/4/2131/39651047/jvaa045.pdf},
}

@article{goldenberg2020collective,
  title={Collective emotions},
  author={Goldenberg, Amit and Garcia, David and Halperin, Eran and Gross, James J},
  journal={Current directions in psychological science},
  volume={29},
  number={2},
  pages={154--160},
  year={2020},
  publisher={Sage Publications Sage CA: Los Angeles, CA}
}

@article{bak2021stewardship,
  title={Stewardship of global collective behavior},
  author={Bak-Coleman, Joseph B and Alfano, Mark and Barfuss, Wolfram and Bergstrom, Carl T and Centeno, Miguel A and Couzin, Iain D and Donges, Jonathan F and Galesic, Mirta and Gersick, Andrew S and Jacquet, Jennifer and others},
  journal={Proceedings of the National Academy of Sciences},
  volume={118},
  number={27},
  pages={e2025764118},
  year={2021},
  publisher={National Academy of Sciences}
}

@article{waller2021quantifying,
  title={Quantifying social organization and political polarization in online platforms},
  author={Waller, Isaac and Anderson, Ashton},
  journal={Nature},
  volume={600},
  number={7888},
  pages={264--268},
  year={2021},
  publisher={Nature Publishing Group UK London}
}

@article{oldemburgo2024twitter,
  title={Twitter (X) use predicts substantial changes in well-being, polarization, sense of belonging, and outrage},
  author={Oldemburgo de Mello, Victoria and Cheung, Felix and Inzlicht, Michael},
  journal={Communications Psychology},
  volume={2},
  number={1},
  pages={15},
  year={2024},
  publisher={Nature Publishing Group UK London}
}

@article{soroka2012gatekeeping,
  title={The gatekeeping function: Distributions of information in media and the real world},
  author={Soroka, Stuart N},
  journal={The Journal of Politics},
  volume={74},
  number={2},
  pages={514--528},
  year={2012},
  publisher={Cambridge University Press New York, USA}
}

@article{buerger2019counterspeech,
  title={Counterspeech: A literature review},
  author={Buerger, Catherine and Wright, Lucas},
  journal={Available at SSRN 3829816},
  year={2019}
}

@article{buerger2021iamhere,
  title={iamhere: Collective counterspeech and the quest to improve online discourse},
  author={Buerger, Catherine},
  journal={Social Media+ Society},
  volume={7},
  number={4},
  pages={20563051211063843},
  year={2021},
  publisher={SAGE Publications Sage UK: London, England}
}

@incollection{bishop2012psychology,
  title={The psychology of trolling and lurking: The role of defriending and gamification for increasing participation in online communities using seductive narratives},
  author={Bishop, Jonathan},
  booktitle={Virtual community participation and motivation: Cross-disciplinary theories},
  pages={160--176},
  year={2012},
  publisher={IGI Global Scientific Publishing}
}

@inproceedings{kobayashi2016tideh,
  title={Tideh: Time-dependent hawkes process for predicting retweet dynamics},
  author={Kobayashi, Ryota and Lambiotte, Renaud},
  booktitle={Proceedings of the international AAAI conference on web and social media},
  volume={10},
  number={1},
  pages={191--200},
  year={2016}
}

@article{wu2007novelty,
  title={Novelty and collective attention},
  author={Wu, Fang and Huberman, Bernardo A},
  journal={Proceedings of the National Academy of Sciences},
  volume={104},
  number={45},
  pages={17599--17601},
  year={2007},
  publisher={National Academy of Sciences}
}

@inproceedings{leskovec2009meme,
  title={Meme-tracking and the dynamics of the news cycle},
  author={Leskovec, Jure and Backstrom, Lars and Kleinberg, Jon},
  booktitle={Proceedings of the 15th ACM SIGKDD international conference on Knowledge discovery and data mining},
  pages={497--506},
  year={2009}
}

@misc{Disqus,
  title = {Disqus platform},
  howpublished = {\url{https://disqus.com/}},
  note = {Accessed: 2025-04-01}
}

@article{bacaksizlar2023group,
  title={Group threat, political extremity, and collective dynamics in online discussions},
  author={Bacaksizlar Turbic, N Gizem and Galesic, Mirta},
  journal={Scientific Reports},
  volume={13},
  number={1},
  pages={2206},
  year={2023},
  publisher={Nature Publishing Group UK London}
}

@article{muchnik2013social,
  title={Social influence bias: A randomized experiment},
  author={Muchnik, Lev and Aral, Sinan and Taylor, Sean J},
  journal={Science},
  volume={341},
  number={6146},
  pages={647--651},
  year={2013},
  publisher={American Association for the Advancement of Science}
}

@article{bowing2024news,
  title={The news comment gap and algorithmic agenda setting in online forums},
  author={B{\"o}wing, Flora and Gildersleve, Patrick},
  journal={arXiv preprint arXiv:2408.07052},
  year={2024}
}

\end{document}



\maketitle
\SItext

\section*{1. Computational model formulation}

  Here, we provide an analytic description of the computational model of the collective mind dynamics of the online news community proposed in the main manuscript. We first initialized the general and community semantic networks and iterated the frequency and weight updates $T$ times to obtain the model's simulated results. The following description aims to formulate a general framework for our computational model; thus, all specific model settings, hyperparameters, and functional forms of the distributions used in our main study are explicitly specified in Table \ref{table:s1} for the reader's convenience. Note that most of the model settings we employed are chosen to reflect empirical findings from our data (see the captions in Table \ref{table:s1} for more details), and one can freely alter the settings of our framework to suit one's data and model different online communities. Also, note that, aside from the configurable settings listed in \ref{table:s1}, our model has exactly two free parameters that characterize individual community characteristics: the filter strength ($\lambda_f$) and the memory strength ($\lambda_m$).

  \subsection*{Semantic network definition} 
  
  At any given time $t$, the \textbf{general semantic network} from the current time $t$ is expressed as $G^g_{t} = (V^g, E^g, F_{t}^g, W_{t}^g)$, where $V^g$ and $E^g$ denotes the set of vertices and edges, respectively. Here, we assume the topic vertices and edges between them are persistent through time, and there are a total of $|V^g| = N$ vertices (topics) and $|E^g| = N(N-1)/2$ edges since the network is complete without self-loops. Each vertex $v_{i}^g \in V^g$ indicates a single topic and has a \textbf{normalized frequency value} $f_{i,{t}}^g \in F_{t}^g$ where $F_{t}^g$ is the set of all normalized frequency values at time ${t}$ and $\sum_i f^g_{i, t} = 1$. From this normalized frequency, we can define a \textbf{normalized frequency ranking} $r_{i,{t}}^g \in R_{i,{t}}^g$ where $r_{i,{t}}^g = \text{rank}(f_{i, t}^g)/N$ and $\text{rank}(f_{i, t}^g)$ denotes a ranking of $f_{i, t}^g$ among $F_{t}^g$. Note that by definition, $r_i$ lies between $0$ and $1$. Each edge $e_{ij,{t}}^g \in E_{t}^g$ indicates the semantic closeness between two topics, and has a \textbf{weight value} $w_{ij,{t}}^g \in W_{t}^g$ where $0 \leq w_{ij,{t}}^g \leq 1$ and $W_{t}^g$ is the set of all weight values at time ${t}$.  Also, we consider $K$ different community with respective \textbf{community semantic network} $G^{k}_{t} = (V^{k}, E^{k}, F_{t}^{k}, W_{t}^{k})$ at time ${t}$, which shares vertices and edges $(V^{k} = V^g, E^{k} = E^g)$ but with (potentially) different values for $F_{t}^{k}$ ($R_{t}^{k}$) and $W_{t}^{k}$. Hence, we drop the superscript for $V$($v$) and $E$($e$) from here for simplicity.

  \subsection*{Semantic network initialization} 
  
  Without loss of generality, we set the general semantic network's initial ranking order to follow the indices, i.e., topic $1$ is the first most frequent, topic $2$ is the second most frequent, and so on ($r_{i,{0}}^g = i$). We achieve this ranking by setting initial frequencies $f_{i, 0}^g = F_{f}(i)$, where $F_{f}$ is a monotonically decreasing \textbf{initial frequency distribution}. In our study, we chose $F_{f} \propto r_{i}^{-1}$, which leads to $f_{i, 0}^g = {r_{i,{0}}^g}^{-1}/C = i^{-1}/C$ where $C=\sum_1^N i^{-1}$. Note that we preserve this initial distribution after the updating (See \ref{sec:ctnu}), so the while the ranking of each topic changes with respect to the frequency at the given time ($f_{i, t}^g = {r_{i,{t}}^g}^{-1}/C$), the frequency distribution remains the same. We also sample the weights $w_{ij, 0}^g$ from the \textbf{initial weight distribution} $F_{w}$, finishing the initialization of the general semantic network. We employed the log-normal distribution for the initial weight distribution, $F_{w} \propto e^{{-\ln^2(\frac{x-a}{b})}/{2s^2}}$, where $a$, $b$, and $s$ are the parameters that control the distribution's shape.

    We further initialize the community semantic network depending on the initial settings. For each community, we first copy the frequencies and weights from the general semantic network and perturb them by adding noise. In our study, we used $f_{i, 0}^{k} = f_{i, 0}^g + \mathcal{N}(0, \sigma_{\text{fp}}) \times F_{f}(i)$ to ensure the noise scale matters for all frequency ranges, where $\sigma_{\text{fp}}$ denotes the standard deviation for the frequency perturbation. Similarly,  $w_{ij, 0}^{k} = w_{ij, 0}^g + \mathcal{N}(0, \sigma_{\text{wp}})$, where $\sigma_{wp}$ denotes the standard deviation for the weight perturbation. We used $\sigma_{\text{fp}}=0$ and $\sigma_{\text{wp}} = 0$ in most of the cases, which assumes the community semantic network is identical to the general semantic network at the beginning (equilibrium state). For the simulation in a non-equilibrium state (e.g., Alignment (Fig. 4a) and Membership turnover (Fig. 5a) scenario), we used $\sigma_{\text{fp}}=1.0$ and $\sigma_{\text{wp}} = 0.05$.

  \subsection{Events generation} 
  
  At each time step $t$, the general semantic network $G^g_t$ generates a new set of \textbf{events} $X_{t}=\{x_{1, t}, \allowbreak x_{2, t}, \cdots, x_{N_x, t}\}$ for the current timestep, where $N_x$ denotes the number of events per each time step. Each event is consists of $N_{w}$ number of topics, $x_{a, t} = \{v_{i, t, 1}, v_{i, t, 2}, \cdots, v_{i, t, N_w}\} = \{v_{z_1}, v_{z_2}, \cdots, v_{z_{N_{w}}}\}$, where the $z_q$ denotes $q$-th tier topic of the event. This definition implies that the $q$-th tier topic of the $i$-th event at time $t$ ($v_{i, t, q}$) is a topic numbered as $z_q$ ($v_{z_q}$). For this work, we choose $N_w=3$, so the event is described as a triplet of topics. For each event, we sample topics for each tier $q$ with a probability proportional to the $v_{i, t, q} \sim F_{ns}(r_{z_q, t}^g)$, where the $F_{\text{ns}}$ denotes the \textbf{event sampling distribution}. In our study, we chose $F_{\text{ns}} \propto -\ln(r_{i, t}^g)$, independent of tier and a monotonically decreasing function of the ranking $r_{i, t}^g$. Also, we ensured the $v_{i, t, q}$ are unique for each tier by sequentially sampling each tier while excluding all the previous tier's topics and renormalizing $F_{\text{ns}}$ accordingly.

    If one needs to ensure the time correlation of the event topics, one can consider the previous event topics as a prior for the current event topics. This can be achieved by first sampling the new title topic frequency distribution from the given distribution $\hat{F}_{ns}(r_{i, t}^g)$ every time step, and creating a set of events by choosing topics from linearly interpolated distribution $F_{\text{ns}}(r_{i, t}^g) = (1-\lambda_e)\hat{F}_{\text{ns}}(r_{i, t}^g) + \lambda_e F_{\text{ns}}(r_{i, t-1}^g)$. To ensure the uniqueness of the topic in the events, we employed rejection sampling, where we repeated the sampling till there were no events with the duplicate topic in the events set. In this study, we adopted this setting with the event memory strength $\lambda_e=0.5$.

  \subsection*{Filter definition} 
  After the event generation, each event $x_{j, t}$ passes through a filter of each community and is determined whether it will be filtered or not, and posted as news, i.e., the filtered event becomes news. We first specify the editors' criteria for the filtering, which is determined by their view on both general and community semantic networks, defined as follows.
    \begin{align} 
    \bar{f}_{i, t}^{k} = \lambda_f f_{i, t}^{k} + (1-\lambda_f) f_{i, t}^g \label{eq:5} \\
    \bar{w}_{ij, t}^{k} = \lambda_f w_{ij, t}^{k} + (1-\lambda_f) w_{ij, t}^g \label{eq:6}
    \end{align}

    Here, both the view on normalized frequency rank $\bar{r}_{i, t}^{k}$ (which is derived from $\bar{f}_{i, t}^{k}$) and weight $\bar{w}_{i, t}^{k}$ are controlled by the filter strength $0 \leq \lambda_f \leq 1$. The filter strength $\lambda_f$ serves as a role of linear interpolation parameter between general and community semantic networks and determines whether the editors' criteria are more inclined to the outside world or their community.

    With these, the filter of each community consists of a two-stage sampling process; one considering the frequency of topics ($\bar{f}_{i, t}^{k}$), and another considering the similarity between topics ($\bar{w}_{i, t}^{k}$). The filtering ratio $0 \leq R_1, R_2 \leq 1$ determines how much of the events will survive for the first and second filtering, respectively. First, we calculate the product of exponentiated frequencies of topics as $\prod_{q} (\bar{r}_{{z_q}, t}^{k})^{\alpha_q}$, where $\alpha_q$ denotes the $q$-th tier \textbf{filter exponent}, and normalize them as a probability for each event. We then sample $R_1$ of the events (without replacement) according to this probability. With this filtered events, we further calculate the product of similarities between topics as $\prod_{q_1, q_2} \bar{w}_{{z_{q_1}}{z_{q_2}}, t}^{k} $ for each event $x_{i, t} = \{v_{z_1}, v_{z_2}, \cdots, v_{z_{N_w}}\}$, and we keep only the top $R_2$ of the filtered event by sorting them based on this sum. Finally, we keep a total of $R_1R_2N_x = \bar{N_x}$ events that pass both filters, and the rest of the events are filtered out. We denote filtered events as $X_{t}^{k} = \{x_{1, t}^{k}, x_{2', t}^{k}, \cdots, x_{\bar{N_x}, t}^{k}\} \subseteq X_t$ for each community $k$ at time $t$. This process is equivalent to considering both the perceived importance (frequency) and inter-topic similarity (weight) of the topics in the event, in order to decide whether the editors accept it as news in their community or not.

    In practice, if we want to calibrate the model with $\alpha_q$, we apply $\alpha_q/R_2$ at the filtering stage. This is because the exponent gets decreased due to the second stage of the filtering, which is effectively random (since there is no correlation between weight and frequency in the beginning). Intuitively, random sampling reduces the steepness of the original distribution, which is equivalent to scaling down the exponent. Strictly speaking, the assumption of the non-correlation between the weight and frequency is not always true, as the correlation slowly builds up as the model evolves because the inter-topic weight(similarity) increases as the co-occurrence between two topics happens, and the topic with higher frequency generally has a higher chance to get this. However, we found that this effect is negligible in practice, and the calibration with $\alpha_q/R_2$ is sufficient to capture the overall behavior of the model, especially in the early stage.

  \subsection*{Comment semantic network generation} 
  
  The filtered events (news) will elicit responses from the collective mind of the community as a form of comments. Based on empirical evidence, we make two model assumptions. First, the frequency of comments that match the subject of the news, which we'll call on-topic comments, increases. Also, the appearance of the specific topic pair in the news increases the weight between those topics. Combining these two, we define the \textbf{comment network} for community $k$ at time $t$ as $A^{k}_{t} = (V, E, \hat{F}_{t}^{k}, \hat{W}_{t}^{k})$, which shares vertices and edges with other semantic networks, but with comment frequency $\hat{f}_{i, t}^{k} \in \hat{F}_{i, t}^{k}$ and comment weight $\hat{w}_{ij, t}^{k} \in \hat{W}_{ij, t}^{k}$. 

    First, we need to construct the \textbf{comment frequency}, which is a direct sum of all comment frequency distributions under the news. For given news $x_{i, t}^{k} = \{v_{z_1}, v_{z_2}, \cdots \allowbreak, v_{z_{N_w}}\}$ (where the $q$-th tier topic is $z_q$), we first assign the relative number of comments under this news by sampling from a comment number distribution, $c_{i, t}^{k} \sim P_c(c, x_{i, t}^{k})$ (In our implementation, we used topic-independent sampler, hence $P_c(c, x_{i, t}^{k}) = P_c(c)$). We then determine whether the comment multiplier would be zero or non-zero (for each tier) by sampling a uniform random number from $0$ to $1$ and comparing it to the tier-wise zero ratio, $Z_q(r)$, and setting it to zero if it is smaller than the ratio. If the value is higher and comment multiplier is determined to be a non-zero value, now we sample tier-wise comment multipliers from a (non-zero) tier-wise \textbf{comment multiplier distribution}, $m_{i, t, q}^{k} \sim P_{m, q}(m, r_{z_q}^{k}, c_{i, t}^{k})$. More precisely, the multiplier distribution is a function of the tier $q$ itself (denoted in the subscript), community topic ranking for each tier $r_{z_q}^k$, and the comment number $c_{i, t}^k$. From these comment multipliers, we get the comment frequency distribution under the news $x_{i, t}^{k}$ as follows.

    \begin{align} 
      \hat{f}_{j, t}^{k}(x_{i, t}^{k}) = \begin{cases}
        c_{i, t}^{k}m_{i, t, q}^{k}f_{j, t}^{k}  &\text{if $v_j = v_{z_q}$}\\
        c_{i, t}^{k}f_{j, t}^{k} / C_{i, t}^k &\text{if $v_j \notin x_{i, t}^{k}$}
        \end{cases} \label{eq:3} \\
        C_{i, t}^k =\frac{1-\min(\sum_q m_{i, t, q}^{k} f_{q, t}^{k}, 1)}{1-\sum_q f_{q, t}^{k}} \label{eq:4}
    \end{align}

    Here, $C_{i, t}^k$ is the normalization constant for off-topic comments to keep the assigned comment number, and the subscripts $i$ and $j$ denote the $i$-th news and $j$-th topic, respectively. Basically, this means that we would like to multiply the frequency of the $q$-th tier on-topic comments by $m_q$, and the rest of the assigned comments simply follow the previous community frequency distribution. Note that this implementation sometimes results in the sum of the frequency of the comments being more than the assigned number of comments. We find that this exception happens rarely in practice (less than $2\%$), hence it does not affect the overall comment number distribution. The overall comment frequency at time $t$ then becomes the sum of all comment frequency distributions under the news, $\hat{f}_{j, t}^{k} = \sum_i \hat{f}_{j, t}^{k}(x_{i, t}^{k})$.

    Referencing the observation from the empirical data (See Fig. \ref{fig:s1} and SI Appendix, section 2), we implemented the comment multiplier distribution by splitting the distribution into two parts: one with zero multiplier and one without. For the zero case, we assign the zero multiplier ratio $Z_q(r)=C_{z, q}r$ for each tier $q$, which is a linear function of normalized comment frequency ranking $r$ and denotes the probability that the multiplier of interest is zero. We perform the \textit{zero-check} by using a Bernoulli trial with $p=Z_q(r)$. If it passes this zero check, we sample the multiplier from the non-zero distribution $P^{nz}_{m, q}$, which is a function of the tier $q$ itself, the community topic ranking for each tier $r_{z_q}^k$, and the comment number $c_{i, t}^k$.

    Now, we need to construct the \textbf{comment weight}. For each news $x_{i, t}^{k} = \{v_{z_1}, v_{z_2}, \cdots \allowbreak, v_{z_{N_w}}\}$, we first define a set of co-occurring pairs between news topics $S$ and assign comment weights as follows,

    \begin{align} 
      S_{i, t}^{k} = \{\, (a, b) \mid v_a \in x_{i, t}^{k} \land v_b \in x_{i, t}^{k}\,\}, \label{eq:51} \\
      \hat{w}_{ab, t}^{k}(x_{i, t}^{k}) = \begin{cases}
        c_{i, t}^k  &\text{if $(a, b) \in S_{i, t}^{k}$}\\
        0 &\text{if $(a, b) \notin S_{i, t}^{k}$},
        \end{cases} \label{eq:61}
    \end{align}

    where $c_{i, t}^k$ is the assigned comment number for the news $x_{i, t}^{k}$. This setting implies that co-occurring topic pairs in news articles with many comments will have a high impact on the increase in inter-topic similarity. Similar to the comment frequency, the overall comment weight at time $t$ becomes the sum of all comment weights under the news, $\hat{w}_{ij, t}^{k} = \sum_i \hat{w}_{ij, t}^{k}(x_{i, t}^{k})$. Note that both comment frequency and weight are not properly normalized at this point, and we will normalize them at the update step.

    \subsection*{Community semantic network update} \label{sec:ctnu} 
    
    From the comment semantic network, we finally update the community semantic network to complete the feedback loop. For the frequency, we adopt a \textbf{memory strength} $0 \leq \lambda_m \leq 1$ to retain the previous frequency distribution and update it as follows. First, we construct a proxy frequency distribution for this time step as 

    \begin{equation}
    \hat{f}_{i, t+1}^{k} = \lambda_m f_{i, t}^{k} + (1-\lambda_m)\hat{f}_{i, t}^{k} / \sum_j \hat{f}_{j, t}^{k}. \label{eq:7}
    \end{equation}

    With this proxy frequency, we update the frequency by first computing the rank according to the proxy frequency and assigning the frequency of that rank, $f_{i, t+1}^{k} = F_f(\text{rank}(\hat{f}_{i, t+1}^{k})/N)$. This effectively quantizes the possible frequency and ensures the initial frequency distribution $F_f$ is preserved after the update. Note that this only enforces the distribution of the frequency (unobservable in real data), not the comment frequency (observable in real data), resulting from an additional sampling process.

    For the weight, we employed a Hebbian learning scheme \cite{hebb2005organization} to update it. For each pair of topics $v_i$ and $v_j$ in the community semantic network, we update the weight as follows.
    \begin{equation}
    w_{ij, t+1}^{k} = \eta(w_{\max}-|w_{ij, t}^{k}|) \hat{w}_{ab, t}^{k} / D_t^k - \gamma w_{ij, t}^{k} + \epsilon_{ij} \label{eq:8}
    \end{equation}

    Here, $\eta$ is a learning rate, $w_{\max}$ is a maximum cap for a weight value, $\gamma$ is a decaying rate, and $\epsilon_{ij}$ is a Gaussian noise with $\mathcal{N}(0, \sigma_{wn}^2)$. Again, the weight is normalized by $D_t^k = \frac{N_w(N_w-1)}{2}\sum_i c_{i, t}^{k} $ before the update, which considers the number of total comments and the possible number of topic pairs based on the event $N_w$. Also, to ensure stability, we used an adaptive decaying rate $\gamma(w_{ij, t}^k, \hat{w}_{ij, t}^k)$ as 
    \begin{equation}
      \gamma(w_{ij, t}^k, \hat{w}_{ij, t}^k)=\eta \frac{\sum_{i, j} (w_{\max}-|w_{ij, t}^{k}|) \hat{w}_{ab, t}^{k} / D_t^k + \epsilon_{ij}}{\sum_{ij}w_{ij, t}^{k}},
    \end{equation}
    which normalizes the decaying rate by the relative scale of the Hebbian learning term.

    With these updated frequencies and weights of each community semantic network, the full iteration is ended, and we repeat this process $T$ times to get the simulated result of the model.

    We mainly calibrated and selected the functional form and constants of our model based on its counterpart in the empirical data, except in some notable cases. For the semantic network, since it's not directly observable, we used the distribution from the comment network in the empirical data (See SI Appendix, section 9 for discussion), and used a Gaussian perturbation to simulate community semantic networks from the general semantic network. Filtering ratios ($0.5, 0.5$) are chosen arbitrarily but feasibly and can be easily modified if one has prior knowledge of the community's filtering behavior (for instance, the survival rate of the initial draft). They only control the effective time scale of a single time step. For the learning rate and weight noise standard deviation, we chose parameters to ensure model stability. We summarized these in the Rationale column in Table \ref{table:s1}, where 1 is for the informed choices from the empirical data (check Fig. 2 in the main manuscript for 1$^*$ and SI appendix section 6 for 1$\dagger$), and 2 is for arbitrary choices for the simulation and model stability.

\section*{2. Hypersensitive filter ($\lambda_f > 1$)}

In the main manuscript, we set our model's filter strength ($\lambda_f$) between $0$ and $1$. However, our formulation enables us to expand this into the case where the filter strength $\lambda_f$ is greater than $1$, which we call a hypersensitive filter. The hypersensitive filter is not only more inclined to the community semantic network but also actively avoids the general semantic network by negatively assessing their frequency and weights during the filtering process. Since it extrapolates from the original linear interpolation range, the criteria frequency and weight (which represents the worldview of the filter) in both equations \ref{eq:5} and \ref{eq:6} can be negative. Although negative frequency and weight are not meaningful in our model, it doesn't matter since they only appear in the intermediate step of the filtering process. Precisely, we only use the rank of those values, which is perfectly valid even if any of the values are negative.

We first investigate the behavior of the model with a hypersensitive filter by varying the filter strength $\lambda_f$ from $0.2$ to $3.0$ and fixing the memory strength $\lambda_m=0.9$ (Fig. \ref{fig:s10}a). We found that the model with a relatively weak hypersensitive filter shows a similar trend as the model with $\lambda_f < 1$; the distance between the general and community semantic network decreases over time. But, as the filter strength increases (typically $\lambda_f > 1.5$), the distance between the two semantic networks increases over time, suggesting that a strong hypersensitive filter can repel the community semantic network from the general semantic network. This is well shown in the t-SNE plot of the comment frequency profile (Fig. \ref{fig:s10}b), where the model with $\lambda_f=3.0$ shows a clear separation from the general semantic network while the model with $\lambda_f=0.2$ is attracted. This separation resembles the behavior of a community with an extreme echo chamber effect, which strongly rejects the conventional norm and reinforces the community-specific view that is drastically different from the rest of society. With these demonstrations, we show that our model is capable of capturing those radical behaviors of the community by simply tuning the filter parameter.

\section*{3. Empirical data preparation}

Here, we provide a detailed description of the empirical data from online news communities used in the main manuscript. 

We collected data from five online news communities, namely, Mother Jones (MJ), Atlantic (AT), The Hill (TH), Breitbart (BB), and Gateway Pundit (GP). The collected data consists of news articles (hereafter 'news') and comments on the respective websites within varying periods. We crawled the data using the \texttt{Disqus} API, which functioned as a common platform for commenting on various websites during the period. The data includes mainly the news title text and comment text along with the timestamp, but other metadata were also collected, such as the number of likes on comments and user ID (which are not used in this study). 

We first preprocessed the whole dataset by applying several cleansing steps to the data. For the news title, we removed all the news that contains HTML addresses (since these are typically not genuine news, but rather corrupted data or a duplicate of another news), and removed all news from further analysis that has equal to or fewer than $\theta_n = 10$ comments. For the comments, we removed all the HTML tags and consecutive spaces for further processing. The summary of the collected empirical data is provided in Table \ref{table:s3}, and the time series of the number of news and comments before and after the filtering is shown in Fig. \ref{fig:s4}.

In this study, we used aggregated data for all analyses where data were pooled and added together over a given period. For the aggregated data, as mentioned in the main manuscript, we merged the news posted during $1$-month intervals, and only the comments made within $7$-days from the news post date were valid to be aggregated. During the process, we also removed news that is classified (in its top-3 classification) as an outlier (topic "$-1$") or contains fewer than $\theta_n$ non-outlier comments, to focus on a more meaningful (non-outlier) distribution. The rationale behind this removal is that articles that only have outlier comments (and fewer than $\theta_n$ non-outlier comments) have a high chance of only containing simple expressions and not significantly contributing to the landscape of the collective mind. Note that we did not remove all outlier comments at this stage, although most of the analysis in this study (unless specified) was done with a non-outlier comments distribution. Finally, after both of the filterings (removing overdue comments and outliers), we further removed all news that had fewer than $\theta_n$ comments. The summary of the filtered data is provided in Table \ref{table:s4}.

\section*{4. Topic modeling with BERTopic}

In this work, we employed BERTopic\cite{grootendorst2022bertopic} for the construction of topic models (TMs). With the given model settings (See the methods in the main manuscript), the construction consists of two steps: (1) the fitting phase, where we fit the model with sampled comments from the full data, and (2) the transforming phase, where the rest of the comments are classified based on the fitted model. We performed the following procedures to construct the global TM, which used data from all $5$ communities combined, and also for the local TM, which used data from each community separately. Note that we mainly used the result from global TM (which is referred to as plain "topic model" in the main manuscript) for the analysis, and the local TM was used for the validation of the overall results.

  \subsection*{Local topic model} For the fitting phase, we sample 2 million comments from each of the five communities using a variant of stratified sampling to better preserve the overall trend of comments without ignoring the influence of smaller news articles. Precisely, given the histogram of comment numbers, we choose the sampling threshold $k'$ that matches the following condition,

  \begin{equation}
    k^* = \arg \max_{k^*} \left\{ k^* \mid \sum_{k=1}^{k^*} k \cdot X(k) + k^* \cdot \sum_{k=k^*}^{k_{\max}} X(k) < S \right\}
  \end{equation}

  where $X(k)$ is a histogram of the number of news articles depending on the number of comments $k$, $k_{\max}$ is the maximum number of comments, and $S$ is a sampling size ($2$ million). Simply, given a threshold $k^*$, we collect all of the comments from the news articles that have fewer than $k^*$ comments and randomly sample $k^*$ comments from the news articles that have more than $k^*$ comments, so every news article has at most $k^*$ sampled comments. The $k^*$ values for each community are $80$ for Mother Jones, $87$ for Atlantic, $4$ for The Hill, $6$ for Breitbart, and $23$ for Gateway Pundit. We repeated the sampling process to construct a $5$ different set of sampled comments (by changing random seeds from $1$ to $5$) for later purposes.

  With the sampled comments and their BERT embeddings, we ran the grid search on the hyperparameter space to find the optimal hyperparameters for the BERTopic model. The hyperparameters we tuned are the number of neighbors (neighbors, $n$) in UMAP, the minimum cluster size for HDBSCAN (cluster size, $c$), and the random seed for the fitting dataset (seed, $s$). We performed a two-stage grid search for each TM, where we first searched the coarse-grained hyperparameter space to find a local peak and then searched the fine-grained hyperparameter space around the optimal hyperparameters found in the first stage. Coarse-grained hyperparameter space is defined as follows: neighbors $\in \{30, 60, 90\}$ and cluster size $\in \{200, 300, 400\}$. If the optimal hyperparameters found in the first stage are called $n_{1}$ (neighbors) and $c_{1}$, respectively, the hyperparameter stage of the second stage is given by neighbors $\in \{n_{1}-10, n_{1}, n_{1}+10\}$ and cluster size $\in \{c_{1}-25, c_{1}, c_{1}+25\}$. For both stages, the random seed is chosen from $\{1, 2, 3, 4, 5\}$.

  For the coarse-grained search, we chose the pair of hyperparameters $(n_{1}, c_{1})$ based on the "outlier$/0$ ratio", which is defined as a frequency ratio between the sum of topic $-1$ (outlier) and topic $0$ (which we found to be quite typical and not very well separated in most of the cases) and rest of the comments. The smaller this ratio is, the better the model is, as it better represents the other topics other than outliers and topic $0$. For each pair of hyperparameters, we averaged this value for the $5$ seeds and chose the best pair of hyperparameters that minimizes the mean outlier$/0$ ratio. Table \ref{table:s5} shows the coarse-grained search results for each community.

  For the fine-grained search, we aim to find the local peak around the $(n_{1}, c_{1})$ as well as the best-performing seed. First, We chose top $5$ triplets of hyperparameters $(n_{2}, c_{2}, s)$ that minimize the outlier$/0$ ratio as intial candidates. We chose the final triplet among the candidates according to the following criteria: (1) First, we sorted them according to the DBCV \cite{moulavi2014density} metric. (2) Next, we chose $4$ significant topics (Guns, Abortion, Vaccine, and Climate) and manually checked whether these topics are well-separated in the final candidate. If the model didn't separate these topics distinctly, we discarded them from the candidates. (3) Finally, from the remaining candidates, the hyperparameter triplet with the lowest DBCV metric was chosen to be the representative model for the community. The final hyperparameters for each community are summarized in Table \ref{table:s6}.
      
  \subsection*{Global topic model} 
  
  For the global TM, we gathered locally sampled comments from $5$ communities (which share the random seed) and further sampled $0.4$ million comments each by using the same random seed, constructing $5$ sets of $2$ million sampled comments (as in the local case). The rest of the procedures are the same as the local TM construction, and both the coarse-grained and fine-grained search results are summarized in Tables \ref{table:s1} and \ref{table:s2}.

\section*{5. Survey results for the topic model quality assessment}

To validate the quality of (both global and local) topic models constructed by BERTopic, we conducted a survey using the crowdsourcing platform \texttt{Prolific} \cite{palan2018prolific}. The survey consists of five tasks with a total of $1,022$ participants representative of the U.S. public. 

In all tasks, we aimed to get $6$ participants per task item, but the number of participants for each item of each task follows a Gaussian distribution centered at $6$ due to the random assignment of tasks to participants. This procedure maximized the diversity of participants assigned to each item, as well as the diversity of items presented to each participant. The aim was to minimize response biases that could occur if sets of items were all evaluated by the same $6$ participants, and maximize the number of items that can be evaluated in total. Participants had to complete a prespecified number of task items to get paid, so there was no differential attrition, and we excluded all responses that did not complete the full task.

Cronbach’s Alpha is not applicable here because it is designed for assessing the internal consistency of items within a psychometric scale, where multiple items are expected to measure the same latent construct. Our task instead involved evaluating whether individual comments and article titles were correctly assigned to topics, which is closer to an accuracy or precision assessment than to a reliability test of a scale. For this reason, we reported task-specific measures of agreement that are appropriate for classification and topic validation rather than internal consistency.

\begin{enumerate}
\item T$1$: Word intrusion: tests whether each of the model-generated topics (for local models: 120 topics for Mother Jones, 219 for The Atlantic, 257 for The Hill, 287 for Breitbart, and 251 for Gateway Pundit; for the global model: 228 topics) is perceived as coherent by human participants. Each topic is described by four most representative words as chosen by the model. Participants receive the four words plus one word randomly sampled from an unrelated topic description, and must identify the unrelated word. Each topic is evaluated by an average of $6$ participants, with each of the total of 103 participants evaluating a different set of 80 randomly assigned topics. Baseline accuracy in this task is $1/5$ = $20\%$, and the average achieved accuracy (percentage of topics with accuracy above the baseline) is $72\%$ ($92\%$), $72\%$ ($91\%$), $58\%$ ($79\%$), $65\%$ ($87\%$), $62\%$ ($85\%$), and $65\%$ ($85\%$) for the local models of Mother Jones, The Atlantic, The Hill, Breitbart, Gateway Pundit, and the global model, respectively (see Fig. \ref{fig:s5} for details).
  \item T$2$: Topic assignment (comment): tests whether a comment from news communities can be correctly assigned to the model-generated topic. A total of 2601 comments are chosen from each of the five sites across time and from more or less frequent topics identified by the local and the global model. Participants receive a comment and four possible topics describing the comment, with each topic characterized by the four most representative words. One topic is the correct one according to the model, and the other three are unrelated topics sampled randomly from all topics within a given model. Participants have to choose the topic that best describes the comment. Each comment is evaluated by an average of $6$ participants, with each of the total of 390 participants evaluating a different set of 40 randomly assigned comments. Baseline accuracy in this task is $1/4$ = $25\%$, and the average accuracy (percentage of comments with accuracy above the baseline) is above $60\%$ for all comments, with 100\% of comments in all models and from all sites having accuracy above the baseline (see Fig. \ref{fig:s6} for details).
  \item T$3$: Topic assignment (title): tests whether a news article title can be correctly assigned to the model-generated topic. A total of 1874 article titles are randomly sampled from all titles appearing at the five news sites across time. Participants receive a title and four topics (each described by four most representative words, and taken from either the local or the global model). Of the four topics, three are relevant for the title (topics tier 1, 2, and 3, see Fig. 1 in the main text), and one is irrelevant. Participants have to evaluate how related each topic is to the title on a scale from 1 (not at all related) to 5 (very related), with the expectation that the tier 1 topic will be judged as the most related, followed by tier 2 and tier 3 topics, and that the unrelated topic will be the least relevant.  Each comment is evaluated by 6 participants on average, with each of the total of 188 participants evaluating a different set of 60 randomly assigned titles. The results show the expected trend (see Fig. \ref{fig:s7} for details).
  \item T$4$: Topic similarity (description): tests whether a cosine similarity between a pair of topic embeddings (averaged BERT embeddings) aligns with the human-evaluated semantic similarity. A total of 1255 topic pairs are randomly generated among all possible topic pairs within each local and the global model. Participants receive topic descriptions in terms of the four most representative words, and have to evaluate the similarity of topics on a scale from 1 (not at all similar) to 5 (very similar).  Each pair is evaluated by 6 participants on average, with each of the total of 126 participants evaluating a different set of 60 randomly assigned pairs of topics. The results suggest a good correspondence of human-evaluated similarity to the cosine similarity of topic embeddings, with average correlation between them over sites and models of $r=0.48$ (see Fig. \ref{fig:s8} for details).
  \item T$5$: Topic similarity (comment): tests whether a cosine similarity between a pair of comment (BERT) embeddings correctly aligns with the human-evaluated semantic similarity. A total of 1719 comment pairs are randomly generated among all possible comment pairs within each local and the global model. Participants receive a pair of comments and have to evaluate their similarity on the scale from 1 (not at all similar) to 5 (very similar). Each pair is evaluated by 6 participants on average, with each of the total of 215 participants evaluating a different set of 48 randomly assigned pairs of comments. The results suggest a good correspondence of human-evaluated similarity to the cosine similarity of comment embeddings, with average correlation between them over sites and models of $r=0.50$ (see Fig. \ref{fig:s9} for details).
\end{enumerate}

In all tasks, we aimed to get $6$ participants per task item, but the number of participants for each item of each task follows a Gaussian distribution centered at $6$ due to the random assignment of tasks to participants. This procedure maximized the diversity of participants assigned to each item, as well as the diversity of items presented to each participant. The aim was to minimize response biases that could occur if sets of items were all evaluated by the same 6 participants, and maximize the number of items that could be evaluated in total. 

\section*{6. Analysis of empirical findings and verification of model assumptions}

  Here, we provide more analysis on statistical features in our data that were used to initialize our model, and empirical evidence to support some of the implicit model assumptions in the proposed computational model.

  \subsection*{Normalized topic rank}
  Throughout the SI Appendix (such as Figs. S9, S10, S14, and S15), we often present various statistical features of topics as a function of their normalized topic rank. To compute this, we first sort topics by their normalized (title or frequency) topic rank. At any given time period, we define normalized topic rank $r \in (0, 1]$ as the ranking of the (title or comment) frequency of a topic divided by the total number of topics ($228$ in this case). This normalization maps the most frequent topic (rank $1$) into $1/228$, and the least frequent topic (rank $228$) into $1$. Because this ranking changes every month, the values in Fig. \ref{fig:2} for certain $r$ indicate the average value from topics with $r$ in each month.

  \subsection*{Title frequency distribution modeling} 
  
  We observed that the title topic frequency follows an interesting distribution, a product of negative log and power-law distribution with tier-specific exponent (Fig. 2 in the main manuscript). Considering that this title topic distribution corresponds to the title topic distribution of the filtered events in our computational model, the distribution should come from the combined effect of both event generation and the filtering process. The event generation and the filtering process are independent in our model, so the most natural assumption is that each process is responsible for one of the two distributions (although a more complex division is not impossible). 
  
  While either combination is mathematically plausible, and both are monotonically decreasing functions with a heavy tail, we chose the negative log distribution for the event generation process and the power-law distribution for the filtering process in this study for the following reasons. We find that the exponents of the power-law distribution for each tier are empirically different for each community (Fig. 2), while the log distribution is universal across communities. This suggests that the filtering process, which is a community-specific process, is more likely to be responsible for the power-law distribution, while the negative log part is more likely to be responsible for the negative log part. Also, note that the choice of filtering process as a power-law implicitly assumes that this process heavily emphasizes the high-frequency topics and is responsible for the extremely high frequency of popular topics (see Fig. 2b, where the differences in exponents are only meaningful for the popular topics), which is a reasonable assumption considering the nature of the filtering process.

  \subsection*{Number of comment distribution} 
  
  Each article in the online news communities has a different number of comments, and the distribution of the number of comments can be an important factor in understanding the dynamics of the collective mind, especially considering that our computational model explicitly samples the number of comments to simulate the comment distribution (by multiplying the sampled number of comments by the normalized topic distribution). In this work, we introduce the concept of the relative number of comments, which is the number of comments divided by the total number of comments in the given period (in this case, we chose $1$ month). With this, we can construct the distribution of the number of comments without dealing with the volumetric change of the comments over time. We show the distribution of the relative number of comments for each community (Fig. \ref{fig:s1}a), which nicely fits the log-normal distribution. The fitting parameters for the empirical data are summarized in Table \ref{table:s2}, and we used $a = 5.7 \times 10^{-6}$, $b = 1.0 \times 10^{-4}$, and $s=1.5$ for the model simulation.

  \subsection*{Comment multiplier distribution} 
  
  In our computational model, we use the concept of comment multiplier to describe the behavior of the comment topic distribution under certain news articles. From the time series of comment topic distribution, we can calculate the comment multiplier for each topic, which is defined as the ratio of the comment topic frequency under the news article to the expected (previous) comment topic frequency. For stability, we use $12$-month average topic distribution as the expected frequency.

  First, we find that a considerable amount of comment multiplier is zero, which indicates that no comment corresponds to the title topic, and the frequency of zero increases as the topic rank gets larger (i.e., less frequent topics). We show the zero multiplier ratio ($Z_q(r)$) for each community and tier in the Fig. \ref{fig:s1}b. We used the linear approximation for the model calibration for simplification, although a more complex fitting function can be used. In our model, we used $Z_1 = 0.7r$, $Z_2 = 0.9r$ and $Z_3 = 0.9r$ where the $r$ is the normalized comment topic rank.

  For the non-zero multipliers, we show that it follows the exponential distribution, with different decay rates $\lambda$ for different topic ranks and tiers (Fig. \ref{fig:s1}b-d). We further show that the comment multiplier in the specific article has both the theoretical upper and lower bounds (Fig. \ref{fig:s1}e). Let $m_{n,q}$ be the comment multiplier for topic $n$ with tier $q$. Due to the power-law distribution of the comment topic frequency, the expected frequency for topic $n$ is proportional to $n^{-\alpha_c}$. In case of $\alpha_c=1$, the normalization constant becomes the harmonic series $H(n) = \sum_{x=1}^{n}\frac{1}{x} = \ln(n) + \gamma + \frac{1}{2}n^{-1} - \frac{1}{12}n^{-2} + \mathcal{O}(n^{-3})$ where $\gamma=0.5772...$ is the Euler–Mascheroni constant. Considering that the expected frequency of topic $n$ is $(1/n)/H(n)$, the (expected) maximum multiplier for topic $n$ is the reciprocal of this, $B_{\max} = nH(n)$ (note that this is the case where all of the comment under that article is topic $n$). 
  
  Conversely, since the number of comments is a natural number, the minimum multiplier happens when there is exactly $1$ comment with topic $n$ (since $0$ comments would yield the zero multiplier, which we handled separately). So, if we know the number of comments $c$ under the article of interest, we can simply calculate the minimum multiplier as $B_{\min}(c) = \frac{1}{c}/\frac{1}{nH(n)} = \frac{1}{c}nH(n)$. Here, we have two problems: (1) the number of comments $c$ is different for each article, and (2) we would like to express this with the \textit{relative} number of comments, $x$. The first issue can be handled by relaxing the boundary to \textit{expected} minimum boundary, using the \textit{expected} number of comments $\mathbf{E}(c)$ by averaging over all news (at a given time interval, $1$ month in this case). $\mathbf{E}(B_{\min}(c)) = \frac{1}{\mathbf{E}(c)}nH(n)$.
  
  Now, we can resolve the second issue by first expressing $x$ with $c$ as
  
  \begin{equation}
    x_i = \frac{c_i}{\sum_i c_i},
  \end{equation}

  where $x_i$ and $c_i$ is the $i$-th article's relative and raw number of comments. The expected number of comments is $\mathbf{E}(c) = \sum_i c_i / \bar{N_x}$ where the $
  \bar{N_x}$ is the number of news articles (notation is aligned with the computational model). Hence, to express this with expected \textit{relative} number of comments $\mathbf{E}(x) = \sum_i x_i / \bar{N_x}$, 

  \begin{equation}
      \mathbf{E}(c) = \sum_i c_i / \bar{N_x} = \sum_i \frac{x_i\sum_j c_j}{\bar{N_x}} = \sum_j c_j \mathbf{E}(x).
  \end{equation}

  Hence, with $C_{\text{com}}$ as the inverse of the total number of comments per month ($C_{\text{com}} = 1/\sum_j c_j$), we can express the expected minimum multiplier as

  \begin{equation}
    \mathbf{E}(B_{\min}) = \frac{1}{\mathbf{E}(c)}nH(n) = \frac{1}{\sum_j c_j \mathbf{E}(x)}nH(n) = \frac{C_{\text{com}}}{\mathbf{E}(x)}nH(n),
  \end{equation}\

  These two lines greatly match with the empirical maximum and minimum values in the Fig. \ref{fig:s1}e, and we used $C_{com} = 1.0 \times 10^{-6}$ for the model calibration (which matches with the scale of a bigger community like The Hill and Breitbart, since it assumes the number of comments per month as $1.0 \times 10^6$).

  All of these findings are reflected in the choice of the comment multiplier sampling distribution in the computational model (See Table \ref{table:s8}).
  
  \subsection*{Off-topic frequency distribution follows previous community frequency distribution} 
  
  We modeled the response of the community to the news by assuming that the off-topic frequency distribution is the same as the previous community semantic network's frequency distribution, which follows a power-law distribution with the exponent of $-1$ (Fig. \ref{fig:s2}). In Fig. \ref{fig:s2}a, we show the relative comment topic frequency distribution for off-topic comments from the online news communities by removing all of the on-topic comments under the news in the data aggregation stage. Considering its fitted power-law exponents $\alpha_c$ (see captions), we can confirm that this off-topic distribution is also roughly a power-law distribution with the exponent of $-1$, which supports our model assumption. Note that a more detailed investigation by comparing the off-topic frequency distribution at time $t$ with the previous community frequency distribution at time $t-1$ is also possible.

  \subsection*{Similarity decays without cooccurrence} 
  
  In our model, we adopted the updated scheme similar to the Hebbian learning for the topic similarity dynamics. This is based on two assumptions: one is that the similarity between two topics increases with the co-occurrence in the news, and the other is that the similarity decays without the co-occurrence. We verified the latter assumption by calculating the difference in topic similarity as a function of the time absent from the news title in the online news communities (Fig. \ref{fig:s2}b). Atlantic, The Hill, and Gateway Pundit show a relatively strong decaying trend. At the same time, Motherjones was relatively weak and Breitbart showed a positive trend (but it was only fitted from merely $6$ datapoints since no topic pair once existed and did not appear for more than $6$ months, which greatly reduces the fidelity of the Breitbart case for this analysis). Overall, (considering the fact that the overall aggregated fitting line shows a strong negative slope), we can confirm that the similarity decays without the co-occurrence, which supports our model assumption.
  
  \subsection*{Inter-topic similarity is enhanced by the cooccurrence in the news} 
  
  We also verified the former assumption (inter-topic similarity increases with the co-occurrence) by comparing the similarity between on-topic and off-topic comments for the same topic pairs. We calculated the average cosine similarity between off-topic and on-topic comment embeddings for the same topic pairs and compared the similarity distribution for each community (Fig. \ref{fig:s3}). The rationale behind this comparison is that the on-topic comments under the news with certain topic pairs are more likely to be similar to each other since there is a much higher chance that the comment is talking about both topics or the relation of those on-topic at the same time, compared to the null-case off-topic comments.

  Note that this averaged pair-wise similarity is not directly comparable to the similarity between the topic pairs used in the model, since the similarity in the model is calculated by first constructing the topic representation by averaging all of the embeddings, and the cosine similarity is calculated from the averaged embeddings. The reason we used average pair-wise cosine similarity here is that of the systematic difference in the number of on-topic and off-topic comments, where on-topic comments for each topic pair are much smaller (sometimes three orders of magnitude) than the off-topic comments, hence the variance in on-topic cosine similarity gets too high. Still, this averaged pair-wise similarity can be used as a proxy to investigate the relative magnitude of similarities for this analysis.

  In Fig. \ref{fig:s3}, we observe that the similarity between on-topic comments (blue) is generally higher than the off-topic comments (orange) for $3$ online news communities, which indicates that the similarity between two topics is enhanced by the co-occurrence in the news. This supports our model assumption that the similarity between two topics increases with the co-occurrence in the news.

\section*{7. Analysis on basic behavior of computational models}

In the main manuscript (especially Fig. 2b-c, Fig. 3b-c, and SI Appendix, Fig. S13), we showed that the comment topic profile is getting closer to or moving away from the topic profile of the general semantic network, depending on its initial state. In this section, we describe these behaviors in more detail and discuss the underlying mechanism. Hereafter, we consider the computational model with $\lambda_m \neq 1$, since the transition of comment topic profile is impossible with an unchanging community ($\lambda_m = 1$).

The general semantic network is the main source of events, hence greatly affecting the topic distribution of the filtered events (news) as well. More precisely, in our model, the $q-$th tier news topic frequency is roughly proportional to $-\ln(r^g_i) (r^g_i)^{\alpha_q/2}$ (factor of $1/2$ comes from the near-randomness of similarity-based second filter), and this proportionality becomes exact in the extreme case of $\lambda_f=0$. Naturally, the high frequency of the news topic will lead to the high frequency of the comment frequency (amplified by the sampled comment multiplier), which will affect the community semantic network's frequency via memory strength. While it is nearly infeasible to analytically solve the full model, with a similar argument as above, we can expect that this effect will lead the community semantic network's frequency closer to the general frequency distribution (and especially the ranks of them) in the long run. A similar argument can be made for weight since weight is updated by the co-occurrence of topics in the news, which is directly affected by the general semantic network's similarity pattern.

But there is another factor that prevents the community semantic network from fully converging to the general semantic network: the randomness in the comment generation process. Since the comment generation process is stochastic, the comment topic profile will not be exactly the same as the general semantic network's topic profile, even if the community semantic network is fully converged to the general semantic network. This randomness then affects the community semantic network and repels it from the general semantic network till the two forces are balanced. This effect is well shown in the SI Appendix, Fig. S13, where the distance between two semantic networks converges to the same non-zero value regardless of its starting position (SD $0.0$ or $1.0$). 

Interestingly, we find that this equilibrium distance is inversely proportional to both filter strength ($\lambda_f$) and memory strength ($\lambda_m$). It is straightforward to see that the distance is inversely proportional to the memory strength since high memory strength suppresses the randomness in the comment generation process and affects the community semantic network. The inverse proportionality to the filter strength is somewhat counterintuitive at first glance, since the low filter strength should lead the community semantic network to be closer to the general semantic network. On closer inspection, we find that the distance of the high filter strength case ($0.8$) from SD $0.0$ in fact decreases over time after the initial soaring (around $t=50$), suggesting that the source of inverse proportionality comes from something that is changing during the iteration. Given that the only thing that changes during the iteration is the community semantic network, we can infer that the community semantic network that is already attracted and has become similar to the general one reinforces its effect, with the aid of high filter strength. This paradoxical trend is well-aligned with the findings described in the effect of influence (in the main manuscript), where the community with high filter and high memory strength is more prone to internalize and keep the influence from the influences. Further analyzing the asymptotic behavior of the coarse-grained, simplified (and thus analytically tractable) version of this framework would be a promising direction for theoretical future work.

\section*{8. Fitting parameters for the empirical data}

  \subsection*{Global topic model} 
  
  Here, we provide a detailed description of the fitting parameters for the empirical data from online news communities used in the main manuscript (Fig. $2$), where the global topic model is used. In Fig. $2$b, the relative title topic frequency of the news ("Title") is fitted to a $y \propto \ln(x)x^{-\alpha_q}$, where $q$ indicates the tier ($1, 2, 3$). In Fig. $2$c, the relative comment topic frequency ("Comment") is fitted to a power-law distribution, $y \propto x^{-\alpha_c}$. In Fig. $2$d and $2$e, both the probability density of the topic similarity histogram ("Similarity") and the relative number of comments histogram ("$\#$ of comment") is fitted to a log-normal distribution, $y \propto e^{{-\ln^2(\frac{x-a}{b})}/{2s^2}}$. All of the parameters for each community are summarized in Table \ref{table:s7}.

  \subsection*{Local topic model} 
  
  In the main manuscript, we have shown that the statistical distribution of the empirical data, which is an outcome of the classification of the global topic model, matches our model results. For further verification and to demonstrate the robustness of the data distribution, we also present empirical data, which is classified by the respective local topic model and their fittings in Fig. \ref{fig:s10}. The fitting parameters for local models are summarized in Table \ref{table:s7}.

\section*{9. Discussion on the correspondence between the empirical semantic network and the comment network}

  In our study, we calibrated the initial frequency (and similarity) distribution of both general and community semantic networks from the empirical data. However, there are some noteworthy points to rigorously address the validity of this approach.

  The point here is that the semantic network is not directly observable from the empirical data; rather, it's a structural concept that we employed to explain the underlying dynamics of the collective mind and to construct the computational model. The only thing we can directly observe are comments, which correspond to the comment network in our model. Hence, we need to justify that the semantic network also follows the same distribution as the empirical comment distribution.

  In the case of the community semantic network, the reason is quite straightforward; if we update our community semantic network to a comment network with memory strength $\lambda_m \neq 1$, the distribution of the community semantic network will eventually converge to the comment network. This can be easily shown by considering the update rule of the community semantic network. For example, if we consider the frequency update rule, the community semantic network's frequency at time $t+1$ is given by 

  \begin{equation}
    f_{i, t+1}^{k} = \lambda_m f_{i, t}^{k} + (1-\lambda_m) \hat{\bar{f}}_{i, t}^{k},
  \end{equation}

  where the term $\hat{\bar{f}}_{i, t}^{k} = \hat{f}_{i, t}^{k} / \sum_j \hat{f}_{j, t}^{k}$ denotes relative comment frequency distribution. If we assume the frequency distribution of the semantic network is stationary, i.e., ${f}_{i, t+1}^{k} = {f}_{i, t}^{k}$, the equation becomes 

  \begin{equation}
    f_{i, t}^{k} = \lambda_m f_{i, t}^{k} + (1-\lambda_m) \hat{\bar{f}}_{i, t}^{k},
  \end{equation}
  
  and therefore $f_{i, t}^{k} = \hat{\bar{f}}_{i, t}^{k}$, which means the community semantic network's frequency distribution will converge to the comment frequency distribution in the long run as a steady state. The same logic applies to the weight update rule as well. More rigorous proof can be done by showing that the distance between probability distributions (either L1 norm or KL divergence) decreases as the iteration goes to infinity, and this is related to concepts like mixing in the Markov process.

  For the general semantic network, if we assume the general semantic network is an averaged version of all existing community semantic networks (since it represents the general popularity and semantic structure of the entire population), the general semantic network's distribution will also converge to the comment network's distribution.


\section*{8. Auxilary analysis}

 \subsection*{Similarity analysis for COVID shock} 

We have conducted a similarity analysis on COVID and compared its qualitative behaviors with the external shock model simulation (similar to Fig. 3b and c in the main manuscript). As Fig.~\ref{fig:S14} shows, in both cases (apart from the issue of different time scales), (1) the sudden shock of influx of a certain topic (Vaccine in this case) also increases its similarity with other topics, and (2) previously more similar topics (top $10\%$) are more benefited by this shock compared to the topics that were not. It is both expected and consistent with our model design, since the topic will have a higher chance of being paired with topics it was already semantically similar to (in the article); hence, a previously more similar topic gets more benefit when there is an exogenous influx of attention.

 \subsection*{Sensitivity analysis on the initial weight distribution} 

 We conducted additional sensitivity analyses on the initial weight distribution: will different initial distributions exhibit the same stability over a long (120 timesteps) simulation, as our current (empirically grounded) log-normal distribution does? We switched our initial calibration into Gaussian normal (with mean $0.5$ and std $0.2$, capped in $[0.2, 0.8]$, and the exponential function (with the decay rate of $0.1$, capped at $[0, 0.8]$). The results are shown in Fig.~\ref{fig:s15}.

As we can see, the Gaussian distribution, which is a unimodal distribution similar to the empirical log-normal, persists for a long time (120 time steps), but if we switch to a qualitatively different one, such as the exponential distribution, its stability completely breaks down. This is because the current weight update rule (similar to Hebbian learning, whose existence in the real-world semantic network is also verified in the SI appendix, section 6) has both reinforcing and decaying terms, effectively pulling the weights toward the mean while allowing for strong volatility. This is also consistent with the behavior of the empirical similarity distribution: individual similarity can change drastically, but the overall distribution does not change for a long time (around 10 years). It demonstrates that the model output is sensitive to the choice of model, but that empirically grounded distributions and model designs were crucial for the successful simulation of the real-world collective mind.

\bibliography{Nature}


  \begin{figure}[t]
  \centering
  \includegraphics[width=\linewidth]{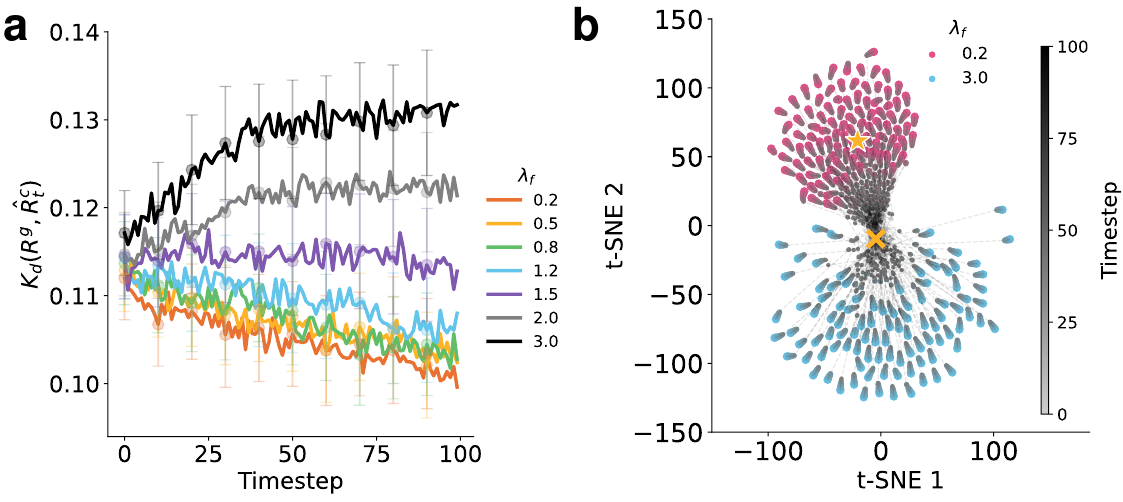}
  \caption{\textbf{Behavior of model with hypersensitive filter ($\lambda_f>1$).} \textbf{a}, Kendall-tau rank distance ($K_d$) between relative topic frequencies of general semantic network ($R^g$) and comment frequencies of community semantic network at time step $t$ ($\hat{R}^c_t$) with various $\lambda_f$ ranging from $0.2$ fo $3.0$, where the initial community frequencies are perturbed from general frequencies by log-normal noise with standard deviation of $0.2$. Data is gathered from $1,000$ iterations, and the errorbar indicates $\pm1$ standard deviation and is plotted every $10$ time step. \textbf{b}, The t-SNE plot of $100$ trajectories of the comment frequency profile for the model simulation with $\lambda_f=0.2$ (red) and $\lambda_f=3.0$ (blue), all started from the same initial frequency (orange cross) and attracted by the same general semantic network (orange star). $\lambda_m=0.9$ was used for all simulations.}
  \label{fig:s11}
\end{figure}

\newpage
  \begin{figure}[t]
  \centering
  \includegraphics[width=0.8\linewidth]{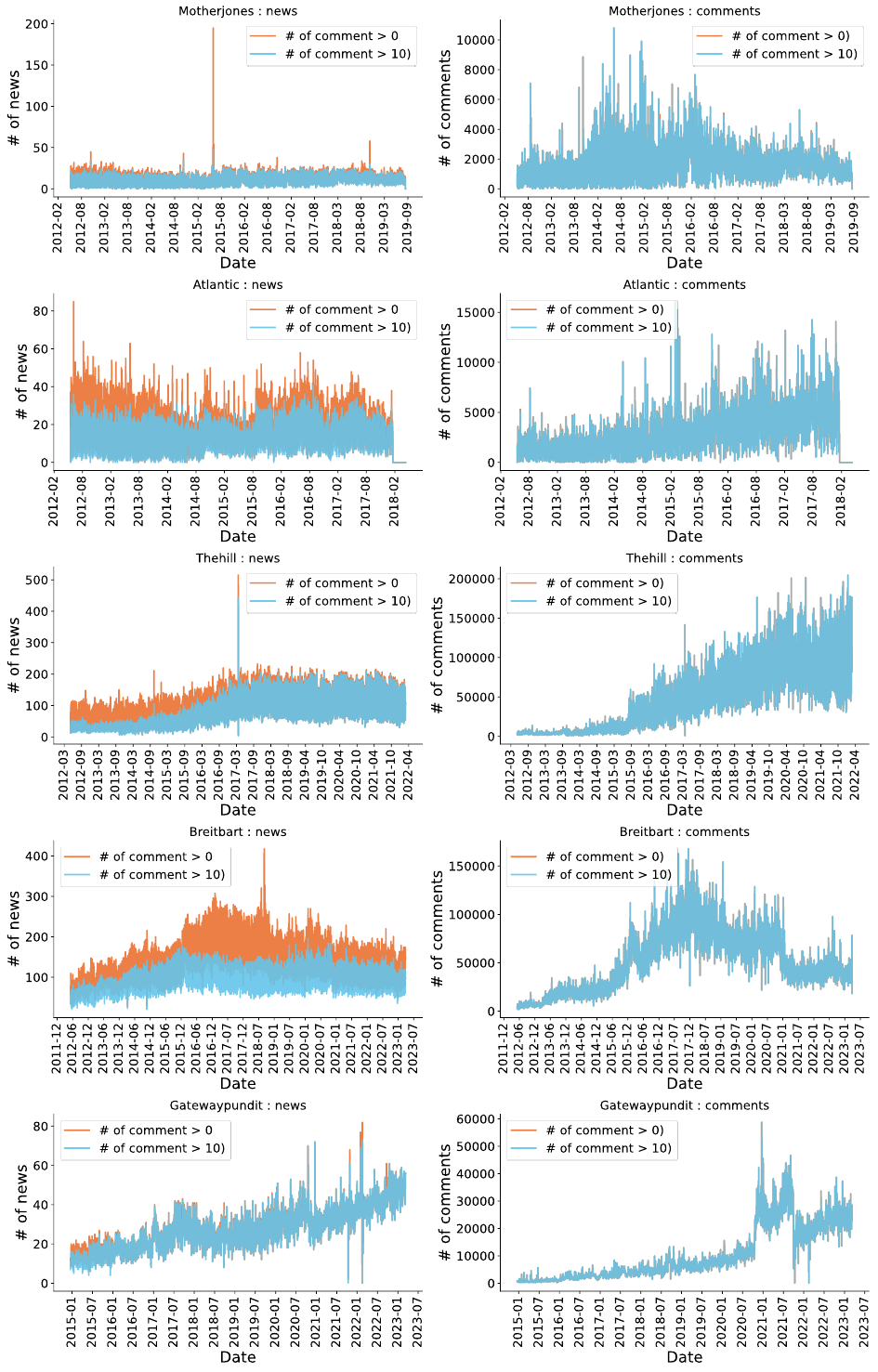}
  \caption{\textbf{Time series of the number of news and comments for $5$ online news community.} Before (orange) and after (blue) the filtering with $\theta_n=10$ is plotted.}
  \label{fig:s4}
\end{figure}

\newpage
\begin{figure}[t]
  \centering
  \includegraphics[width=0.9\linewidth]{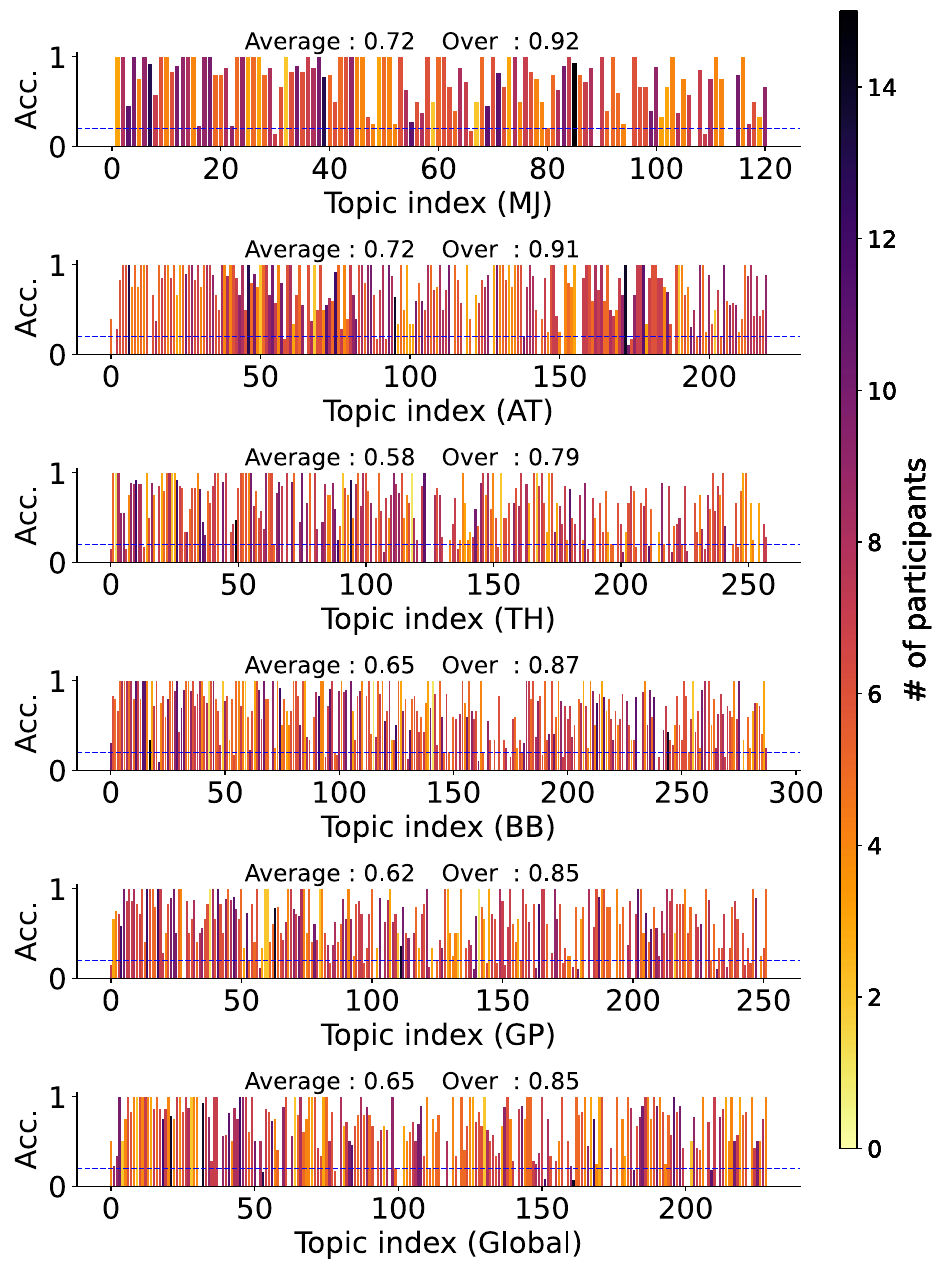}
  \caption{\textbf{Results from the topic model survey, word intrusion task (T$1$).} The average indicates the mean accuracy of all topics' results, while Over indicates the percentage of topics that have an accuracy over $20\%$ (chance level).}
  \label{fig:s5}
\end{figure}

\newpage
\begin{figure}[t]
  \centering
  \includegraphics[width=0.9\linewidth]{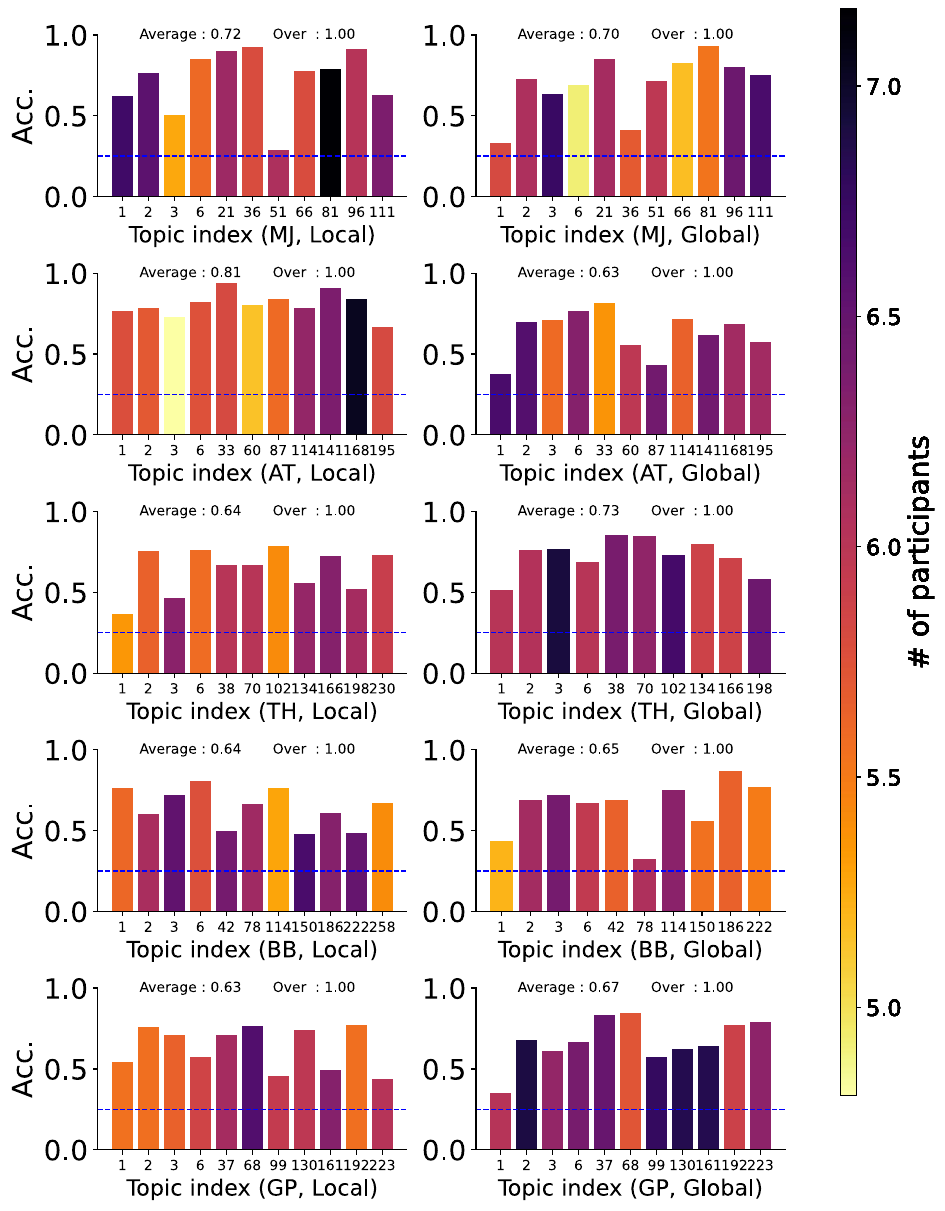}
  \caption{\textbf{Results from topic model survey, topic assignment (comment) task (T$2$).} Topics from both the local topic model and the global topic model from the same site are displayed next to each other. The average indicates the mean accuracy of all topics' results, while Over indicates the percentage of topics that have an accuracy over $25\%$ (chance level).}
  \label{fig:s6}
\end{figure}

\newpage
\begin{figure}[t]
  \centering
  \includegraphics[width=\linewidth]{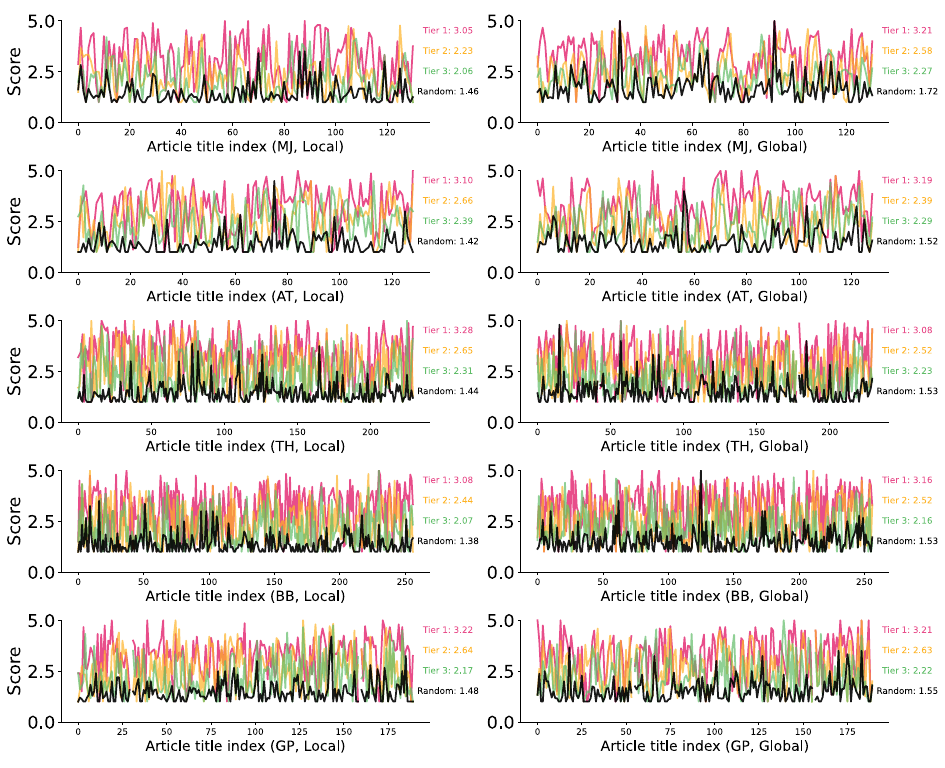}
  \caption{\textbf{Results from topic model survey, topic assignment (title) task (T$3$).} Red, yellow, green, and black lines indicate the averaged scores for the (correct) tier $1$, $2$, and $3$ topics, and a random topic, respectively. Topics from both the local topic model and the global topic model from the same site are displayed next to each other. Average scores for each category are shown on the right side of the plot.}
  \label{fig:s7}
\end{figure}

\newpage
\begin{figure}[t]
  \centering
  \includegraphics[width=0.8\linewidth]{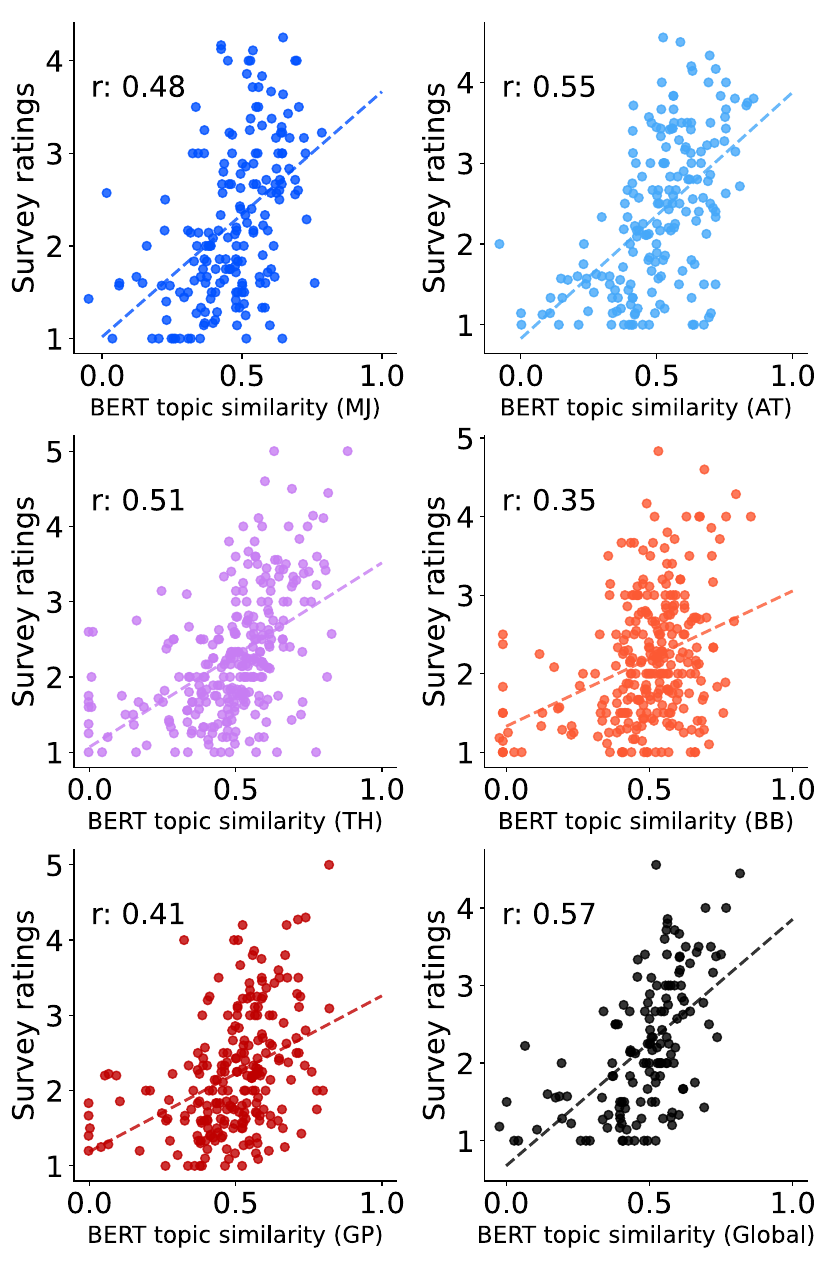}
  \caption{\textbf{Results from topic model survey, topic similarity (description) task (T$4$).} Pearson correlation $r$ is shown in each plot, and the linear fit is shown as a dashed line.}
  \label{fig:s8}
\end{figure}

\newpage
\begin{figure}[t]
  \centering
  \includegraphics[width=0.7\linewidth]{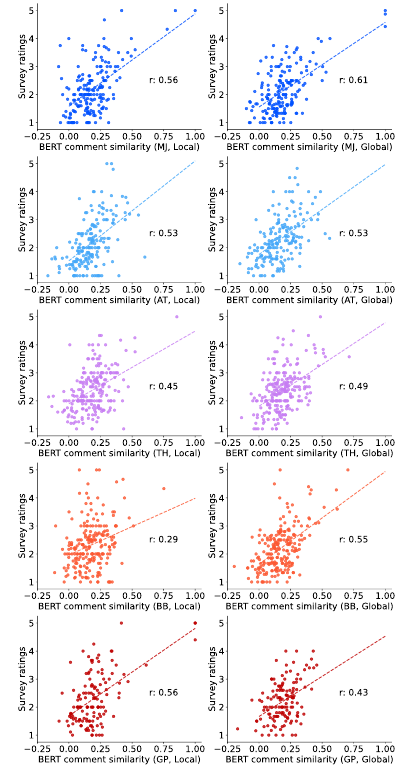}
  \caption{\textbf{Results from topic model survey, topic similarity (comment) task (T$5$).} Pearson correlation $r$ is shown in each plot, and the linear fit is shown as a dashed line.}
  \label{fig:s9}
\end{figure}

\begin{figure}[h!]
	\centering
	\includegraphics[width=\linewidth]{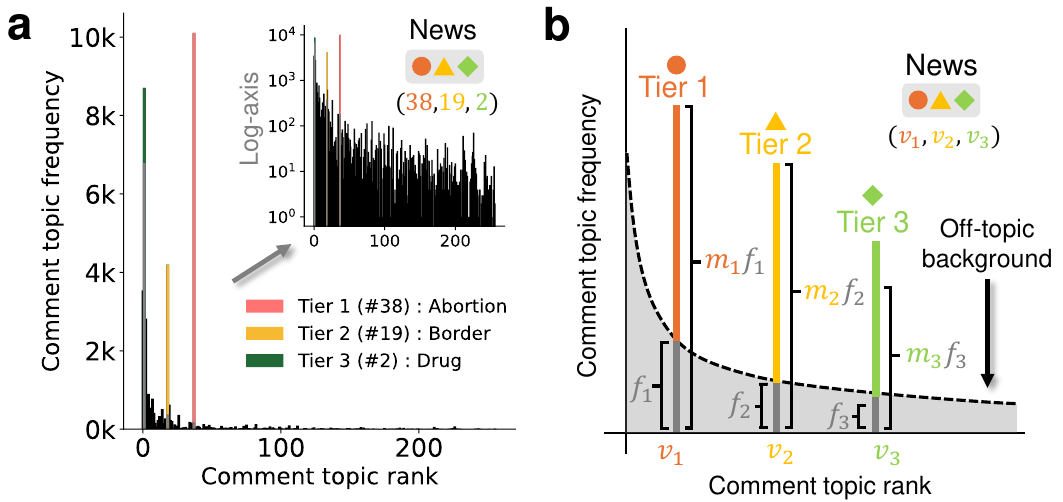}
	\caption{\textbf{Response of community semantic network to the news.} \textbf{a}, Illustrative example of the topic frequency distribution of the comments posted under all news characterized by certain topics (here, by the triplet $(38, 19, 2)$) and its log scale plot (inset). Gray bars represent the expected frequencies of these topics when they are not in the news. Data is taken from The Hill's all-time aggregated data. \textbf{b}, Conceptual diagram that shows the modeling of the community semantic network's response (comment frequency) to the news $(v_1, v_2, v_3)$ in the proposed model. Each of the on-topic comments' frequencies ($f_i$) is multiplied by the comment multiplier $m_i$, while other off-topic comments' frequencies follow the previous comment frequency distribution as a common background signal.}
	\label{fig:e1}
	\end{figure}

\newpage
\begin{figure}[t]
    \centering
    \includegraphics[width=\linewidth]{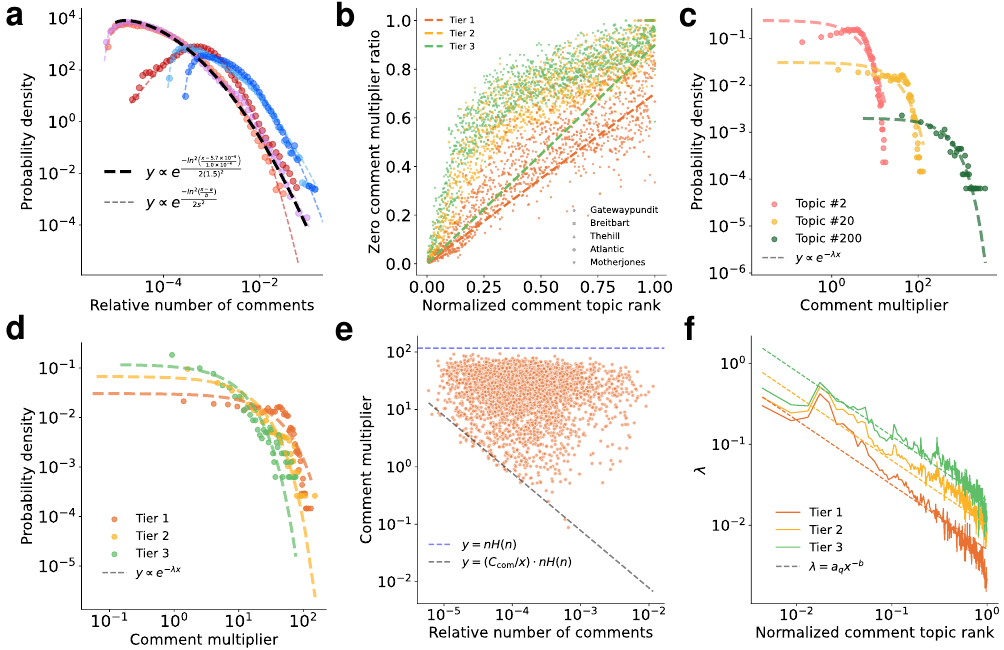}
    \caption{\textbf{Empirical data distribution for computational model calibration.} \textbf{a}, Histograms of the relative number of comments for each community and their fittings. Individual fittings are drawn in dotted lines, and thick dashed lines indicate the distribution used to calibrate the computational model. \textbf{b}, Zero comment multiplier ratio ($Z_q$) for community and tier, with the linear approximation used for the model calibration. Note that we used the same linear function for $Z_2$ and $Z_3$. \textbf{c}, comment multiplier histogram for different topic ranks ($2, 20, 200$) and their fittings. \textbf{d}, comment multiplier histogram for different tiers ($1, 2, 3$) and their fittings. \textbf{e}, Scatter plot between the comment multiplier of topic $20$, tier $1$ and their relative number of comments. Two theoretical boundaries are drawn in dashed lines, where $H(n)$ is a harmonic series to $n$ and $C_{\text{com}}$ is the inverse of the mean total number of comments (in the time interval of $1$ month). \textbf{f}, Fitting exponent $\lambda$ for the exponential fitting of comment multiplier distribution with respect to their normalized topic rank (and their fittings). The data in \textbf{c-f} is aggregated from the all-time data of The Hill.}
    \label{fig:s1}
  \end{figure}

  \newpage
  \begin{figure}[t]
    \centering
    \includegraphics[width=\linewidth]{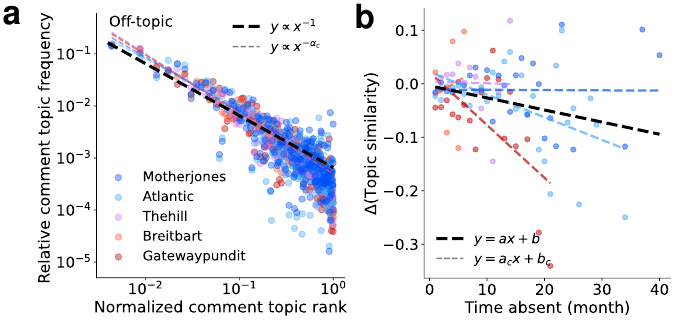}
    \caption{\textbf{Verification of the model assumptions.} \textbf{a}, Relative comment topic frequency distributions for off-topic comments from online news communities and their best-fitting lines. The fitted exponents $\alpha_c$ are $1.03$ for Motherjones, $0.95$ for Atlantic, $1.15$ for Thehill, $1.10$ for Breitbart, and $1.14$ for Gatewaypundit. \textbf{b}, Difference in topic similarity as a function of the time absent from the news title in online news communities and their best linear fitting lines. For each point, all data from instances of topic pairs that were missing for the same months in the same community were averaged to highlight the dependence between the time absent and the similarity difference. The slope for individual linear fitting lines are $-5.35 \times 10^{-5}$ for Motherjones, $-4.23 \times 10^{-3}$ for Atlantic, $-3.01 \times 10^{-4}$ for Thehill, $6.43 \times 10^{-3}$ for Breitbart, and $-9.86 \times 10^{-3}$ for Gatewaypundit. Black dashed line indicates the aggregated fitting line for all $5$ communities, where its slope is $-2.27 \times 10^{-3}$. The legend from panel \textbf{a} is shared with panel \textbf{b}.}
    \label{fig:s2}
  \end{figure}

  \newpage
    \begin{figure}[t]
    \centering
    \includegraphics[width=0.9\linewidth]{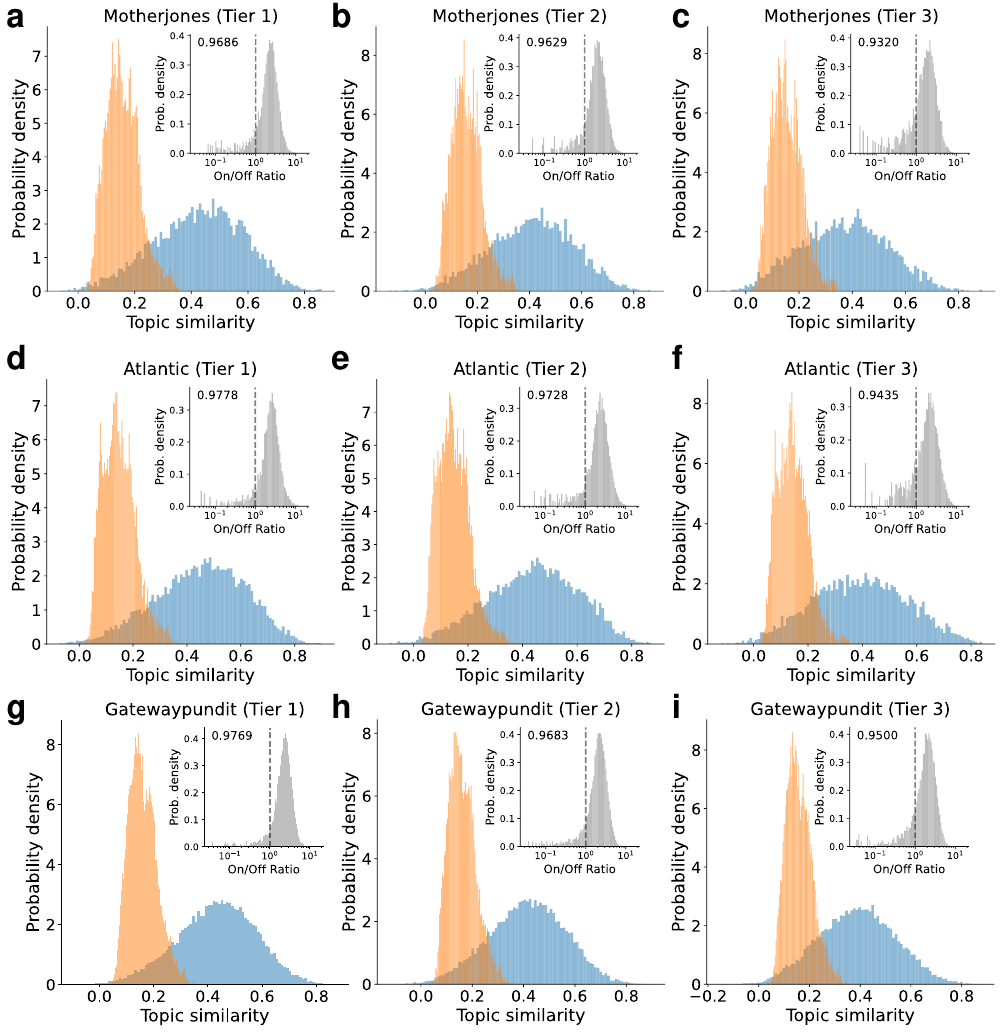}
    \caption{\textbf{Comparison between on-topic and off-topic comment similarity for the same topic pairs.} For each panel, the orange and blue histograms indicate the distribution of average cosine similarity between off-topic and on-topic comment embeddings for the same topic pairs, aggregated from the entire data of the respective community. The Inset histogram shows the ratio between on-topic and off-topic average cosine similarity for the same topic pairs, where the dashed line indicates the ratio of $1$ and the annotated number indicates that the ratio is greater than $1$. Each panel is titled with the community name and the tier of the title topic that is used to determine on-topic comments.}
    \label{fig:s3}
  \end{figure}

  \newpage
\begin{figure}[t]
  \centering
  \includegraphics[width=\linewidth]{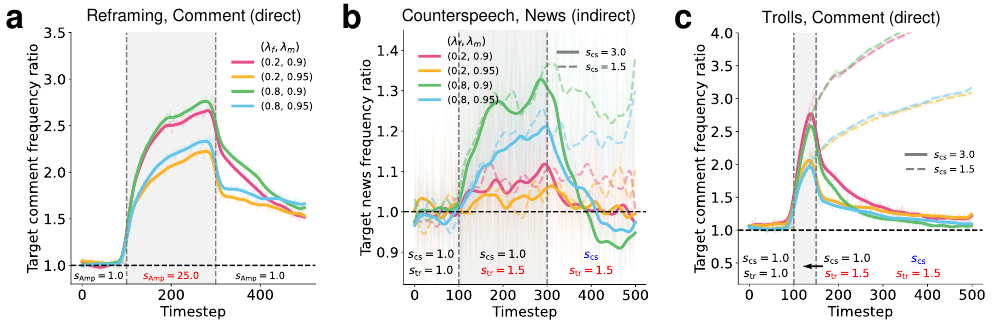}
  \caption{\textbf{Auxiliary plots from the influence result.} \textbf{a}, Ratios between baseline and influenced case of target topic frequency in the comment (reframing). \textbf{b}, Ratios between baseline and influenced case of target topic frequency in tier $1$ news topic (counterspeech). \textbf{c}, Ratios between baseline and influenced case of target topic frequency in the comment but with earlier removal of trolls with $t=150$ (trolls). All of the other details are the same as the corresponding plots in Figs. 4 and 5 in the main manuscript.}
  \label{fig:s12}
\end{figure}

\begin{figure}[h!]
	\centering
	\includegraphics[width=0.5\linewidth]{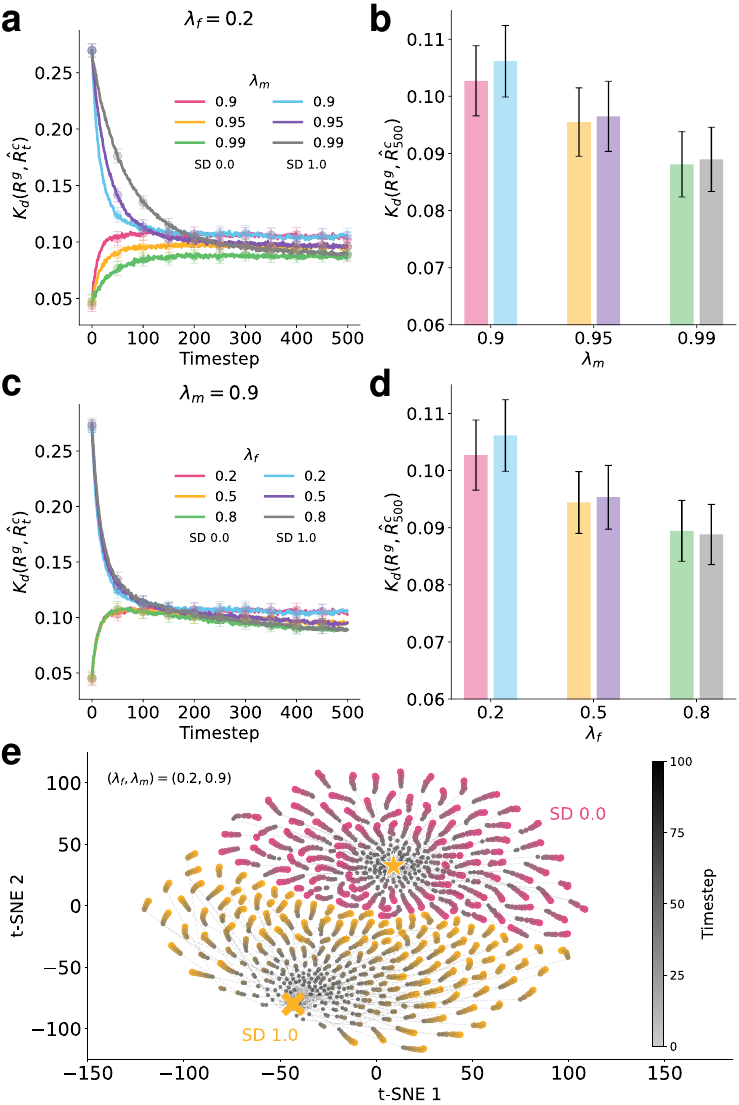}
	\caption{\textbf{Behavior of the comment topic profile from model simulations.} \textbf{a}, Kendall-tau rank distance ($K_d$) between relative topic frequencies of general semantic network ($R^g$) and comment frequencies of community semantic network at timestep $t$ ($\hat{R}^c_t$) for various $\lambda_m$ ($0.9, 0.95, 0.99$) and different initial state (SD $0.0$, $1.0$) with fixed filter strength ($\lambda_f=0.2$). The final distance at $t=500$ is highlighted in \textbf{b}, where data with the same $\lambda_m$ are grouped. \textbf{a}, Kendall-tau rank distance ($K_d$) between relative topic frequencies of general semantic network ($R^g$) and comment frequencies of community semantic network at timestep $t$ ($\hat{R}^c_t$) for various $\lambda_f$ ($0.2, 0.5, 0.8$) and different initial state (SD $0.0$, $1.0$) with fixed memory strength ($\lambda_m=0.9$). The final distance at $t=500$ is highlighted in \textbf{d}, where data with the same $\lambda_f$ are grouped. Data is gathered from $1,000$ iterations, and the error bar indicates $\pm1$ standard deviation. For \textbf{a} and \textbf{c}, error bars are plotted every $50$ timestep. \textbf{a} shares the legend with \textbf{b} and \textbf{c} shares the legend with \textbf{d}. \textbf{e} t-SNE plot of the trajectories of the comment frequency profile for $100$ model simulations each, where the initial community frequencies are perturbed from general frequencies by log-normal noise with standard deviation (SD) of $0.0$ (red) and $1.0$ (yellow). All trajectories started from the same initial frequency (yellow cross) and were attracted by the same general semantic network (yellow star), and $(\lambda_f, \lambda_m) = (0.2, 0.9)$.}
	\label{fig:e3}
	\end{figure}

\newpage

\begin{figure}[t]
  \centering
  \includegraphics[width=\linewidth]{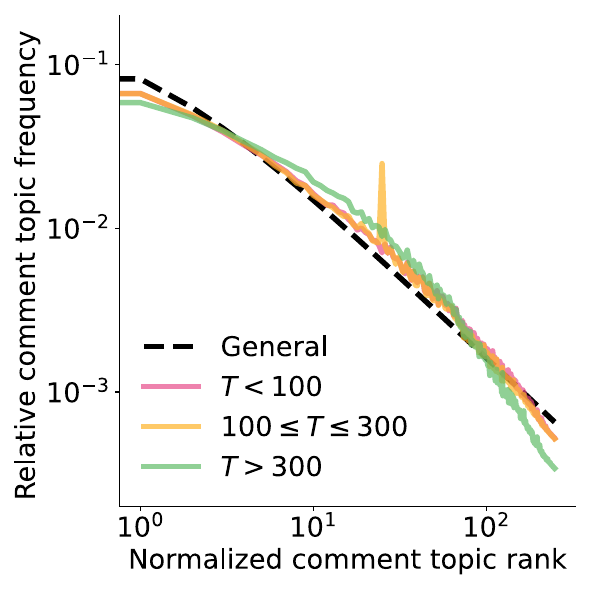}
  \caption{\textbf{Relative comment topic frequency distribution after the counterspeech.} The red, yellow, and green lines indicate the relative frequency of the target topic in the comment topic profile before the trolls, after the trolls, and after the counterspeech (with trolls), respectively. $s_{\text{tr}}=1.5$ and $s_{\text{tr}}=3.0$ is used. The green line (with both trolls and counterspeech present) does not match the original red line (before the trolls); instead, it overrepresents high-rank topics ($r>100$, except for the top few) while underrepresenting low-rank ($r<100$) topics.}
  \label{fig:s13}
\end{figure}

\begin{figure*}
    \centering
    \includegraphics[width=0.95\linewidth]{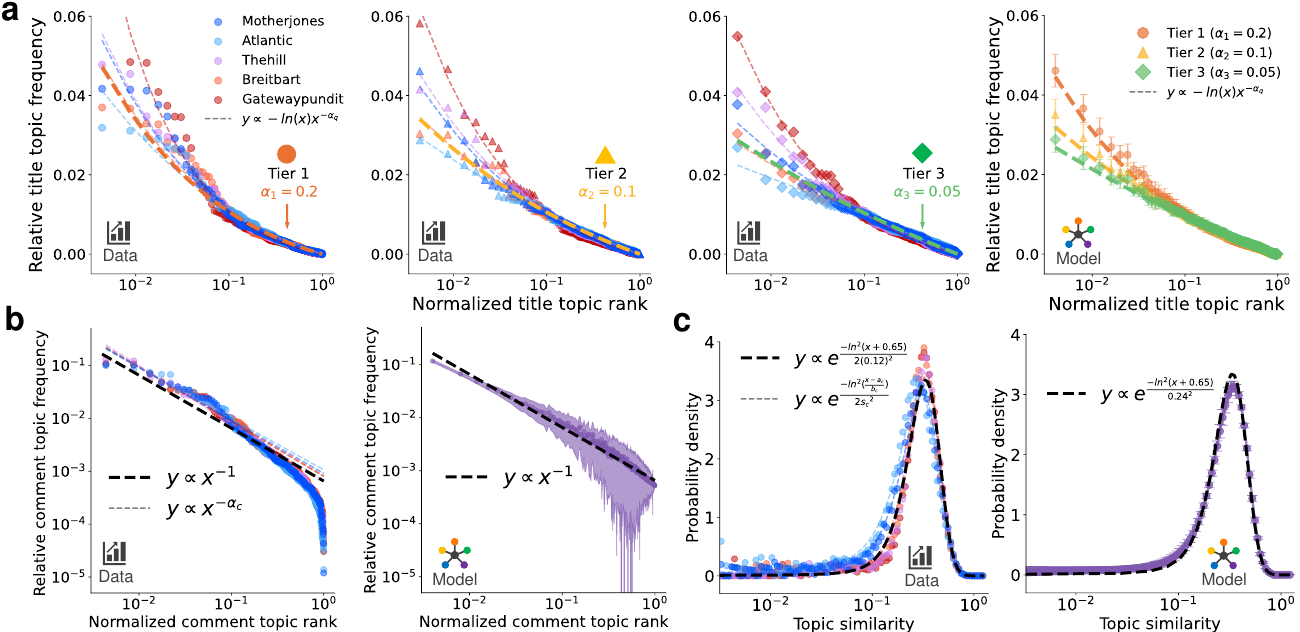}
    \caption{\textbf{Quantitative comparison of data and model output.} \textbf{a}, Relative frequency distributions of topics in article titles in tiers 1, 2, and 3, in each community, compared with the model simulation (right-most panel). \textbf{b}, Relative frequency distributions of topics in comments in each community (left), compared with the model simulation (right panel). \textbf{c}, Topic similarity distributions in each community (left), compared with the model simulation (right panel). \textbf{f}, Topic similarity distributions from the model simulation. The topic frequency distributions (\textbf{a-b}) are sorted by their topic ranks, normalized by the total number of topics. Thick dashed lines indicate the distributions used to calibrate the computational model, and thin dashed lines indicate the best-fitting lines for individual communities (see SI Appendix, Table S3 for best-fitting parameters). Here, the individual fittings were performed by the ordinary least squares (OLS) method, and the representative exponent was chosen manually to be both simple and close to the average of the individual exponents. Error bars indicate $\pm1$ standard deviations across $10,000$ simulations of $120$ time steps (equivalent to $10$ years).}\label{fig:2}
\end{figure*}

   \begin{figure}[h!]
	\centering
	\includegraphics[width=0.8\linewidth]{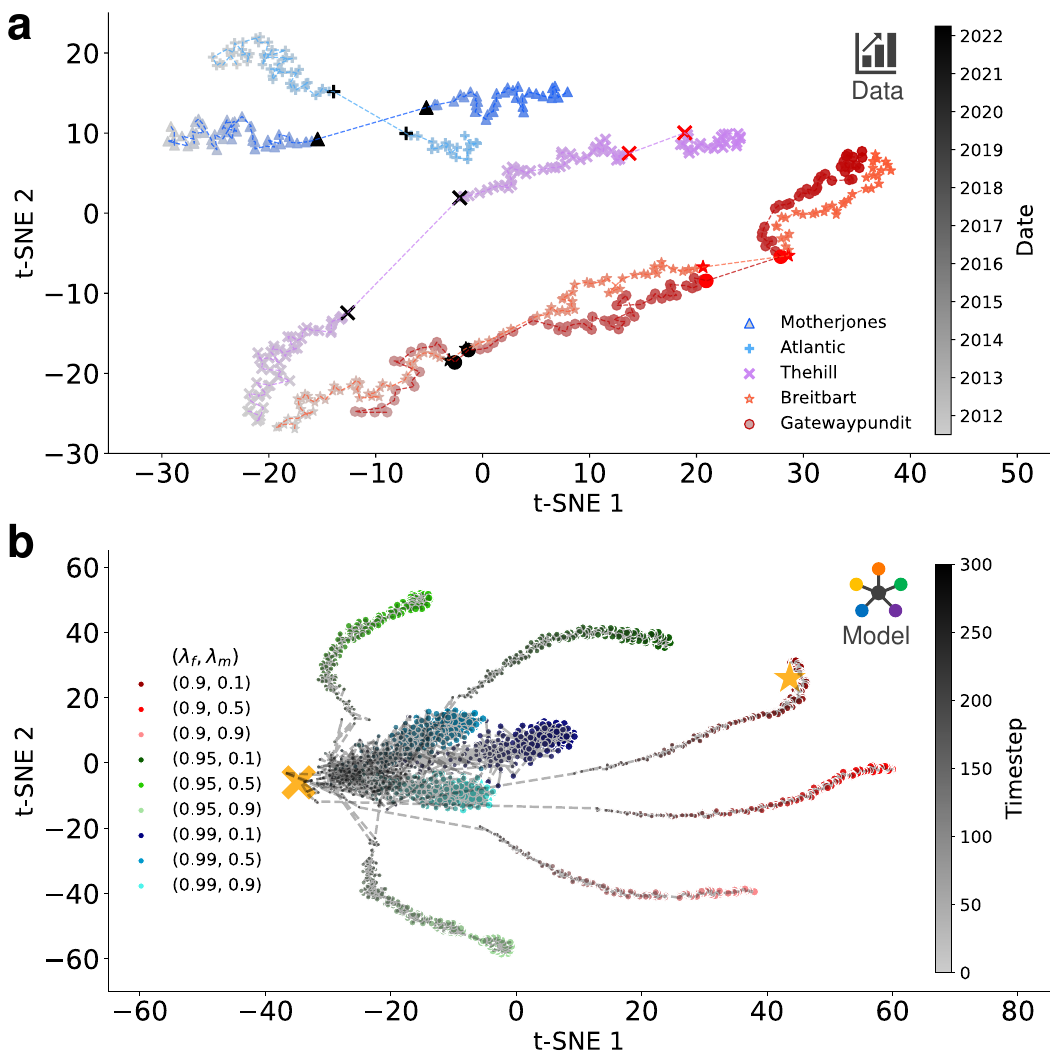}
	\caption{\textbf{Trajectories of comment topic profile} \textbf{a}, Empirical t-SNE plot of the trajectories of the comment topic profile for online news communities. t-SNE algorithm is employed to visualize the trajectory of the $228$-dimensional comment topic profile in the $2$D space. Two notable jumps in trajectories that affected all communities significantly are denoted as black (Jun/Jul 2016, Orlando mass shooting) and red (Feb/Mar 2020, COVID-19) markers. \textbf{b}, Model-based t-SNE plot of the trajectories of the comment frequency profile with diverse filter strength and memory strength, all started from the same initial frequency (orange cross) and attracted by the same general semantic network (orange star). In the model, t-SNE algorithm is applied to the $250$-dimensional model comment topic profile, which corresponds to the comment topic profile of the empirical data. We can observe that the lower the filter strength and the memory strength, the more the trajectory is affected by the general semantic network and quickly converges to it.}
	\label{fig:e2}
	\end{figure}

\newpage
  \begin{figure}[t]
    \centering
    \includegraphics[width=\linewidth]{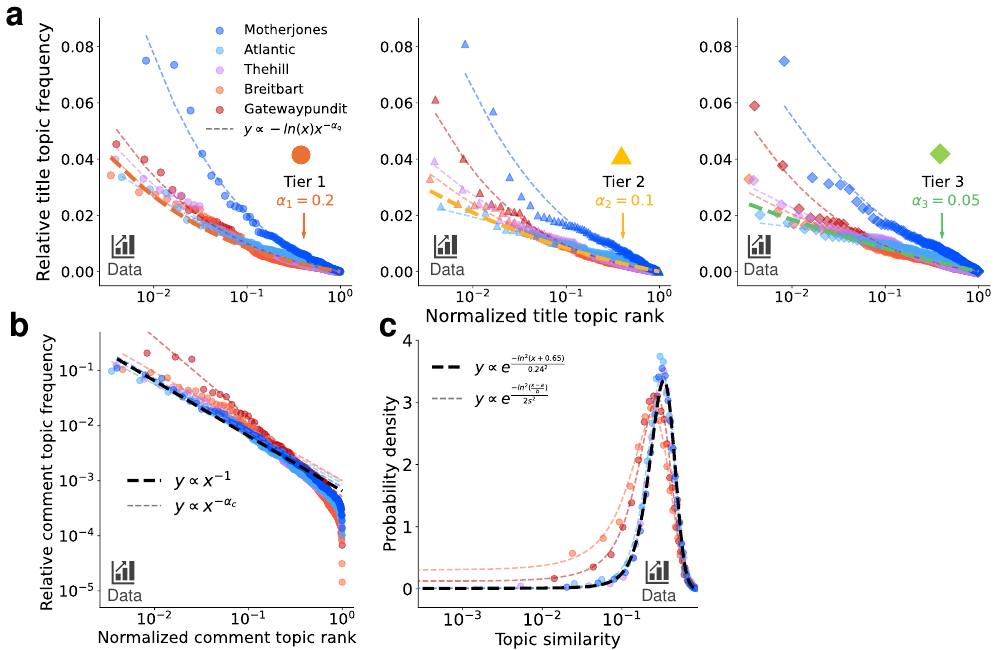}
    \caption{\textbf{Quantitative comparison of the real data (local TM) and the model output.} \textbf{a}, Relative article title topic frequency of tier $1$ (left), tier $2$ (middle) and tier $3$ (right) from each online news communities. \textbf{b}, Relative comment topic frequency distribution from online news communities. \textbf{c}, Topic similarity histogram from online news communities. Individual fittings are drawn in dotted lines, and thick dashed lines indicate the distribution used to calibrate the computational model. All of the topic frequency distributions (\textbf{a, b}) are sorted by their normalized topic rank. A legend in \textbf{a}(left) shows the color scheme used to represent the data from each online community, which is applied consistently across panels \textbf{b} and \textbf{c}. All of the fitting parameters for real data ($\alpha_q, \alpha_c, a_c, b_c, s_c$) are listed in Table \ref{table:s8}.}
    \label{fig:s10}
  \end{figure}

  \begin{figure*}
\centering
\includegraphics[width=\linewidth]{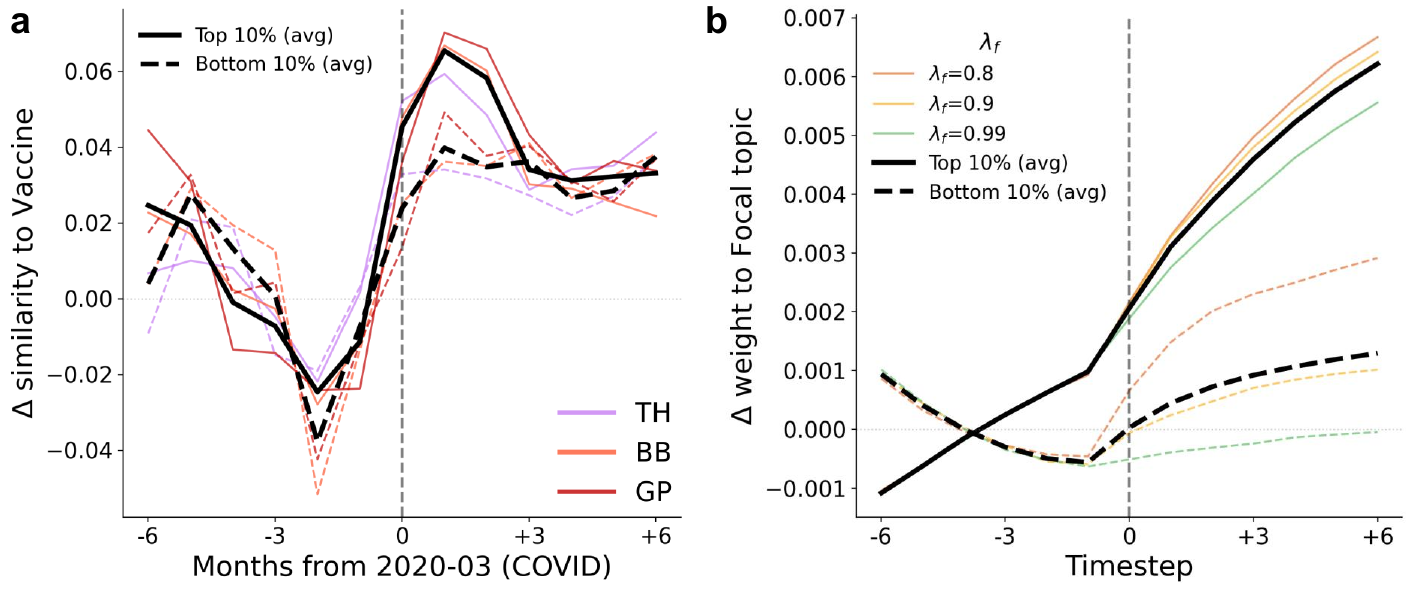}
\caption{ \textbf{a} Similarity difference ($\Delta$) between other topics and the Vaccine topic, for the online news community The Hill (TH), Breitbart (BB), and Gateway Pundit (GP).(b) Results from the model simulation with the external shock are similar to those shown in Fig. 3c of the main manuscript. In both panels, the baseline for differences and top/bottom similar topics is calculated by averaging the previous 6 months of similarity from the month of the COVID-19 pandemic. Similarities from top/bottom $10\%$ topics are averaged and plotted as thick (top) and dotted (bottom) lines. Model results are averaged from $1,000$ simulations.}\label{fig:s14}
\end{figure*}

  \begin{figure*}[h!]
    \centering
    \includegraphics[width=0.9\linewidth]{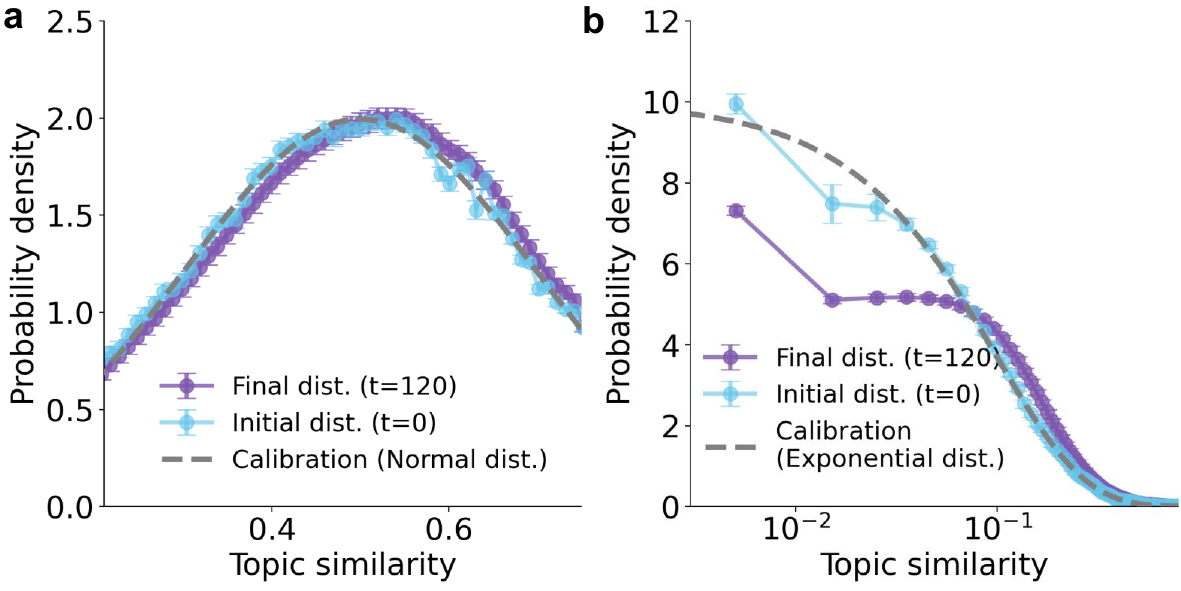}
    \caption{\textbf{Topic similarity distributions in the model simulation with different initial distribution.} \textbf{a}. Gaussian distribution (mean $0.2$, standard deviation $0.2$). \textbf{b}. Exponential distribution $y \propto e^{-10x}$.}\label{fig:s15}
\end{figure*}

\begin{table}[t]
\footnotesize
\centering
\caption{Summary of online news community datasets after data cleansing.}
\begin{tabular}{lcccccc}
\toprule
\multirow{2}{*}{Name} & \multirow{2}{*}{Inclination} & \multirow{2}{*}{Data period (months)} & \multicolumn{2}{c}{$\#$ of news (k)} & \multicolumn{2}{c}{$\#$ of comments (k)} \\
\cmidrule(lr){4-5} \cmidrule(lr){6-7}
 & & & Before & After (\%) & Before & After (\%) \\
\midrule
Mother Jones & Far-left & 12/06--19/09 (87) & 35.968 & 31.510 (87.61) & 4783.86 & 4763.04 (99.56) \\
Atlantic & Left & 12/06--18/05 (71) & 46.262 & 32.144 (69.48) & 6736.16 & 6675.60 (99.10) \\
The Hill & Center & 12/06--22/03 (117) & 380.62 & 313.67 (82.41) & 176263.19 & 175989.96 (99.84) \\
Breitbart & Right & 12/06--23/04 (130) & 591.04 & 400.03 (67.68) & 205816.32 & 205280.91 (99.74) \\
Gateway Pundit & Far-right & 15/01--23/04 (99) & 85.20 & 83.77 (98.32) & 31279.42 & 31271.54 (99.97) \\
\bottomrule
\end{tabular}
\label{table:s3}
\end{table}

\begin{table}[t]
\footnotesize
\centering
\caption{Online news communities data summary after overdue / outlier filtering.}

\begin{tabular}{lccccc}
\toprule
\multirow{2}{*}{Name}  & \multicolumn{2}{c}{$\#$ of news (k)} & \multicolumn{3}{c}{$\#$ of comments (k)} \\
\cmidrule(lr){2-3} \cmidrule(lr){4-6}
& Before & After ($\%$) & Before & After ($\%$) & Non-outlier ($\%$) \\
\midrule
Mother Jones & 31.510 & 23.92 (75.91) & 4763.04 & 3707.93 (77.85) & 2027.33 (42.56) \\
Atlantic & 32.144 & 25.10 (78.10) & 6675.60 & 6223.72 (93.23) & 3030.48 (45.40) \\
The Hill & 313.67 & 284.86 (90.81) & 175989.96 & 172172.40 (97.83) & 88812.39 (50.46) \\
Breitbart & 400.03 & 360.94 (90.23) & 205280.91 & 199875.94 (97.37) & 103869.43 (50.60)\\ 
Gateway Pundit & 83.77 & 79.49 (94.89) & 31271.54 & 30439.70 (97.34) & 15306.64 (48.95)\\
\bottomrule
\end{tabular}
\label{table:s4}
\end{table}

\begin{table}[t]
\footnotesize
\centering
\caption{Fitting parameters for the empirical data in Fig. $2$ (Global TM)}
\begin{tabular}{lccccccc}
\toprule
\multirow{2}{*}{Name} & \multicolumn{3}{c|}{Title} & {Comment} &\multicolumn{3}{c|}{Similarity} \\
\cmidrule(lr){2-4} \cmidrule(lr){5-5} \cmidrule(lr){6-8}
& $\alpha_1$ & $\alpha_2$ & $\alpha_3$ & $\alpha_c$ & $a$ & $b$ & $s$ \\
\midrule
Mother Jones & 0.2269 & 0.1754 & 0.1024 & 1.0026 & 0.1315 & -0.6535 & 0.9846 \\
Atlantic & 0.1399 & 0.0597 & -0.0114 & 0.9665 & 0.1322 & -0.6595 & 0.9733 \\
The Hill & 0.2363 & 0.1863 & 0.1706 & 1.0893 & 0.0989 & -0.8295 & 1.1807 \\
Breitbart & 0.1849 & 0.1061 & 0.0519 & 1.0203 & 0.1164 & -0.6515 & 0.9942 \\
Gateway Pundit & 0.3785 & 0.2847 & 0.2578 & 1.0359 & 0.0948 & -0.8778 & 1.2302 \\
\bottomrule
\end{tabular}
\label{table:s7}
\end{table}

\begin{table} 
  \centering
  \caption{Computational model implementation and the list of functional forms and constants that are used to specify the model configuration, which is mainly inferred and adopted from the empirical data (see Fig. 2 in the main manuscript). The normalization constants are omitted for simplicity. Items with $^*$ indicate that the parameters are selectively used depending on the specific scenario. $\bar{H}(n)$ denotes $nH(n)$, where $H(n)=\sum_i(1/n)$. Note that, aside from these basic settings, our model has exactly two free parameters: the filter strength ($\lambda_f$) and the memory strength ($\lambda_m$). For the Rationale column, 1 is for the informed choices from the empirical data (check Fig. 2 in the main manuscript for 1$^*$ and SI appendix section 6 for 1$\dagger$), and 2 is for arbitrary choices for the simulation and model stability.}

  \begin{tabular}{lccccc}
  \toprule
  Process & Components & Functional form & Constants & Rationale\\
  \midrule
  \multirow{4}{2cm}{\centering{Network initialization}} & Initial frequency dist. ($F_f$) & $F_f(i) \propto r_i^{-\alpha_c}$ & $\alpha_c=1.0$ & 1$^*$  \\
  & Initial weight dist. ($F_{w})$ & $F_{w}(w) \propto e^{{-\ln^2(\frac{w-a}{b})}/{2s^2}}$ & $a=-0.65$, $b=1.0$, $s=0.12$ & 1$^*$ \\
  & Frequency perturbation s.d. ($\sigma_{\text{fp}}$) & Const. & $\sigma_{\text{fp}}=0.0, 1.0^*$ & 2\\
  & Weight perturbation s.d. ($\sigma_{\text{wp}})$ & Const. & $\sigma_{\text{wp}}=0.0, 0.05^*$ & 2\\
  \midrule
  \multirow{2}{2cm}{\centering{Events generation}} & Event sampling dist. ($F_{\text{ns}}$) & $F_{\text{ns}}(r_{i, t}^g) \propto -\ln(r_{i, t}^g)$ & - & 1$^*$  \\
  & Event memory strength ($\lambda_e$) & Const. & $\lambda_e=0.5$ & 2 \\
  \midrule
  \multirow{2}{2cm}{\centering{Filter definition}} & Filtering ratio ($R_1, R_2$) & Const. & $R_1=0.5$, $R_2=0.5$ & 2\\
  & Filter exponent ($\alpha_q$) & Const. & $\alpha_1=0.4$, $\alpha_1=0.2$, $\alpha_3=0.1$ & 1$^*$ \\
  \midrule
  \multirow{5}{2cm}{\centering{Response generation}} & Comment number dist. ($P_c$) & $P_c(c, x_{i, t}^{k}) \propto e^{{-\ln^2(\frac{c-a}{b})}/{2s^2}}$ & $a=5.7 \times 10^{-6}$, $b=1.0\times10^{-4}$, $s=1.5$ & 1$^\dagger$ \\
  & Zero multiplier ratio ($Z_q$) & $Z_q(r)=C_{z,q}r$ & $C_{z,1}=0.7$, $C_{z,2}=0.9$, $C_{z,3}=0.9$ & 1$^\dagger$ \\
  & \multirow{2}{2cm}{Non-zero distribution} ($P_{m, q}$) & $P_{m, q}(m, r_{z_q}^{k}, c_{i, t}^{k}) \propto e^{-\lambda_{q}(r_{z_q}^{k})m},$ & $a= C_{\text{com}}\bar{H}(r_{z_q}^{k})/c_{i, t}^k$, $b=\bar{H}(r_{z_q}^{k})$ & 1$^\dagger$ \\
  & & $m \in [a, b]$ & $C_{\text{com}}=1.0\times10^{-6}$ & 1$^\dagger$\\
  & Non-zero exponent ($\lambda_q$)& $\lambda_q(r_{z_q}^{k}) = a_qe^{-b(r_{z_q}^{k})}$ & $a_1=0.005$, $a_2=0.01$, $a_3=0.02$, $b=0.8$ & 1$^\dagger$\\
  \midrule
  \multirow{3}{2cm}{\centering{Network update}} & Learning rate ($\eta$) & Const. & $\eta=10.0$ & 2\\
  & Maximum weight ($w_{\max}$) & Const. & $w_{\max}=0.8$ & 1$^\dagger$\\
  & Weight noise s.d. ($\sigma_{\text{wn}}$) & Const. & $\sigma_{\text{wn}}=0.001$ & 2\\
  \bottomrule
  \end{tabular}
  \label{table:s1}
  \end{table}

\begin{table}[t]
\footnotesize
\centering
\caption{Outlier$/0$ ratio of TM hyperparameter grid search (coarse-grained), bold-faced values indicate the lowest ratio for each community.}

\begin{tabular}{lcccccccccc}
\toprule
\multirow{2}{*}{Name}  & \multicolumn{3}{c|}{$c=200$} & \multicolumn{3}{c|}{$c=300$} & \multicolumn{3}{c|}{$c=400$}\\
\cline{2-4} \cline{5-7} \cline{8-10}
& $s=30$ & $s=60$ & $s=90$ & $s=30$ & $s=60$ & $s=90$ & $s=30$ & $s=60$ & $s=90$ \\
\midrule
Global & 0.7199 & 0.8976 & 0.8172 & \textbf{0.6373} & 0.7971 & 0.8875 & 0.8581 & 0.8888 & 0.7105 \\
Mother Jones & 0.8214 & 0.9905 & 0.9881 & 0.8073 & 0.8299 & 0.8807 & 0.8474 & 0.8341 & \textbf{0.7123} \\
Atlantic & 0.8876 & 0.8381 & \textbf{0.7757} & 0.8879 & 0.8056 & 0.8993 & 0.8887 & 0.9508 & 0.8836 \\
The Hill & 0.6954 & 0.6915 & 0.8501 & 0.6597 & 0.6758 & \textbf{0.6116} & 0.6883 & 0.7051 & 0.7588  \\
Breitbart & \textbf{0.5291} & 0.6151 & 0.5929 & 0.6153 & 0.6308 & 0.7253 & 0.7199 & 0.6720 & 0.6707 \\
Gateway Pundit & 0.8314 & 0.7894 & 0.8257 & 0.7850 & 0.8255 & 0.8313 & \textbf{0.6875} & 0.8221 & 0.7885 \\
\bottomrule
\end{tabular}

\label{table:s5}
\end{table}

\begin{table}[t]
  \footnotesize
  \centering
  \caption{Outlier$/0$ ratio of TM hyperparameter grid search (coarse-grained). The DBCV rank is calculated among the top $5$ candidates. Cases, where the DBCV rank is not $1$st, indicate that the higher rank models were discarded due to the topic separation check.}

  \begin{tabular}{lcccc}
  \toprule
  Name & coarse-grained result $(n_1, c_1)$ & fine-grained result $(n_2, c_2, s)$ & DBCV (rank) \\
  \midrule
  Global & $(300, 30)$ & $(325, 20, 1)$ & 0.2901 ($1$st) \\
  Mother Jones & $(400, 90)$ & $(425, 90, 5)$ & 0.3918 ($1$st) \\
  Atlantic & $(200, 90)$ & $(200, 80, 4)$ & 0.1958 ($2$nd) \\
  The Hill & $(300, 90)$ & $(300, 80, 2)$ & 0.2695 ($1$st) \\
  Breitbart & $(200, 30)$ & $(225, 20, 3)$ & 0.2496 ($1$st)\\
  Gateway Pundit & $(400, 30)$ & $(400, 30, 4)$ & 0.2305 ($3$rd) \\
  \bottomrule
  \end{tabular}

  \label{table:s6}
\end{table}

\begin{sidewaystable} 
\centering
\begin{tabular}{cccc}
\toprule
Rank & Topic & Size & Representative keywords \\
\midrule
1 & youre & 24.6M & youre, apple, stupid, im, dont, just, know, thats, like, lol \\
2 & election & 13.9M & obama, trump, democrats, election, republicans, president, gop, vote, party, republican \\
3 & tax & 10.9M & tax, taxes, economy, money, income, debt, pay, jobs, rich, spending \\
4 & school & 8.95M & school, students, schools, education, college, teachers, kids, student, teacher, parents \\
5 & shes & 7.98M & shes, hillary, pelosi, clinton, woman, nancy, palin, clintons, sarah, trump \\
6 & black & 6.87M & black, white, racist, blacks, police, racism, race, whites, cops, slavery \\
7 & muslims & 6.35B & muslims, islam, muslim, saudi, isis, islamic, allah, afghanistan, taliban, terrorists \\
8 & god & 5.88M & god, church, jesus, catholic, pope, christian, bible, religion, christ, christians \\
9 & news & 5.59M & news, fox, cnn, media, breitbart, fake, propaganda, msnbc, press, journalism \\
10 & california & 5.28M & california, texas, city, state, states, blue, mayor, red, cities, detroit \\
\end{tabular}
\caption{Top $10$ topics and their respective representative keywords for the global TM, ranked by their frequency after excluding outliers.}
\label{table:st}
\end{sidewaystable}

\begin{table}[t]
\footnotesize
\centering
\caption{Additional fitting parameters for the empirical data (Global TM)}
\begin{tabular}{lcccc}
\toprule
\multirow{2}{*}{Name} &\multicolumn{3}{c|}{$\#$ of comment} \\
\cmidrule{2-4} 
& $a (\times 10^{-5})$ & $b (\times 10^{-4})$ & $s$ \\
\midrule
Mother Jones & 26.6973 & 29.9659 & 1.7790 \\
Atlantic & 12.3054 & 14.4338 & 2.0085 \\
The Hill & 0.5928 & 1.0383 & 2.1505 \\
Breitbart & 0.6471 & 1.0575 & 2.2161 \\
Gateway Pundit & -1.8909 & 8.8054 & 0.7850 \\
\bottomrule
\end{tabular}
\label{table:s2}
\end{table}

\begin{table}[h!]
\footnotesize
\centering
\caption{Fitting parameters for the empirical data in Fig. \ref{fig:s10} (Local TM)}
\begin{tabular}{lccccccc}
\toprule
\multirow{2}{*}{Name} & \multicolumn{3}{c|}{Title} & {Comment} &\multicolumn{3}{c|}{Similarity} \\
\cmidrule{2-4} \cmidrule{5-5} \cmidrule{6-8} 
& $\alpha_1$ & $\alpha_2$ & $\alpha_3$ & $\alpha_c$ & $s$ & $a$ & $b$ \\
\midrule
Mother Jones & 0.2670 & 0.2373 & 0.1603 & 1.4487 & 0.1365 & -0.6650 & 0.9586 \\
Atlantic & 0.0951 & -0.0239 & -0.0946 & 0.9831 & 0.1798 & -0.5377 & 0.8108  \\
The Hill & 0.1908 & 0.1471 & 0.0765 & 0.9076 & 0.1307 & -0.5675 & 0.9146  \\
Breitbart & 0.1838 & 0.1491 & 0.2533 & 0.8983 & 0.1177 & -0.6546 & 0.9869 \\
Gateway Pundit & 0.2321 & 0.2854 & 0.2533 & 0.9870 & 0.1178 & -0.6501 & 0.9985 \\
\bottomrule
\end{tabular}
\label{table:s8}
\end{table}